\documentclass[12pt]{article}\usepackage[]{graphicx}\usepackage[]{xcolor}
\makeatletter
\def\maxwidth{ %
  \ifdim\Gin@nat@width>\linewidth
    \linewidth
  \else
    \Gin@nat@width
  \fi
}
\makeatother

\usepackage{Sweave}

\usepackage{times}
\usepackage{mystyle}
\IfFileExists{upquote.sty}{\usepackage{upquote}}{}
\begin{document}
\maketitle
\vspace{-30pt}
\begin{abstract}
% no more than 200 words
Spatial misalignment arises when datasets are aggregated or collected at different spatial scales, leading to information loss. We develop a Bayesian disaggregation framework that links misaligned data to a continuous-domain model through an iteratively linearised integration scheme implemented with the Integrated Nested Laplace Approximation (\inla). The framework accommodates different ways of handling observations depending on the application, resulting in four variants: (i) \textit{Raster at Full Resolution}, (ii) \textit{Raster Aggregation}, (iii) \textit{Polygon Aggregation} (PolyAgg), and (iv) \textit{Point Values} (PointVal). The first three represent increasing levels of spatial averaging, while the last two address situations with incomplete covariate information. For PolyAgg and PointVal, we reconstruct the covariate field using three strategies — \textit{Value Plugin}, \textit{Joint Uncertainty}, and \textit{Uncertainty Plugin} — with the latter two propagating uncertainty.

We illustrate the framework with an example motivated by landslide modelling, focusing on methodology rather than interpreting landslide processes. Simulations show that uncertainty-propagating approaches outperform \textit{Value Plugin} method and remain robust under model misspecification. Point-pattern observations and full-resolution covariates are therefore preferable, and when covariate fields are incomplete, uncertainty-aware methods are most reliable. The framework is well suited to landslide susceptibility modelling and other spatial mapping tasks, and integrates seamlessly with \inla-based tools.

% This study is motivated by landslide modelling, where occurrences are often aggregated into slope unit counts, reducing spatial detail. Results on simulated data show that point pattern observations and full-resolution covariate fields should be prioritised, while for incomplete fields, methods that propagate uncertainty are preferred. This framework supports landslide susceptibility and other spatial mapping, integrating seamlessly with \inla \ and its extension packages.\\
\end{abstract}
% \Keywords{INLA, data misalignment, approximate Bayesian computation, uncertainty quantification}
% \Plainkeywords{INLA, data misalignment, approximate Bayesian computation, uncertainty quantification}

% !TEX root = paper.Rnw

\section{Introduction}

Information may exist everywhere, but data are rarely available at all locations in space and time. Depending on collection methods, data are measured at individual point locations and various spatial scales, such as partitioning the domain of interest. This leads to spatial misalignment, and when inference is needed at a different scale, it results in a change-of-support problem (COSP; \citet{gelfand2001change}). Traditionally, misaligned data are aligned to a chosen spatial partition, with each element termed a “block”, which often requires aggregation or disaggregation, and modelled using block-kriging \citep{gotway2002combining}.  % Spatial misalignment is a broader data integration issue, while COSP is a specific modelling problem arising due to differing supports. 

Early work includes \citet{fuentes2005model}, who proposed a Bayesian hierarchical framework for fusing areal and point-referenced data, extended by \citet{berrocal2010spatio} through hierarchical downscaling for multi-resolution data, and by \citet{wilkie2019nonparametric} to spatio-temporal supports using nonparametric methods. However, these approaches requires spatially varying coefficients, resulting in high computational burden due to the large number of hyperparameters estimated via Markov Chain Monte Carlo. Recent applications across diverse domains increasingly rely on approximate Bayesian inference through Integrated Nested Laplace Approximation (INLA) \citep{rue2009approximate}, implemented in packages such as \texttt{R-INLA} and \inlabru{} \citep{lindgren2024inlabru}, for continuous-domain models \citep{moraga2017geostatistical, cameletti2019bayesian, forlani2020joint, villejo2023data, zhong2024bayesian}. These models yield a continuous intensity field, which characterises the spatial (or spatio-temporal) variation in the expected density of events across a domain.

When it comes to spatial point process and continuous domain model, inference at block resolution require averaging misaligned data within each block which leads to the issues of aggregation and specification biases raised by \citet{konrad1982effects}. The main problem with these methods is that averaging over blocks results in loss of point-level information, making the underlying signal less accessible. Moreover, both modelling and inference are constrained by the chosen block structure, often defined by grid cells, which limits flexibility. At fine resolutions, spatial misalignment can render computations intractable and obscure the interpretation of results.  Additionally, these methods are not easily transferable to unseen study regions.

%%%%%%%
% In this context, the use of \texttt{R-INLA} and \inlabru{} has grown rapidly owing to their computational efficiency.
% including applications in earthquake analysis \citep{bayliss_data-driven_2020, naylor_bayesian_2022}.
%%%%%%%%%%%

A key motivation for this study is its relevance to landslide susceptibility modelling, where landslide polygons are often reduced to centroids and both point locations and covariates are subsequently aggregated into areal counts and averages \citep{lombardo_point_2018, lombardo_3_2019, loche_landslide_2022, opitz_high-resolution_2022}. A major challenge in this setting is the spatial misalignment between point-pattern observations and covariates measured at varying spatial resolutions. A common strategy is to aggregate landslide centroids into counts within predefined blocks, such as ``slope units" \citep{alvioli_automatic_2016, alvioli_parameter-free_2020}, which are a series of finite width polygons bounded by topographic ridges at the top and rivers at the bottom such that they define regions within which landslides are typically geometrically constrained. This type of model could, for example, considering the counts and sizes of landslides aggregated within each slope unit. However, this aggregation inherits the limitations noted above: specification bias arises because aggregated count models rely on proxy covariates confounded with slope units, while aggregation bias arises from discarding landslide location information, adding uncertainty. Retaining point locations is fundamental in point pattern analysis. Although landslides are not inherently point events and their manually registered locations carry potential errors, avoiding aggregation prevents further uncertainty when constructing continuous susceptibility maps. Moreover, aggregation renders uncertainty propagation intractable, as exact event locations are lost within counts, and individual event uncertainties and covariate resolution effects are blended into summary statistics, e.g.\ means or medians.

 % Moreover, defining slope-unit scales and mapping landslides require manual adjustments to terrain and area parameters, which are not consistent or transparent across large domains. Consequently, assuming that slope units remain constant and unbiased for post-seismic landslide modelling may be unrealistic. Recent research shows that slope units do not systematically preserve post-seismic landslide susceptibility patterns \citep{kincey_modelling_2023}, which describe the likelihood of landslides across space or space–time, modelled as spatial random fields conditioned on triggering factors (see Section \ref{sec:def}).

%%%%%%%%%%%%%%%%%%%%%%%%%%%%%%%%%%%%%%%%%%%%%%%%%%%%%%%%%%%%%%%%%%%%%

In practice, only a limited subset of the covariate field is often observed, requiring either joint reconstruction with the intensity field (one-stage) or sequential estimation (two-stage). The one-stage approach is less common in the literature due to the undesirable feedback in joint estimation for spatial models. Examples include autoregressive disturbances with spatial autoregressive models \citep{kelejian1998generalized}, spatial disaggregation with environmental feedback on land-use transport models \citep{spiekermann2008environmental}, and two-stage approach on incorporating spatial lag autocorrelation \citep{lambert2010two}. However, it remains unclear in what situation and how the one-stage and two-stage approach would affect the model performance in the more flexible \inla-SPDE setting since the neighbouring structure is much finer in scale that depends on the mesh \citep{rue_advanced_2018}.

%%%%%%%%%%%%%%%%%%%%%%%%%%%%%%%%%%%%%%%5

In this paper, we address how to 
\begin{enumerate*}[label=(\roman*\upshape)]
    \item strike a balance between computation and model accuracy in aligning the point patterns with areal data, i.e.\ data misalignment, via linearisation;
    \item model aggregated count and covariate data, incomplete covariate fields (e.g.\ missing data and downscaling) and incorporate the uncertainty of these fields into the model; and,
    \item handle model misspecification for nonlinear transformations of incomplete covariate fields.
\end{enumerate*}

\section{Definition of Poisson Point and Count Processes}\label{sec:def}

In this paper, we consider unmarked point process and a bounded domain $\Omega \subset \R^d,  \, d \in \mathbb{N}$. However, our framework for aggregated data is extendable to marked point process, as well as space and time $\Omega \times T$ settings.
\begin{definition}[Poisson Point Process]\label{def:ppp}
% or Let $Y$ be a countable set of points in $\R$ and let N_Y (A) = \textsf{card} (Y \cap A)$ be the counts of points in bounded domains $A \subset \R^d$. The counts are 
Let $Y \subset \Omega$ be a countable random subset of points. We define a Poisson point process for $Y$ with an intensity function $\lambda: \Omega \to [0, \infty)$ via the counts of points for every Borel measurable set $A \subseteq \Omega$, $N_Y (A) = \textsf{card} (Y \cap A)$, where $\textsf{card}(\cdot)$ denotes the cardinality of a set. The counts are Poisson distributed with mean $\Lambda (A)$, which is assumed to be non-negative and finite,
\begin{equation}
N_Y(A) \sim \textsf{Pois} \left[ \Lambda(A) \right] \text{, and }
\Lambda (A) = \int_{\s \in A} \lambda (\s) \, d\s, \label{eq:point}
\end{equation}
where $\lambda(\cdot)$ is the intensity function. 
\end{definition}
If the log intensity function of a spatial point process is a Gaussian random field, this point process is called log-Gaussian Cox process (LGCP) \citep{moller_log_1998, diggle_spatial_2013}. In the case of aggregated count observations, we extend the Poisson point process to a Poisson count process.
\begin{definition}[Poisson Count Process]\label{def:pcp}
Let $\{A_p\}_{p = 1}^{n_A}$ be disjoint and bounded aggregation sets (regions/volumes) and $N_p$ be the count of points in $A_p \subset \Omega$ ,  
\begin{equation}
 N_p := N_Y(A_p) \sim \textsf{Pois}\left(\int_{A_p} \lambda(\s) \, d\s\right). \label{eq:npois}
\end{equation}
We define a Poisson count process $\mathcal{N}:= \{N_p\}$ with intensity function $\lambda: \Omega \to [0, \infty)$. 
\end{definition}
Suppose that we have a point pattern observation $\mathcal{Y} = \{y_1, \dots, y_N\}$, where $N$ is a random variable, the density of $\mathcal{Y}$ for $N=n$ given the intensity function $\lambda$ is (see Appendix \ref{appendix:intensity} for derivation), 
\begin{equation*}
    p_Y ( \mathcal{Y}, N = n | \lambda) = \frac{ e^{-\Lambda(\Omega)}}{n!}\prod_{i=1}^n \lambda(y_i). 
\end{equation*}
\begin{definition}[Log-likelihood of Poisson Point Process]\label{def:loglik}
The likelihood can be defined as a ratio with respect to a homogeneous Poisson point process, i.e.\  $\lambda\equiv 1$,
\begin{align}
    \mathcal{L}(\lambda|\mathcal{Y}) &:= \frac{p_Y(\mathcal{Y}, N = n|\lambda)}{p_Y(\mathcal{Y}, N=n|\lambda\equiv 1)}= e^{|\Omega| - \Lambda(\Omega)} \prod_{i=1}^n \lambda(y_i). \label{eq:lik}
    % \\ &\propto e^{-\Lambda(A)} \prod_{i=1}^n \lambda(y_i)
\end{align}
The log-likelihood of a Poisson Point Process is
\begin{equation}
    \ell(\lambda| \mathcal{Y}) = \log \mathcal{L}(\lambda| \mathcal{Y}) =
    % cant use \propto, because ... TODO read inlabru vignette as well
    \underbrace{|\Omega|}_{\text{domain volume}}\underbrace{-\int_{\Omega} \lambda (\s) \,d\s}_{\text{domain contribution}} + \underbrace{\sum_{i=1}^n \log \lambda (y_i)}_{\text{observed point contribution}}.
    \label{eq:loglik}
    % \text{observation part of the log likelihood}
\end{equation}
    % where $\bm{x}$ is the observed data. 
\end{definition}
\begin{definition}[Log-likelihood of Poisson Count Process]
We define the log-likelihood of a Poisson count process (See Appendix \ref{appendix:intensity} for derivation)
\begin{equation}
    \ell(\lambda| \mathcal{Y}, \{A_p\}) =
    \underbrace{|\Omega|}_{\text{domain volume}}\underbrace{-\int_{\Omega} \lambda (\s) \,d\s}_{\text{domain contribution}} + \underbrace{\sum_{p=1}^{n_A} \log \lambda (A_p)}_{\text{observed count contribution}}.
    \label{eq:loglik_count}
\end{equation}
\end{definition}
It is straightforward to compute the observed contribution once we define a log-intensity function as a linear model and extendable to a monotonic increasing nonlinear model. The domain contribution is analytically intractable and would be ideally approximated in a computationally efficient manner \citep{simpson_going_2016}. Having a triangulation over the domain, we can approximate the spatial random effect in a latent field $\s \in \R^d $ with $K$ vertices via a basis expansion,
 \begin{equation}
         u(\s) = \sum^{K}_{k=1} \psi_k(\s) u_k, \text{ and } \bm{u} = (u_1, \dots, u_k) \sim \textsf{N}(\bm{0},  % , \label{eq:u} 
         \bm{Q}^{-1}), \label{eq:psi} 
 \end{equation}
where $\{\psi_k\}$ is a set of piecewise linear basis function and $\bm{u}$ are Gaussian distributed weights with a sparse covariance matrix $ \bm{Q}^{-1}$. For illustration, we define the linear predictor in a purely spatial setting that is extendable to space-time setting, 
\begin{align}
\log \lambda(\s) \approx \eta(\s)
&:= \bm{1}\beta_0 + \beta_x x(\s)
% f\big(X_1(\s)\big)
+ u(\s), \text{ and} \label{eq:eta1}\\  
\left[ \log \lambda(s_i) \approx \eta(s_i), i = 1,\dots, n \right]^\intercal &:= \bm{A}\bm{v}, \label{eq:discret}
\end{align}
where  $\beta_0$ is an intercept, $\beta_x$ is the coefficient of the linear fixed effect covariate, $x(\s)$ is the covariate field; $\bm{A}$ is a projector matrix, and $\bm{v}= [\beta_0, \beta_x, \bm{u}^\intercal]^\intercal$. The product $\bm{Av}$ represents the evaluation of the field defined by the mesh at any given finite set of locations. Given the data, the model can be fitted via the Integrated nested Laplace approximations (INLA) \citep{rue_advanced_2018}. Here the predictor $\eta(\s)$ approximates $\log \lambda (\s)$ and this will be explained in Section \ref{sec:com}. 

For theory, $u(\cdot)$ is defined as a stationary Gaussian Random Field (GRF) with zero mean and  Mat\'ern covariance function or kernel,
\begin{equation}
    \varrho(\s,\s') = \frac{\sigma^2}{\Gamma(\nu)2^{\nu-1}}(\kappa \|\s-\s'\|)^{\nu} K_\nu (\kappa \|\s-\s'\|), \label{eq:matern1}
\end{equation}
where $\sigma^2 > 0$ is the marginal variance, $\nu > 0$ is the smoothness parameter, $\kappa > 0$ is the the range parameter and $K_\nu$ is the modified Bessel function of second kind. We evaluate the field $u(\cdot)$ via mapping to the stochastic partial differential equation (SPDE) \citep{whittle1954stationary, whittle1963stochastic, lindgren_explicit_2011}. We follow the re-parametrisation of the Penalised Complexity (PC) prior in \citet{fuglstad2015interpretable} by setting $\nu = \alpha - \frac{d}{2}>0$ and $\kappa = \frac{\sqrt{8\nu}}{\rho}$, with $\rho$ as the spatial range, i.e.\
\begin{equation}
  (\kappa^2 -  \Delta)^{\frac{\alpha}{2}} \left[\sqrt{\tau} u(\s)\right] = \mathcal{W}(\s), \label{eq:spde}
\end{equation}
where $\Delta = \sum_{i=1}^{d} {\partial^2}/{\partial s_i^2}$ is the Laplacian operator, $\tau = \Gamma(\nu)\left[(4\pi)^{\frac{d}{2}}\Gamma(\nu+\frac{d}{2})\sigma^2\kappa^{2\nu}\right]^{-1} > 0$ is the scaling parameter of $u(\cdot)$ with $d$ denoting the dimension of the space and $\mathcal{W}(\s)$ as spatial white noise. In this formulation, the covariance function in equation \eqref{eq:matern1} can be rewritten as
\begin{equation}
    \varrho(d_s := \|\s-\s'\|) = \frac{\sigma^2}{\Gamma(\nu)2^{\nu-1}}\left(\frac{\sqrt{8\nu}d_s   }{\rho}\right)^{\nu} K_\nu \left(\frac{\sqrt{8\nu}d_s}{\rho}\right). \label{eq:matern2}
\end{equation}

\section{Computation}\label{sec:computation}
We construct a Taylor approximation of the intensity contribution, to address spatial misalignment through disaggregation (Section~\ref{sec:taylorint}), then consider disaggregation from a domain partitioning perspective (Section~\ref{sec:partition}), and finally assess the effects of linearisation (Section~\ref{sec:com}).

\subsection{Taylor Approximation for Intensity Contribution}\label{sec:taylorint}
% We partition the sample space $\Omega$ (defined in Section \ref{sec:def}) into disjoint subsets $\Omega_j$, such that $\Omega = \bigcup_{j=1}^J \Omega_j, \textrm{ and } \Omega_j \cap \Omega_{j'} = \varnothing, \; j\neq j'$. The flexibility on how to define $\Omega_j$ based on different scenarios will be explained in Section \ref{sec:partition}. 
We partition the sample space $\Omega$ (Section \ref{sec:def}) into disjoint subsets $\Omega_j$, such that $\Omega_j \cap \Omega_{j'} = \varnothing$ for $j \neq j'$, $\Omega = \bigcup_{j=1}^J \Omega_j$, with further specifications discussed in Section \ref{sec:partition}. We consider aggregated count data over regions defined by sets $A_p$ in equation \eqref{eq:npois}, for each $\Omega_j$, $\Omega_j \subseteq A_p$.
% and for $p \neq p'$ , $\Omega_j \cap A_{p'} = \varnothing$. TODO nice thing to have in discussion
Let $M_j(\bm{u})=\log \left\{ \int_{\Omega_j} \exp[ \eta (\s) ] \,d\s \right\}$ in the $j$-th subset, we approximate the logarithmic integral of the intensity function for each subset $\Omega_j$ with the first-order Taylor series,
% define normal M and say that in the text had tilde M and bar M
\begin{equation}
     M_j(\bm{u}) \approx M_j(\bm{u}_*) + \bm{J}^{(j)}(\bm{u}_*)^\intercal (\bm{u}-\bm{u}_*):= \ol{M}_j(\bm{u}),\label{eq:M_j}
    \end{equation}
where $\bm{u_*}$ are the linearisation points of the Taylor approximation, and $\bm{J}^{(j)}(\cdot)$ is the Jacobian matrix, i.e.\ first-order derivative matrix for $M_j$ with respect to $\bm{u}$. Although we ignore the remaining higher-order terms here, we can assess the approximation by checking the leading error term.

Given a stable integration scheme (see Appendix \ref{appendix:int}), we discretise the integrals for each $\Omega_j$ with weights $w_{jk} > 0$ at the $k$-th knot. Similarly, let $m_j(\bm{u}) =\log \left\{\sum_{k=1}^{n_j} w_{jk} \exp \left[ \eta (\s_{jk}) \right] \right\}$, with the first-order Taylor series, 
    \begin{equation}
    m_j(\bm{u}) \approx m_j(\bm{u}_*) + \bm{J}^{(j)}(\bm{u}_*)^\intercal (\bm{u}-\bm{u}_*) =: \ol{m}_j(\bm{u}). \label{eq:m_j}
\end{equation}
Derivations of $\ol{M}_j(\bm{u})$ and $\ol{m}_j(\bm{u})$ in equations \eqref{eq:M_j} and \eqref{eq:m_j} can be found in Appendix \ref{appendix:mj}.

\begin{theorem}[Taylor approximation for the continuous and discretised linearisation]\label{thm:taylor}
Assume both $M_j(\bm{u})$ and $m_j(\bm{u})$ are twice differentiable at some points $\bm{u}_* \in \R$; and the linear predictor $\eta(\cdot)$ is an affine function of $\bm{u}$. For the continuous linearisation, we have 
\begin{equation}
    M_j (\bm{u}) - \ol{M}_j (\bm{u}) = \frac{1}{2} (\bm{u}-\bm{u}_*)^\intercal \bm{H}_{M}^{(j)} (\bm{u}-\bm{u}_*) + \mathcal{O}(\| \bm{u} - \bm{u}_*\|^3), \label{eq:taylor_c}
\end{equation}
where $\bm{H}_{M}^{(j)}$ is the Hessian of $M_j(\bm{u})$ and $\mathcal{O}(\cdot)$ is the higher-order terms with respect to $(\cdot)$.
For the discretised linearisation, we have
\begin{equation}
    m_j (\bm{u}) - \ol{m}_j (\bm{u}) = \frac{1}{2} (\bm{u}-\bm{u}_*)^\intercal \bm{H}^{(j)} (\bm{u}-\bm{u}_*) + \mathcal{O}(\| \bm{u} - \bm{u}_*\|^3), \label{eq:taylor_d}
\end{equation}
where $\bm{H}^{(j)}$ is the Hessian of $m_j(\bm{u})$. Derivations for \eqref{eq:taylor_c} and \eqref{eq:taylor_d} can be found in Appendices \ref{appendix:taylor_c} and \ref{appendix:taylor_d}. 
 \end{theorem}
 
Hence, the domain contribution can be approximated as 
\begin{equation}
    %     -\int_{\Omega} \lambda (\s) \,d\s 
    % &\approx - \sum_{j=1}^J \exp  \left( \ol{M}_j(\bm{u}) \right) \label{eq:eta} \\ 
    %     &\approx - \sum_{j=1}^J \exp \left( \ol{m}_j(\bm{u}) \right).\label{eq:mj}
        -\int_{\Omega} \lambda (\s) \,d\s \approx
  \begin{cases}
    - \sum_{j=1}^J \exp  \left[ \ol{M}_j(\bm{u}) \right]  \\ 
    - \sum_{j=1}^J \exp \left[ \ol{m}_j(\bm{u}) \right]. \label{eq:mj}
    % \label{eq:eta}
  \end{cases}
\end{equation}
We will compare the accuracy between $\ol{M}_j(\bm{\cdot})$ and $\ol{m}_j(\bm{\cdot})$ approximations in equation \eqref{eq:mj} in Section \ref{sec:com}. See Appendix \ref{appendix:agg} for the implementation of the logarithmic sum, i.e.\ $\ol{m}_j(\bm{\cdot})$ approximation.
 
\subsubsection{Taylor Approximation for Discretised Linearisation}\label{sec:discret}

In the \inlabru \ package, the linearisation approach in equation \eqref{eq:mj} requires a sequence of runs of the \inla \ method, each followed by a line search to locate the optimal linearisation point. If the approximation of the domain contribution is not large enough to counter the observed point contributions in equation \eqref{eq:loglik}, the approximation of the likelihood can go to infinity and cause numerical instability (see Appendix \ref{appendix:dof}). To avoid the observed contribution overpowering the likelihood, we need to bound the domain contribution. Hence, Theorem \ref{thm:taylor} shows that the Jacobian $\bm{J}^{(j)}$ and Hessian $\bm{H}^{(j)}$ terms of $m_j(\bm{u})$ converge in equation \eqref{eq:taylor_d}. The line search continues until the given tolerance is met and refer to \citet{lindgren2024inlabru} with regard to setting the initial linearisation point and the stopping rule to optimise the fit.

\subsection{Domain Partition}\label{sec:partition}
The subsets $\{\Omega_j\}$ from Section \ref{sec:taylorint} define the resolution of the intensity field in the outcome prediction. Their volumes have to be smaller than the corresponding $|A_p|$ (see Definition \ref{def:pcp}). There is no benefit in using subsets smaller than the finest data resolution. Here, we define the subsets using the user-specified mesh, which serves as a computational tool for solving the SPDE in equation \eqref{eq:spde} via the Finite Element Method (FEM) \citep{lindgren_explicit_2011,lindgren_spde_2022}. The Mat\'ern field term $u(\bm{s})$ in equation \eqref{eq:eta1} has to be bounded by the covariate field in the resolution scale since there may be no observation in the mesh element and the Gaussian prior is not informative enough without any data input. Essentially, the degrees of freedom in the field should be respected (see Section \ref{sec:lim}). Nonetheless, users may downscale the resolution of the covariate data and refine the mesh size accordingly. Under these constraints, the flexibility to define the subsets then takes into account the observed point locations, the data quality, the integration scheme and the computation costs. 

\subsubsection{Shape of Mesh Elements}
In light of the tessellation of mesh elements over the domain, we want the subsets to be regularly shaped, i.e.\ triangles, squares and hexagons. We do not consider mesh-free methods which treat the data points as nodes. Mesh-free methods can handle particle-like finite element analysis, which can be useful for animal tracking since animals interact with one another \citep{liu2009meshfree}. However, due to these interactions and dynamical nodes, more computation time is required. Additionally, further handling is needed for a stable integration scheme with the current construction of the likelihood. 
% Overlapping of meshes may smooth out the edge effects because there are weighted contributions from different meshes towards the overlapping region. However, we do not consider it here because it complicates the computation.
Mixed-shaped meshes are not considered because their construction can be too flexible and thus more complicated to implement.  

Discretisation and orientation effects are undesirable. In this regard, circles do not suffer from orientation bias. However, circles do not tessellate. Hexagon meshes suffer less from orientation effects since they are almost circular, and each hexagon has six neighbours and smooths out the discrepancy between nearby hexagons. In general, hexagonal meshes fit curved boundary domains well, and have good accuracy and solving time. We take advantage of this by constructing equilateral triangles joining the centres of neighbouring hexagons. These are then joined with irregular triangular elements between the hexagon centres and the domain boundary.  To avoid spatial misalignment with raster data, square elements can alternatively be used to match the raster grid (e.g.\ satellite imagery) and can be constructed from triangular mesh elements.

\subsubsection{Data resolution and mesh element size}\label{sec:datares}
To define the mesh element size, we take the data resolution and spatial correlation length as baselines.
In order to accurately represent the correlation structure, the mesh elements have diameter smaller than the correlation length. As an anectodal rule of thumb, diameters a factor 5--10 smaller than the correlation length gives highly accurate approximations, but in practice, larger elements can be sufficient. Given that constraint, further refining the elements finer than the data does not improve the accuracy of point estimates, but can improve posterior uncertainty estimates on fine scales. Choosing the mesh resolution is a compromise between computation time and approximation error. Finer elements may increase accuracy but also computational cost and time. In real applications, we do not know the exact spatial range and anisotropicity of the data a priori, and we often only observe a limited set of data. Hence, we may not know the optimal balance between mesh resolution and computation time. In addition, any adjustments on maximum element edge in the meshing construction criteria changes the structure of the mesh.  Since there are multiple factors, we do not know exactly how the mesh refinements affect accuracy. As a general rule, we need to make sure that the resolution of the mesh is sufficiently fine within the domain of interest to represent the between-observation location variability.

%; in practice, a reasonable rule of thumb is one third in length for the ratio between mesh element and data resolution
% (see Appendix \ref{appendix:rast}). We will demonstrate this in a simulation study in Section \ref{sec:sim}.

We can consider uniform and non-uniform meshes. 
A uniform mesh places uniformly sized elements across and is easy to implement.  However, in order to avoid boundary effects from the SPDE construction \citep{lindgren_explicit_2011}, we need to extend the mesh beyond the domain of interest. A uniform extension is computationally wasteful, so it is normally preferable to use an extension with non-uniform elements of increasing size. This minimises the computational cost, while keeping a high accuracy inside the domain of interest.
An non-uniform mesh, i.e.\ with varying element shapes and sizes, can also be used
to adapt the mesh resolution to geographical features of the desired domain space.
In landslides, we mostly care about regions with human activities. For instance, we would like
to know the landslides susceptibility for planning road
construction. Adaptivity can be achieved through mesh refinement
\citep{rue_advanced_2018}, with the target mesh quality varying across space.
In some cases, we have higher resolution
data i.e.\ less uncertainty and more information, at observation
sites, such as hospital data in urban areas compared to rural areas, and
animal tracking in ecology. To reflect this need, we can place finer
elements at these locations without creating nonconforming meshes
during refinement. As discussed by \citet{lindgren_explicit_2011},
the mesh quality controls how closely the discretised spatial field
representation lies to the continuous domain model, which can in principle be used to
give error bounds. There is further discussion on computing on fine lattices in
Sections 2 and 7.3 of \citet{simpson_going_2016}, as well as a detailed error analysis.
As seen in Table~\ref{tab:cellsize}, the computational mesh resolution is here chosen to
be coarser than the finest covariate raster resolution, but finer than the aggregate data resolution.

For integration, the $k$-th knot and weight ($w_{jk}$ from Section \ref{sec:taylorint}) in the $j$-th subset is defined through an integration scheme that is aligned with the mesh triangle, but can have a denser resolution than the mesh itself. As the total number of knots goes to infinity, the accuracy eventually does not continue to improve given the data and modelling mesh resolutions. Hence, we choose the integration scheme resolution so that the likelihood approximation is stable, accurate and computationally efficient.

\subsection{Assessment of Effects of Linearisation}\label{sec:com}
% \subsection{Nonlinear case for Taylor approximation}\label{sec:eta}
If $\eta$ is a nonlinear expression, i.e.\ non-linear predictor, say $f(x) = x^{-2}$, there would be further expansion of $\bm{A}$ from $\bm{b}+\bm{Au}$ which can be grouped into higher-order terms of $\|\bm{u}-\bm{u}_*\|$ after differentiation (see Appendix \ref{appendix:nonlinear}). Hence, the first-order Taylor expansion remains adequate.
% We will look into the difference of the non-linearisation and linearisation in the Taylor series later.

% \subsection{Accuracy in Taylor approximation}\label{sec:diff}
Before we proceed to the assessment of nonlinearity, we take a detour to introduce the structure of hierarchical models in \inla. We denote $\bm{y}$ as the realisations of $\mathcal{Y}$ which are conditionally independent given $\bm{\eta}$ and $\bm{\theta}$ in the \inla \ setting. The \inla \ implementation requires incorporating the covariance parameters $\bm{\theta}$ into equation \eqref{eq:psi} such that 
$\bm{\theta} \sim p(\bm{\theta}); \,
\bm{u}|\bm{\theta} \sim \mathsf{N}\!\left(\bm{\mu}_u, \bm{Q}_{\bm{\theta}}^{-1}\right) \text{, and } 
\bm{y}|\bm{u},\bm{\theta} \sim p(\bm{y}|\eta(\bm{u}),\bm{\theta}).$

We aim to assess how well the linearisation of the predictor works. To this end, we expand the second-order Taylor polynomial terms and compute the bias. We define $\log \mathbb{\wt{P}}(\bm{u}|\bm{y},{\bm{\theta}})$ as the log-density of the true model expressed in the second-order Taylor expansion and higher-order terms $ \mathcal{O}(\| \bm{u} - \bm{u}_*\|^3)$, and $\log \mathbb{\ol{P}}(\bm{u}|\bm{y},{\bm{\theta}})$ as the log-density of the linearised model yielded from our proposed method. We assess the Kullback-Leibler (KL) divergence between the $\mathbb{\wt{P}}(\bm{u}|\bm{y},{\bm{\theta}})$ and $\mathbb{\ol{P}}(\bm{u}|\bm{y},{\bm{\theta}})$. Indeed, the KL divergence serves as posterior non-linearity evaluation since it is not easy to come up with a distance metric. With Bayes' theorem, we have
$\mathbb{P}(\bm{u}|\bm{y},{\bm{\theta}}) = \frac{\mathbb{P}(\bm{u},\bm{y}|{\bm{\theta}})}{\mathbb{P}(\bm{y}|{\bm{\theta}})} = \frac{\mathbb{P}(\bm{y}|\bm{u},{\bm{\theta}}) \mathbb{P}(\bm{u}|{\bm{\theta}})}{\mathbb{P}(\bm{y}|{\bm{\theta}})}.$
Since $(\bm{u}|{\bm{\theta}})$ is Gaussian and $\mathbb{P}(\bm{y}|{\bm{\theta}})$ is the normalising factor, we focus on the likelihood factor $\log \mathbb{P}(\bm{y}|\bm{u},{\bm{\theta}})$.

\begin{theorem}[Expectation on the difference of the linearisation on the observation log-density]\label{thm:exp_diff} Using the notations from Theorem \ref{thm:taylor}, 
    \begin{equation}
  \log \frac{ \wt{\mathbb{P}}( \bm{y} | \bm{u}, \bm{\theta})}{ \ol{\mathbb{P}} ( \bm{y} | \bm{u}, \bm{\theta}) }
            = - \sum_j e^{\ol{m}_j(\bm{u}_*)}\left[\frac{1}{2} (\bm{u}-\bm{u}_*)^\intercal \bm{H}^{(j)} (\bm{u}-\bm{u}_*) + \mathcal{O}(\| \bm{u} - \bm{u}_*\|^3) \right]. \label{eq:lin_diff}
    \end{equation}
The derivation of equation  \eqref{eq:lin_diff} can be found in Appendix \ref{appendix:lin_diff}. Then taking expectation, we have
    \begin{align}
     \E_{\bm{u} \sim \mathsf{N}(\bm{\mu_\theta}, \bm{Q}_{\bm{\theta}}^{-1})} \left(\log \frac{ \wt{\mathbb{P}} ( \bm{y} | \bm{u}, \bm{\theta})}{ \ol{\mathbb{P}} ( \bm{y} | \bm{u}, \bm{\theta}) }\right) 
     &= - \frac{1}{2}\sum_j e^{\ol{m}_j(\bm{u}_*)} \left[\mathsf{tr} (\bm{H}^{(j)}\bm{Q}_{\bm{\theta}}^{-1}) + (\bm{\mu_\theta}-\bm{u}_*)^\intercal \bm{H}^{(j)} (\bm{\mu_\theta}-\bm{u}_*)\right] \nonumber \\ &\quad + \mathcal{O}\left[ \E (\| \bm{u} - \bm{u}_*\|^3) \right]. \label{eq:Eu}
    \end{align}
The derivation of equation \eqref{eq:Eu} can be found in Appendix \ref{appendix:exp_diff}.
\end{theorem}

The expectation in equation \eqref{eq:Eu} depends on the $\bm{H}^{(j)}$ yielded from the true likelihood, which in turns relates to both the estimated intensity function $\hat{\lambda}$, and the partial derivative of the nonlinear predictor with respect to the linearisation point $\bm{u}_*$. Both $\hat{\lambda}$ and $\bm{u}_*$ can be evaluated in \inla. In other words, $\bm{H}^{(j)}$ depends on the nonlinearity in the predictor in $l$ and $l'$ dimensions. When $\bm{H}$ is large, the nonlinear terms deviate more from the estimated mode computed from \inla. We can view $\bm{H}^{(j)}\bm{Q}_{\bm{\theta}}^{-1}$ in equation \eqref{eq:Eu} as a correction term, which is composed of the Hessian term and the covariance structure of the Gaussian components $\bm{u}$. In the case of $\bm{\mu_\theta}-\bm{u}_*=0$, it serves as a sanity check as this would minimise the second term. 

% \begin{align*}
% \bm{G} &= - \sum_i \exp(\ol{m}_i) H^{(i)} \\
% \mathsf{E}_{\ol{p}}\left[ \nabla_{\bm{u}}  \left\{\log \wt{p}(\bm{y}|\bm{u},\bm{\theta}) -
%     \log \ol{p}(\bm{y}|\bm{u},\bm{\theta})\right\}\right]
%  &\approx \bm{G}(\bm{m}-\bm{u}_*) 
% \end{align*}

Now we have all the ingredients to compute the KL divergence. Since the KL divergence is asymmetric, we will compute both $\mathsf{KL}\left[\wt{\mathbb{P}}(\boldsymbol{u}|\bm{y},\boldsymbol{\theta}) \,\middle\|\,\ol{\mathbb{P}}(\boldsymbol{u}|\bm{y},\boldsymbol{\theta})\right]$ and $\mathsf{KL}\left[\ol{\mathbb{P}}(\boldsymbol{u}|\bm{y},\boldsymbol{\theta}) \,\middle\|\,\wt{\mathbb{P}}(\boldsymbol{u}|\bm{y},\boldsymbol{\theta})\right]$. We follow the proof from \citet{lindgren2024inlabru}. 

\begin{theorem}[Kullback-Leibler (KL) divergence for the first and second Taylor expansion]\label{thm:KL}
Denote $\bm{G} = - \sum_j e^{\ol{m}_j} \bm{H}^{(j)}$ and the linearised and approximate nonlinearised  posterior distribution as $\textsf{N}(\ol{\bm{\mu}}_{\bm{\theta}},\ol{\bm{Q}}_{\bm{\theta}})$ and $\textsf{N}(\wt{\bm{\mu}}_{\bm{\theta}},\wt{\bm{Q}}_{\bm{\theta}})$ respectively, we have 
\begin{align}
\mathsf{KL}\left[\wt{\mathbb{P}}(\boldsymbol{u}|\bm{y},\boldsymbol{\theta}) \,\middle\|\,\ol{\mathbb{P}}(\boldsymbol{u}|\bm{y},\boldsymbol{\theta})\right] &= \frac{1}{2}
\{
\log\det(\ol{\bm{Q}}_{\bm{\theta}}-\bm{G})-
\log\det(\ol{\bm{Q}}_{\bm{\theta}})
+\tr\left[\ol{\bm{Q}}_{\bm{\theta}}(\ol{\bm{Q}}_{\bm{\theta}}-\bm{G})^{-1}\right] \nonumber \\
&\phantom{= }
-d 
+ (\ol{\bm{\mu}}_{\bm{\theta}}-\wt{\bm{\mu}}_{\bm{\theta}})^\top \ol{\bm{Q}}_{\bm{\theta}} (\ol{\bm{\mu}}_{\bm{\theta}}-\wt{\bm{\mu}}_{\bm{\theta}}) 
\} + \mathcal{O}\left[\E (\| \bm{u} - \bm{u}_*\|^3 ) \right] \label{eq:KL1} , \text{ and} 
\\
\mathsf{KL}\left[\ol{\mathbb{P}}(\boldsymbol{u}|\bm{y},\boldsymbol{\theta}) \,\middle\|\,\wt{\mathbb{P}}(\boldsymbol{u}|\bm{y},\boldsymbol{\theta})\right]
&= 
\frac{1}{2}
\{
\log\det(\ol{\bm{Q}}_{\bm{\theta}})-
\log\det(\ol{\bm{Q}}_{\bm{\theta}}-\bm{G})
+\tr\left[(\ol{\bm{Q}}_{\bm{\theta}}-\bm{G})\ol{\bm{Q}}_{\bm{\theta}}^{-1}\right] \nonumber \\
&\phantom{= }
-d
+
(\ol{\bm{\mu}}_{\bm{\theta}}-\wt{\bm{\mu}}_{\bm{\theta}})^\top(\ol{\bm{Q}}_{\bm{\theta}}-\bm{G})(\ol{\bm{\mu}}_{\bm{\theta}}-\wt{\bm{\mu}}_{\bm{\theta}})
\} + \mathcal{O}\left[\E (\| \bm{u} - \bm{u}_*\|^3 ) \right], \label{eq:KL2}
\end{align}
where $d$ is the dimension of the vector $\bm{u}$. The derivation can be found in Appendix \ref{appendix:KL}.
\end{theorem}

We can see the KL divergences have similar structure as in equation \eqref{eq:Eu}. We wish to minimise the difference between the linearised and nonlinearised means $(\ol{\bm{\mu}}_{\bm{\theta}}-\wt{\bm{\mu}}_{\bm{\theta}})$, which minimises the both divergences. The term $(\ol{\bm{Q}}_{\bm{\theta}}-\bm{G})$ is the sum of the linearised precision matrix and the Hessian term, and can therefore be viewed as an approximation of $\wt{\bm{Q}}_{\bm{\theta}}$.

Indeed, these divergences provide insights for the bias. Once we can compute them in the \inlabru \ package, we consider different integration schemes based on different mesh construction for the intensity field $\lambda$. Although we do not have control over the precision matrix of the Gaussian components $\bm{u}$, a more linear predictor would imply a smaller $\|\bm{G}\|$. The implications of nonlinearity is two-fold. Nonlinearity can be induced by the aggregation and/or in the predictor expression. The former can be alleviated by the mesh construction with respect to the aggregation structure. For the latter, mesh resolution can be increased adaptively in regions where the predictor exhibits greater nonlinearity. Despite the greater number of elements, we aim to minimise $\|\bm{G}\|$ using smaller $\bm{H}^{(j)}$, with respect to the effect of the aggregation.

\section{Model and Method}\label{sec:mod}

We introduce the models in a general setting, illustrate them in the simulation study (Section \ref{sec:sim}), and assess their performance (Section \ref{sec:res}). The true underlying model follows the linear predictor in equation \eqref{eq:eta1}, with observations as either point patterns $\mathcal{Y}$ or aggregated counts $\mathcal{N}$ over domain partitions defined in Definitions \ref{def:ppp} and \ref{def:pcp}, respectively.

\subsection{Models with covariate Observation Plugin (OP)}\label{sec:agg}
We formulate the models using Observation Plugin (OP) with the linear predictor
\begin{equation}
\eta(\s) = \beta_0 + \beta_x z(\s) + u(\s), \label{eq:bench}
\end{equation}
where $z(\s)$ is the observed covariate at location $\s$, and $u(\s)$ is defined in equation \eqref{eq:psi} with a mesh. The resolution of $z(\s)$ depends on data availability; for example, it may be an aggregated raster covariate field (RastAgg) or an aggregated field over polygons (PolyAgg), both forming partitions of the domain.

\subsection{Covariate Field Estimation from Incompletely Observed Data}\label{sec:icap}
When $z(\s)$ in equation \eqref{eq:bench} is an aggregated covariate field over a partition of the domain, it is only partially observed. Beyond OP, the continuous covariate field can be estimated and incorporated into the linear predictor $\eta(\cdot)$ via Joint Uncertainty (JU), i.e.\ a one-stage approach (Section \ref{sec:1stage}), or via two-stage approaches: Value Plugin (VP) and Uncertainty Plugin (UP) (Section \ref{sec:2stage}). Specifically, we assume the continuous covariate field $x(\cdot)$ is aggregated in an unbiased manner such that the observed averages $z_p$ over the aggregation sets $A_p$ (see Definition \ref{def:pcp}),
\begin{equation}
z_p
= \frac{1}{|A_p|} \int_{A_p} x(\s) + \epsilon(\s) \, d\s
= \frac{1}{|A_p|} \int_{A_p} x(\s) \, d\s + \underbrace{\frac{1}{|A_p|} \int_{A_p} \epsilon(\s) \, d\s}_{=:\epsilon_p}, \label{eq:icag}
\end{equation}
where $\epsilon(\cdot)$ is a Gaussian field such that $\epsilon_p$ are independent $\textsf{N}(0,\sigma^2_{z_p|x(\cdot)})$ distributions. For simplicity, we assume all $\sigma^2_{z_p|x(\cdot)}$ share a common value $\sigma^2_z$, though this can be extended to allow variances scaled by functions of $|A_p|$. We treat the covariate field $x(\cdot)$ as a random variable $X(\cdot)$, modelled with the prior $X(\cdot) \sim \textsf{GRF} \left(\mu_x(\cdot), \varrho_x(\cdot, \cdot) \right)$, yielding the posterior $X(\cdot)|\{z_p\} \sim \textsf{GRF}\left(\mu_{x|\{z_p\}}(\cdot), \varrho_{x|\{z_p\}}(\cdot, \cdot)\right)$. We use the posterior mean $\hat{x} = \mu_{x|\{z_p\}}$ as the continuous estimate of the field.

When $z(\s)$ in equation \eqref{eq:bench} are Point Values (PointVal) scattered across the domain (e.g.\ station data), $\{z(s_i)\}_{i=1}^{n_z}$, the estimation is the same as that for aggregated covariate field, except replacing $\{z_p\}$ with $\{z(s_i)\}$.

\subsection{Joint Uncertainty (JU) Method }\label{sec:1stage}

The Joint Uncertainty (JU) approach models the covariate field and parameters simultaneously,
\begin{align}
\begin{cases}
    z(\cdot) &= \hat{x}(\cdot) + \epsilon(\cdot), \quad \bm{\epsilon} \sim \textsf{N}(\bm{0}, \bm{Q}^{-1}_{\epsilon}), \\
    \eta(\cdot) &= \beta_0 + \wt{\beta}_x \hat{x}(\cdot) + u(\cdot), \quad \wt{\beta}_x \sim \textsf{Exp}(\cdot), \label{eq:exp}
\end{cases}
\end{align}
where $\bm{\epsilon}$ uses the same construction of $\bm{u}$ in equation \eqref{eq:psi}. This may bring identifiability or multimodal concerns between $\hat{x}(\cdot)$ and $u(\cdot)$. Another concern is convergence, which largely depends on whether the coefficient of $\hat{x}(\cdot)$ can attain an opposite sign, resulting in a bimodal distribution. Nonetheless, as soon as it picks a "side" for $\beta_x$, the linearisation makes it unimodal. But different \inla \ runs, or with different initial values, might pick the opposite sign. However, with some direct observations of the field $x(\cdot)$ (i.e.\ $\{z_p\}$ or $\{z(s_i)\}$), it is unlikely to happen. In practice, we often have prior information on whether positive or negative effect, this allows us to introduce an exponential prior distribution over the covariate coefficient. This avoids non-identifiability of the product $\beta_x \hat{x}(\cdot)$, which would arise if $\beta_x$ were given a Gaussian prior.

\subsection{Value Plugin (VP) and Uncertainty Plugin (UP) Methods }\label{sec:2stage}

To simplify notation, we denote $\mathcal{Z}$ as $\{ z_p \}$ for the aggregated covariate field case and $\{ z(s_i) \}$ for the PointVal case from Section \ref{sec:icap}. For both cases, the aggregated counts and point patterns follow
$N_p|\mathcal{Z} \sim \textsf{Pois}\left(\int_{A_p} \exp[\eta(\s)] \, d\s \right)$,
and
$\mathcal{Y}|\mathcal{Z} \sim \textsf{PoPr}\bigl(\exp[\eta(\cdot)]\bigr)$
respectively, where \textsf{PoPr} denotes a Poisson process. When aggregated values of $\hat{x}(\cdot)$ are needed, with
$X_j = |\Omega_j|^{-1} \int_{\Omega_j} X(\s) \, d\s$,
we estimate the covariate field via
$\E\bigl(X_j \mid \mathcal{Z}\bigr) = |\Omega_j|^{-1} \int_{\Omega_j} \hat{x}(\s) \, d\s$ on a mesh. The smoothness of the estimated field depends on the mesh resolution, i.e.\ the volume $|\Omega_j|$.

In the estimation step, we compute $\hat{x}(\cdot)$ from Section \ref{sec:icap}. The Value Plugin (VP) method inputs the prediction into the full model as in equation \eqref{eq:exp}, i.e.\ $\eta(\cdot)  = \beta_0 + \wt{\beta}_x \hat{x}(\cdot)  + \epsilon_{y}(\cdot)$. The Uncertainty Plugin (UP) method incorporates $\hat{x}(\cdot)$ and a spatial uncertainty term $\epsilon_x(\cdot)$ in the second step,
\begin{equation}
  \begin{cases}
  \eta(\cdot) = \beta_0 + \wt{\beta}_x\left[\hat{x}(\cdot) + \epsilon_x (\cdot)\right] + \epsilon_y(\cdot),\\
  \wt{\beta}_x \; \, \sim \textsf{Exp}(\cdot), \;
  \bm{\epsilon}_{x} \sim \textsf{N}(\bm{0}, \hat{\bm{Q}}^{-1}_{\epsilon_x}), \text{ and } 
  \bm{\epsilon}_{y} \sim \textsf{N}(\bm{0}, \bm{Q}^{-1}_{\epsilon_y}),
  \end{cases} \label{eq:xhat}
\end{equation}
where $\bm{\epsilon}_{x}$ and $\bm{\epsilon}_{y}$ uses the same construction of $\bm{u}$ in equation \eqref{eq:psi}. From a modelling perspective, treating the noise $\bm{\epsilon}_x$ as independent and identically distributed with a diagonal covariance matrix leads to identifiability issues. To address this, we explicitly provide the estimated precision matrix $\hat{\bm{Q}}_{\epsilon_x}$ (see Appendix \ref{appendix:pmatrix}). For computational efficiency, separate meshes can be used for $\bm{\epsilon}_x$ and $\bm{\epsilon}_y$ (Section \ref{sec:sim}).

The estimation step smooths the covariate field, even when observational noise is present. Unless there are discontinuities in the field or model mismatches (e.g.\ omitted nonlinear effects), this first step typically produces a reasonable covariate model for the second step. Moreover, the estimated covariate field can always be checked before proceeding further.

The VP method shifts the uncertainty ${\epsilon}_x$ to the Mat\'ern field noise ${\epsilon}_y$, allowing estimation of only one noise term and thus significantly speeding up computation. In contrast, ${\epsilon}_x$ and ${\epsilon}_y$ in the UP method are inevitably confounded, requiring a sufficiently fine mesh to minimise nonlinearity within mesh elements and reduce computational strain from linearisation. For uncertainty quantification, we aim to infer and incorporate ${{\epsilon}_x}$ in the second step, as in equation \eqref{eq:xhat}. This uncertainty can be propagated by incorporating the precision matrix from the first step (see Appendix \ref{appendix:pmatrix}).

\subsection{Model Mismatch under Nonlinear (NL) Misspecification}\label{sec:modelmis}
The nonlinear misspecification is introduced by replacing the linear predictor $\eta(\s)$ in equation \eqref{eq:bench} with a nonlinear function $f(\cdot)$, i.e.\ $\check{\eta}(\cdot) = \beta_0 + \beta_x f(z(\cdot)) + u(\cdot)$. The nonlinear function $f(\cdot)$ can be any nonlinear function, such as exponential or logarithmic. The nonlinear misspecification can be viewed as a model mismatch, where the true model is not linear in the covariate field.
We reuse the model frameworks, i.e.\ JU, UP and VP methods, specified in Sections \ref{sec:1stage} and \ref{sec:2stage}, to adapt the nonlinear (NL) misspecification in the simulation study in Section \ref{sec:nl}. We aim to check the robustness of these frameworks for the mismatched nonlinear predictor.

\section{Simulation Study}\label{sec:sim}
% curly N collection of count
% Robustness of nonlinearity 20240627 model vs methods

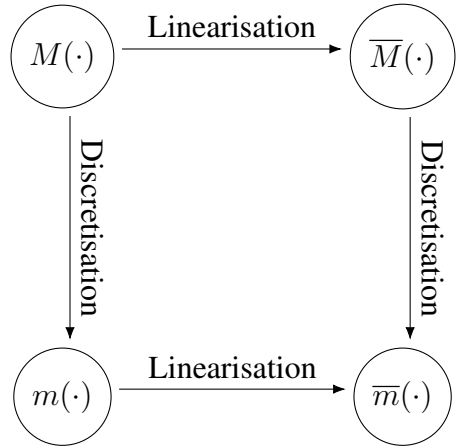
\begin{figure}[htp]
    \centering
    \begin{tikzpicture}[
        shorten < =  1mm, shorten > = 1mm,
    node distance = 45mm, on grid, auto,
    every path/.style = {-Latex},
    sx+/.style = {xshift=1 mm},
    sy+/.style = {yshift=1 mm},
    sx-/.style = {xshift=-1 mm},
    sy-/.style = {yshift=-1 mm},
                        ]
    \node[state] (A) {$M(\cdot)$};
    \node[state] (B) [right=of A] {$\ol{M}(\cdot)$};
    \node[state] (C) [below=of A] {$m(\cdot)$};
    \node[state] (D) [right=of C] {$\ol{m}(\cdot)$};
    \path[->]   ([sy+] A.east)  edge ["Linearisation"] ([sy+] B.west)
                ([sx+] B.south) edge [sloped, anchor=center, above, "Discretisation"] ([sx+] D.north)

                ([sx+] A.south)  edge [sloped, anchor=center, above, "Discretisation"] ([sx+] C.north)

                ([sy+] C.east)   edge ["Linearisation"] ([sy+] D.west);
    \end{tikzpicture}
    \caption{Conceptual diagram of the linearisation and discretisation process. The workflow in the \inlabru \ package: $M(\cdot)\to m(\cdot) \to \ol{m}(\cdot)$. The diagram is not communicative because $\ol{M}(\cdot)\to \ol{m}(\cdot)$ depends on the data.}
    \label{fig:concept}
\end{figure}

We aim to investigate how mesh design, data resolution, and model formulation influence computational cost and model accuracy, using the linearisation approach. Here we simulate data to illustrate the workflow $M(\cdot)\to \ol{M}(\cdot) \to \ol{m}(\cdot)$ (see Figure \ref{fig:concept}). More precisely, we cannot evaluate the discretisation workflow $\ol{M}(\cdot) \to \ol{m}(\cdot)$ analytically as this depends on the data. This is the reason why we assess this workflow through the simulation study. 

\subsection{Data}\label{sec:data}
For a fair model comparison, we construct two meshes: \textit{mesh (i)} --- a hexagonal mesh composed of equilateral sub-triangles (edge length $3.958\,\text{km}$), offset inward from the smoothed, buffered inner boundary by two triangle edge lengths, covering the entire study area; and \textit{mesh (ii)} --- with $1.961$ times edge lengths to that of \textit{mesh (i)}.  See Table~\ref{tab:cellsize} for a comparison of the mesh triangles and covariate raster cells, and Appendix~\ref{appendix:rast} for code.

The elements along the boundary are constructed according to the standard \texttt{fmesher} meshing algorithm~\citep{lindgren_2025}.
As discussed in~Section~\ref{sec:datares}, hexagonal meshes are optimal in the stationary isotropic random field setting since it minimises the discretisation effect. Both the inner and outer boundaries are buffered to smooth the mesh construction (see Appendix \ref{appendix:fmext}). 

The integration scheme takes account of the integration points of the triangular meshes whose density matches that of the RastFull raster and excludes points falling outside the boundary of each $A_p$ (interior boundary) by removing their weights from the integration (see Appendix \ref{appendix:int}). A more accurate result could be achieved by constructing the mesh to conform to the interior sub-boundary of each $A_p$. This would ensure that none of the integration points fall outside the interior boundary. However, this approach (available in experimental via \texttt{fmesher::fm\_intersect()} from version 0.6.1) is not used here because the high-resolution interior boundary complicates mesh construction, making the computation expensive. 

This simulation study is inspired by a simplified landslide occurrence model in which landslide polygons are represented by their centroids, as centroids provide a consistent and well-defined point-based approximation of polygonal features. The covariate field reflects ground shaking from multiple seismic sources over a time window, while the Mat\'ern field captures underlying geomorphological structures. We simulate point pattern observations over Nepal as a test of method as in equation \eqref{eq:point} with the linear predictor
\begin{equation}
    \eta(\s) = \beta_0 + \beta_x x(\s) + u(\s), \label{eq:eta_linear}
\end{equation}
where $x(\s)$ is a continuous covariate field formulated as $({s_1}^2 - {s_2}^2)\exp\left[-\frac{1}{2}({s_1}^2 +{s_2}^2)\right]; \; s_1 \in [-4,4],\, s_2 \in [-2,2]$ projected to Nepal, and $u(\s)$ is defined in equation \eqref{eq:psi}. We set $\beta_0 = -7$, $\beta_x = -6$ and $u(\s)$ with $\rho = 50$ and $\sigma = 0.5$. Both $\beta_0$ and $\beta_x$ are chosen to keep coefficient estimation computationally manageable in this toy example, ensuring that the covariate signals are distinguishable from the Mat\'ern noise component. We also aggregate count observations $N_{\mathcal{Y}}(A_p)$ where $A_p$ is the $p$-th administrative region in Nepal for $p=1,\dots,777$. This results in $139$ simulated observed points. The Coordinate Reference System (CRS) of the data is converted into Universal Transverse Mercator (UTM) zone 44N and are converted into kilometre (km) units. We simulate point pattern observations based on the intensity field evaluated pointwise at the \textit{mesh (i)} nodes (see code and plots in Appendices \ref{appendix:sample.lgcp} and \ref{appendix:ptsobs} respectively).

\subsection{Simulation Scenarios}\label{sec:sim_scenarios}
\begin{figure}[ht]
\centering
\includegraphics[draft=FALSE, width=\textwidth, trim={ 0 0cm 0 0cm},clip]{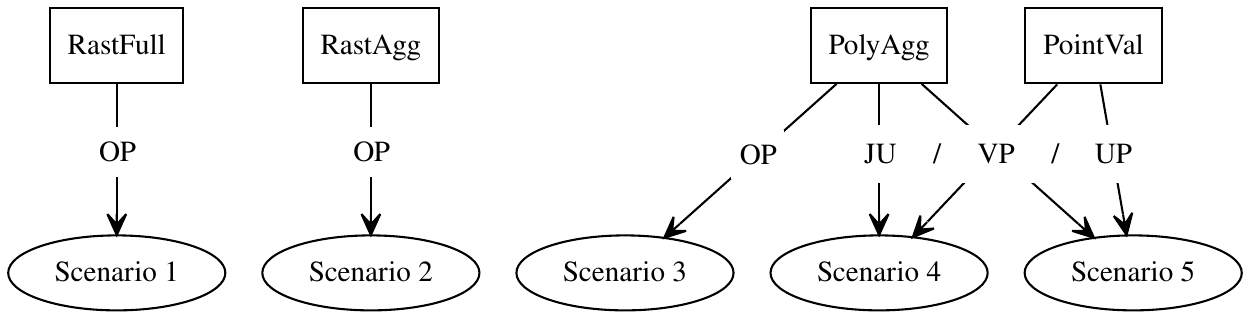}
\caption{Flowchart of the simulation scenarios. The aggregation scenarios consist of: Raster at Full Resolution (RastFull), Raster Aggregation (RastAgg) and Polygon Aggregation (PolyAgg). The incomplete covariate field scenario is divided into two cases: Point Values (PointVal) and PolyAgg. OP: Observation Plugin, JU: Joint Uncertainty, VP: Value Plugin, UP: Uncertainty Plugin.}
\label{fig:scenario}
\end{figure}

We consider both point pattern or aggregated count observations with the following scenarios for the observed covariate field:
% [label=\itshape\alph*\upshape)]
\begin{enumerate*}[label=(\arabic*\upshape)]
    \item \label{itm:perfect} Perfect detection of every covariate raster at full resolution (RastFull) of $0.858 \text{km} \times 0.859 \text{km}$ (Section \ref{sec:raspol});
    \item \label{itm:mean} Covariate raster aggregation (RastAgg) by taking the mean of the RastFull, at a resolution of $8.584 \text{km} \times 8.592 \text{km}$ (Section \ref{sec:raspol});
    \item \label{itm:poly} Polygon aggregation (PolyAgg) by taking the mean of the RastFull across administrative regions in Nepal, i.e.\ $n_A=777$ piecewise-constant polygons (Section \ref{sec:raspol}); 
    \item \label{itm:incomplete} Incomplete covariate field of PolyAgg and Point Values (PointVal) (Section \ref{sec:ic}); and
    \item \label{itm:nlpoly} nonlinear model misspecification of PolyAgg and PointVal.
\end{enumerate*}
We refer to Scenarios \ref{itm:perfect}–\ref{itm:poly} as aggregation scenarios, and Scenario \ref{itm:incomplete} as the incomplete covariate field scenario. Under the aggregation scenarios, the covariate input methods in RastFull, RastAgg, and PolyAgg are termed Observation Plugin (OP), yielding six models for point pattern and aggregated count observations. The RastFull model with point pattern observations serves as a benchmark, with $\bm{u}$ defined on \textit{mesh (i)} (Section \ref{sec:agg}).For the incomplete covariate field scenario, we consider three methods - Joint Uncertainty (JU) and the two-stage approaches, Value Plugin (VP) and Uncertainty Plugin (UP) - for both PointVal and PolyAgg, giving twelve models in total, with the estimated covariate field $\hat{x}(\cdot)$ with the rasterised mesh elements ($1.616$ km $\times 1.616$ km; see Appendix \ref{appendix:fm_pixels} for code), $\bm{\epsilon}_x$ on \textit{mesh (i)} and $\bm{\epsilon}_y$ on \textit{mesh (ii)} (see Sections \ref{sec:1stage} and \ref{sec:2stage}). We then extend Scenarios \ref{itm:poly} and \ref{itm:incomplete} to Scenario \ref{itm:nlpoly}, which introduces nonlinear (NL) misspecification. Specifically, this involves model mismatch where linear models are used to fit observations generated from a nonlinearly transformed covariate field under the incomplete covariate setting (Section \ref{sec:nl}). This again results in twelve models in total. The components $\bm{u}$, $\bm{\epsilon}$, $\bm{\epsilon}_x$, and $\bm{\epsilon}_y$ are all assigned PC priors (see Appendix \ref{appendix:pcprior}; \citet{fuglstad2019constructing}). We summarise above in a flowchart in Figure \ref{fig:scenario}.  

The function on the mesh is defined to be piecewise linear, hence the resolution of the mesh should be similar to that of the data in the optimal case; otherwise we do not fully uilitise the data information. Having said that, the \textit{mesh (i)} resolution in the simulation is roughly $4$ times larger than the covariate raster data at full resolution in Scenario \ref{itm:perfect}. Here we compromise the information from the data to make the computation more efficient. Therefore, it provides insight into the case when we slightly relax the rule of thumb mentioned in Section \ref{sec:datares}.

% flowchart TD
% 
%     A1[RastFull] -->|OP| B1(Scenario 1)
%     A2[RastAgg]  -->|OP| B2(Scenario 2)
%     A3[PolyAgg]  -->|OP| B3(Scenario 3)
%     A3[PolyAgg]  -->|JU/VP/UP| C1(Scenario 4)
%     A4[PointVal] --> |JU/VP/UP|C1(Scenario 4)
%     A3[PolyAgg]  --> |JU/VP/UP| C2(NL misspecification)
%     A4[PointVal] --> |JU/VP/UP| C2(NL misspecification)

\subsubsection{Rasterisation and Polygon Aggregation (PolyAgg) of Covariate Field}\label{sec:raspol}
We compare the effect of the scales of raster aggregations in the covariate field. In Scenario \ref{itm:perfect}, the covariate field is evaluated at the centre of each raster cell. This corresponds to satellite data that can be obtained everywhere and is treated as constant within each cell. For Scenario \ref{itm:mean}, we aggregate covariate raster in Scenario \ref{itm:perfect} into super-cells by a factor of $10$. In this case, reconstruction of the continuous covariate field can be easily done in the same spirit as in Section \ref{sec:icap}. Under Scenario \ref{itm:poly}, we aggregate the covariate by averaging the rasters from Scenario \ref{itm:perfect} over administrative regional polygons. Rasterisation and polygon aggregation can be viewed as a limited set of observed field as the field is not fully observed, and we wish to quantify the accuracy and uncertainty from the resulting estimation procedures.

\subsubsection{Incomplete Covariate Field for Point Values (PointVal)}\label{sec:ic}

For PointVal, we observe a limited set of measured values of the continuous covariate field $x(\s)$ at a finite set of locations, $\{s_i\}$. We assume Gaussian additive noise for the observed covariate values and thus construct an extra layer in the Bayesian hierarchy to model the covariate field. Let $z_i$ denote the observation of $x(s_i)$ with Gaussian additive noise term $\epsilon_{x_i}$ so that, 
\begin{equation}
    z_i = x(s_i) + \epsilon_{x_i}, \;\epsilon_{x_i} \sim \textsf{N}(0, \sigma^2_{\epsilon_x}), \; i=1,\dots,n_z. \label{eq:zobs}
\end{equation}
We randomly sample $n_z = 777$ observed covariate points, with spatially uniform sampling density within each polygon, and then add noises $\sigma^2_{\epsilon_x} = 0.1$ to these stratified samples (see Appendix \ref{appendix:z_locs}). This contrasts with the PolyAgg case where only the means of each polygon are used. 

\subsubsection{Nonlinear (NL) Misspecification}\label{sec:nl}
\begin{figure}[ht]
\centering
\includegraphics[draft=FALSE, width=\textwidth, trim={ 5 0 5 0},clip]{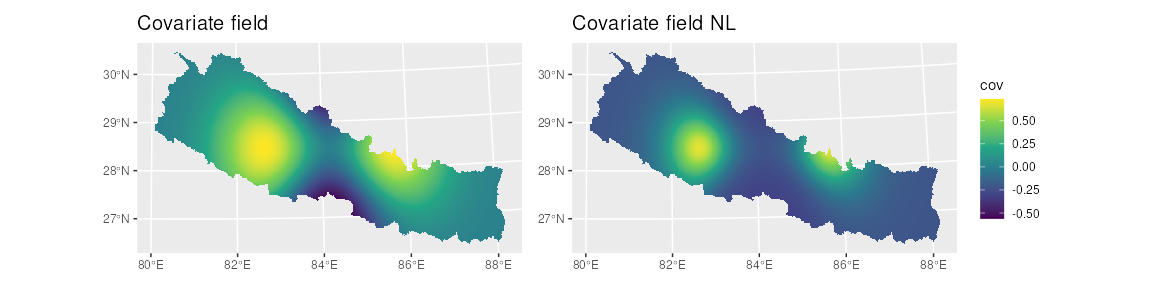}
\includegraphics[draft=FALSE, width=.5\textwidth, trim={ 2.5 0 2.5 0},clip]{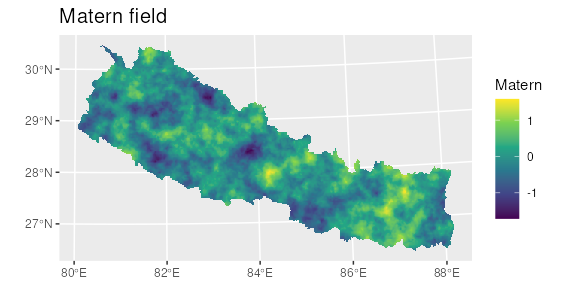}
\includegraphics[draft=FALSE, width=\textwidth, trim={ 5 0 5 0},clip]{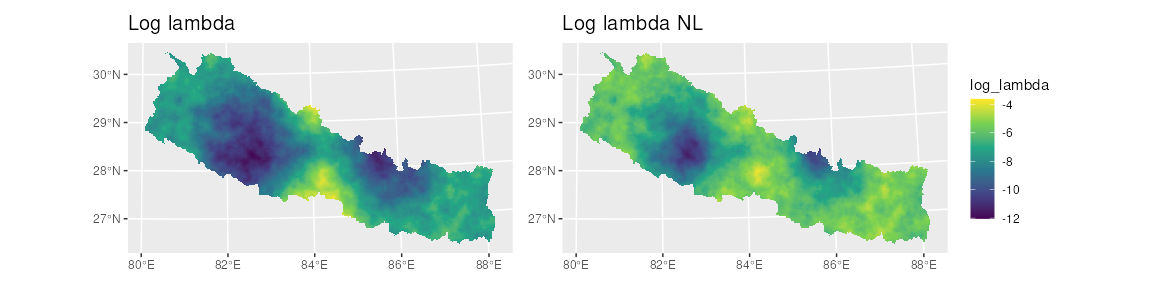}
\caption{Covariate fields $x(\cdot)$ and $f(x(\cdot))$ (top row), Gaussian Mat\'ern field $u(\s)$ (middle) and the intensity fields $log(\lambda)$ and $log(\check{\lambda})$ (bottom row); where top two and bottom two share the same legend respectively. NL: Nonlinear misspecification.}
\label{fig:eta}
\end{figure}

We extend the framework to examine model robustness under nonlinear (NL) misspecification and mismatch conditions. For the data generating mechanism, we transform covariate field via a nonlinear transformation $f(\cdot)$, i.e.\ $\log \check{\lambda}(\cdot) = \check{\eta}(\cdot) = \beta_0 + \beta_x  f\left[x(\cdot)\right] + u(\cdot),$ where $f\left[x(\cdot)\right] = b^{-1} \exp{\left[a x(\cdot)\right]}-c,$ and we set $a = 3$, $b = 9$ and $c = 3$. We then simulate point patterns $\check{\mathcal{Y}}$ and aggregated count observations $\check{\mathcal{N}}$ accordingly. This formulation results in $\check{n} = 149$ simulated point observations. The resulting shift in the covariate and intensity field compared to the original simulation is shown in Figure \ref{fig:eta}.

\subsection{Profile Likelihood for Uncertainty Propagation}\label{sec:plikup}
% TODO show adding the uncertainty bit does not destroy the estimate, how much bias is introduced.

Before presenting the results for the full two-dimensional cases in Sections \ref{sec:data} and \ref{sec:sim_scenarios}, we first assess the robustness of the JU, VP, and UP methods using a one-dimensional toy example with the profile likelihood, noting that JU and UP incorporate uncertainty propagation. Specifically, we estimate the optimal values of $\beta_0$, $\beta_x$, and $\epsilon_{x_i}$ from the log-posterior density. We consider the extreme case where $\{X(s_i)\}_{i = 1}^{n_z}$ are independent of each other, so that the estimation of $X(s_i)$ cannot borrow strength from another location, $s_i \neq s_{i'}$. We formulate the observed covariate field as
$    z(s_i) = X(s_i) + \epsilon_{x_i}, \; X(s_i) \sim \mathsf{N}(0, \sigma_x^2), \; \epsilon_{x_i} \sim \mathsf{N}(0, \sigma^2_{\epsilon_x}). $
Assuming no missing observations, 
\begin{align*}
  \begin{cases}
    X(s_i)|z(s_i) &\sim \mathsf{N}(\hat{x}_i, \sigma^2_{x_i|z(s_i)}), \\
    \mathcal{Y}|\hat{x}_i : \eta(s_i) &= \beta_0 + \beta_x(\hat{x}_i + \epsilon_{x_i}) + \epsilon_{y_i}, \; \epsilon_{x_i} \sim \mathsf{N}(0, \sigma^2_{x_i|z(s_i)}).  
  \end{cases}    
\end{align*}
We aim to check the asymptotic behaviour when $n_z \to \infty$, 
\begin{align}
  \mathfrak{p} &:= \log \mathbb{P} \left[(\beta_0, \beta_x, \epsilon_{x_i})|\{X(s_i) \} \right] \nonumber \\
  &= \sum_i (-e^{\eta_i} + \eta_i y_i) -\frac{1}{2}\left[ \left(\frac{\beta_0^2}{\sigma_{\beta_0}}\right)^2  + \left(\frac{\beta_x}{\sigma_{\beta_x}}\right)^2 + \sum_{i=1}^{n_z} \left(\frac{\epsilon_{x_i}}{\sigma_{x_i|z(s_i)}}\right)^2 \right]. \label{eq:logpost}
\end{align}
We have not found a closed form expression for the optimal value of the derivative from equation \eqref{eq:logpost} (see Section \ref{sec:analytical} in Appendix \ref{appendix:uncertainty}). For this reason, we instead use numerical optimisation to find the optimal $(\beta_0, \beta_x, \epsilon_{x_i})$. Since our focus is on $\beta_x$, we use the profile likelihood to optimise $\beta_x$ with $n_z \in \{2, 4, 8, 16, 32, 64, 128, 256\}$, $ \argmax_{\beta_x} \left(\argmax_{(\beta_0, \epsilon_x)}(\mathfrak{p})\right). $

The plots of the log-posterior density against $\beta_x$ in Appendix \ref{sec:plik} show that all the profile posterior modes converge to the true value of $\beta_x$ as $n_z$ increases. Although, in the plots for UP method (Figure \ref{fig:plik_uncertainty}, Appendix \ref{sec:plik}), the profile likelihood is not a strictly convex function though the global maximum is distinct from the local maximum as $\beta_x$ approaching to infinity. It might pose problem for the optimisation algorithm and one should be careful about the initial starting points. However, we did not encounter any issues in our simulation (see Sections \ref{sec:int_ic} and \ref{sec:int_nl}).

% plugin in the table to method
% 777 points instead of 100 points, it is easier to compare for the story telling, one point in each region plus noise, stratified sampling

% refer to the section and spell the model equation in rasterisation in sim.rnw 
% 10 to -5 in the table 
% left is point and right is count for plots

\section{Results}\label{sec:res}
% \subsection{Proper scoring rules}
A prediction score $S(F,\lambda)$ evaluates some measure of closeness between a pointwise estimate identified by a predictive distribution $F$, and an observed value $\lambda$. We define the expectation of a score as $S(F,G) = \pE_{\lambda \sim G}[S(F,\lambda)]$. A negatively oriented score (i.e.\ the lower the score, the better) is proper if it fulfils $S(F,G) \geq S(G,G).$ This implies that any predictive distributions with a higher or lower prediction uncertainty, other than the true one, will not lead to a better score on average \citep{gneiting2007strictly}. The Squared Error (SE), $S_\text{SE}(F,\lambda) = (\lambda - \wh{\lambda}_F)^2$, and Dawid-Sebastiani (DS) score, $S_\text{DS}(F,\lambda)= {\pVar_F(\wh{\lambda})}^{-1}(\lambda - \wh{\lambda}_F)^2
 + \log[\pVar_F(\wh{\lambda})]$, are common proper scoring rules \citep{dawid1999coherent}. The Mean Squared Error (MSE) and Mean Dawid-Sebastiani (MDS) score for the points across a space are defined as $S(\{F_i\},\{\lambda_i\}) = n^{-1}\sum_{i=1}^n S(F_i,\lambda_i)$.

We compare the prediction accuracy via the proper scoring rules, namely the SE and DS scores for the intensity field. The lower the score, the better the model is. We evaluate the mean scores for the intensity field pointwise at $n=56,809$ evenly distributed point locations across Nepal. We compute these scores of the intensity field for the simulation models because the true simulated intensity field allows us to compare across point and aggregated count observation models. 

\subsection{Intensity Estimation under Aggregation Scenarios}

\begin{figure}[!hb]
\centering
\includegraphics[draft=FALSE,width=\textwidth]{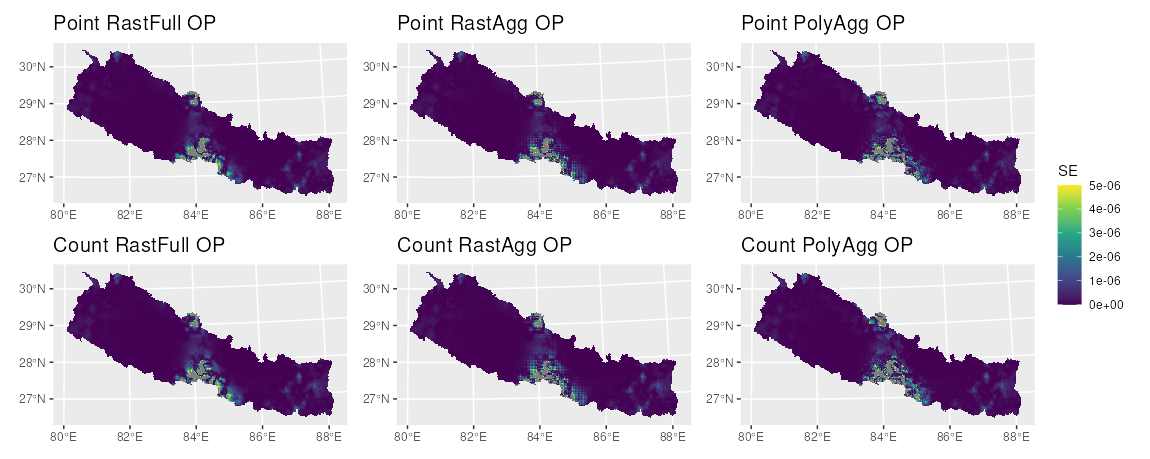}
\caption{Squared Error (SE) for the intensity field $\lambda$ across the models under aggregation scenarios (truncated at $5\times 10^{-6}$). Point Pattern (top row) and Aggregated Count (bottom row) observation models. $\Bar{\lambda} = 2.456 \times 10^{-3}$. RastFull: Raster at Full Resolution, RastAgg: Aggregated Raster, PolyAgg: Polygon Aggregations, OP: Observation Plugin.} 
\label{fig:se}
\end{figure}

\begin{figure}[!hb]
\centering
\includegraphics[draft=FALSE,width=\textwidth]{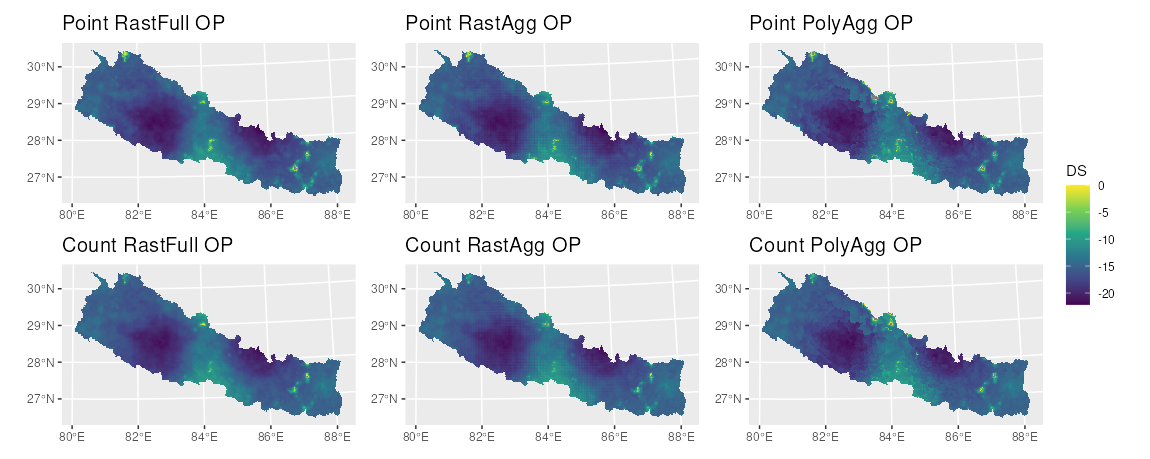}
\caption{Dawid-Sebastiani (DS) score for the intensity field $\lambda$ across the models under aggregation scenarios (truncated at 0). Point Pattern (top row) and Aggregated Count (bottom row) observation models. RastFull: Raster at Full Resolution, RastAgg: Aggregated Raster, PolyAgg: Polygon Aggregations, OP: Observation Plugin.}
\label{fig:ds}
\end{figure}

We compare the performances of the RastFull, RastAgg and PolyAgg OP models for point patterns and aggregated count observations here. In Table \ref{tab:score}, the point pattern models outperform the corresponding aggregated count models in both scores. The RastFull point pattern models achieves the best scores. As the aggregation scale and irregularity increase, the scores worsen across each observation category. The MDS scores for the RastFull and RastAgg models are similar due to well-justified variances. The score distributions are shown in Figures \ref{fig:se} and \ref{fig:ds}. The plots retain a piecewise linear pattern for both the RastAgg and PolyAgg models. In Figure \ref{fig:se}, most SEs are concentrated in the central region, where the majority of observed points are located. In contrast, the aggregated count models exhibit better DS scores in the smaller provinces due to averaging effects. This observation leads us to consider projecting point observations onto mesh nodes, which will be discussed in Section \ref{sec:ver}. Despite this, the overall MDS scores remain lower for point pattern models. Figure \ref{fig:ds} shows that the DS plots for aggregated count models display more consistent variances compared to point pattern models. 

% \afterpage{\clearpage}

\subsection{Covariate Estimation under Incomplete Covariate Field Scenario}\label{sec:cov_ic}

The PolyAgg and PointVal models are not directly comparable under incomplete covariate field scenario (see Table \ref{tab:score_cov} in Appendix \ref{appendix:score_cov}) as the latter includes Gaussian additive noise term $\epsilon_{x_i}$ from equation \eqref{eq:zobs}, while the former does not. However, this noise term negatively impacts the scores though stratified sampling for each polygon provides more precise point locations of the covariate field. This may partly explain why the PolyAgg models achieve better scores than PointVal models in predicting the covariate field. Under both PolyAgg and PointVal categories,  the JU models outperform the VP and UP models. This suggests that jointly modelling the intensity and covariate fields offers more informative predictions than modeling them separately. For the PointVal category, the point pattern JU model slightly outperforms the corresponding aggregated count model in both scores. 

Consequently, we will see how the predicted covariate fields may have an knock-on effect on the intensity field prediction in the following subsection given the covariate field is only part of the story to predict the intensity field. It is worth mentioning that there are visible hotspots in DS plots (see Appendix \ref{appendix:score_cov} for DS and other plots) due to lower uncertainty where the observed points are. 

\subsection{Intensity Estimation under Incomplete Covariate Field Scenario}\label{sec:int_ic}

We compare the performances of the PolyAgg and PointVal models with OP, JU, VP and UP methods for point pattern and aggregated count observations here. In Table \ref{tab:score}, in terms of MSE scores for the intensity field estimation under PolyAgg, the JU and UP models perform better than the OP model while the VP models perform worse. Hence, these results underscore the critical role of uncertainty quantification in intensity field prediction and downscaling the covariate field may not improve the model. The JU models perform better than the UP models in terms of MSE score and very close in terms of MDS score. This suggests that the joint estimation provides a slight edge for the robustness against the additional noises. However, the MDS scores of the JU and UP models are similar to the OP models, which might be due to the extra uncertainty introduced via the covariate field estimation (see discussion in Section \ref{sec:ver}).

\begin{figure}[!hb]
    \centering
    \includegraphics[draft=FALSE,width=\textwidth, trim={0 0 0 0},clip]{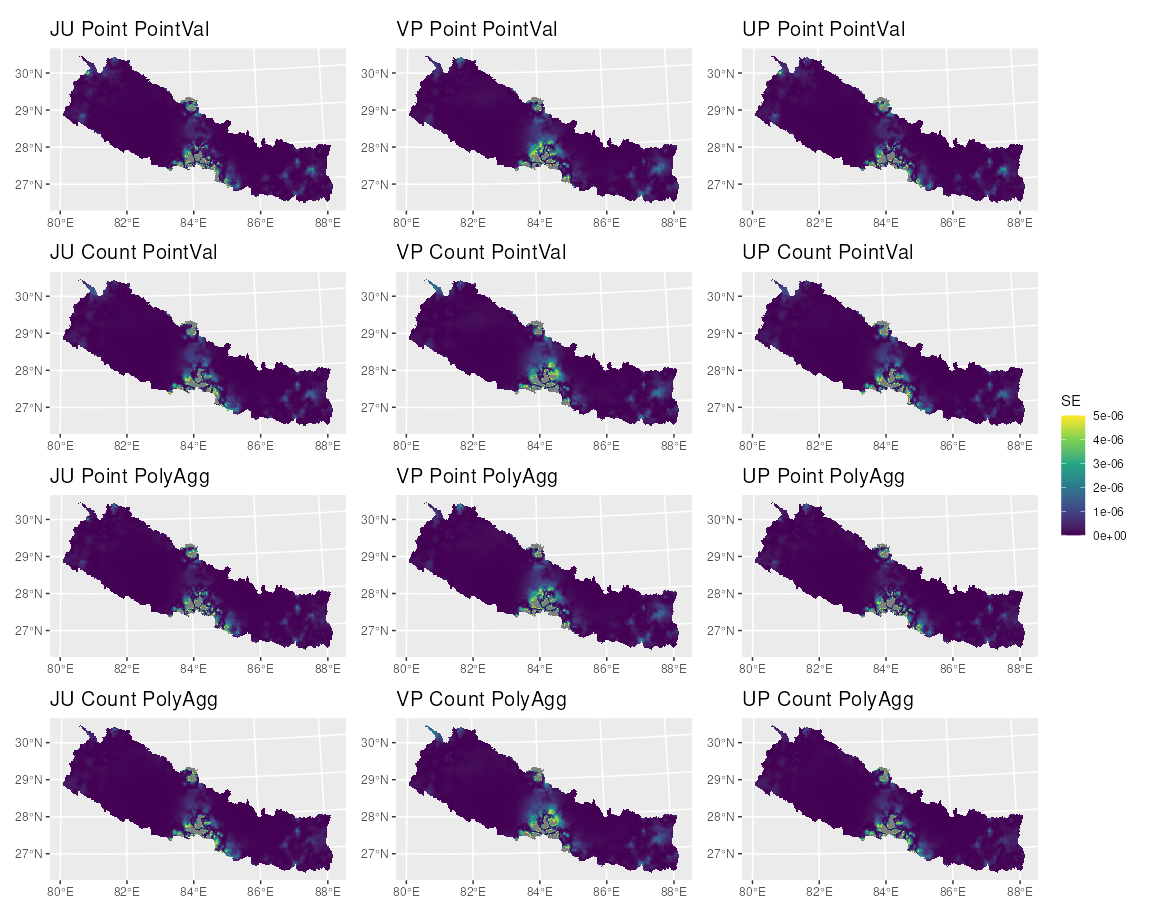}
    \caption{Squared Error (SE) for the intensity field $\lambda$ across the models under incomplete covariate field scenario (truncated at $5 \times 10^{-6}$). JU: Joint Uncertainty, VP: Value Plugin, UP: Uncertainty Plugin, Point: Point Pattern, Count: Aggregated Count, PolyAgg: Polygon Aggregation, PointVal: Point Values. $\Bar{\lambda} = 2.456 \times 10^{-3}$.}
    \label{fig:se_ic}
\end{figure}

\begin{figure}[!hb]
    \centering
    \includegraphics[draft=FALSE,width=\textwidth, trim={0 0 0 0},clip]{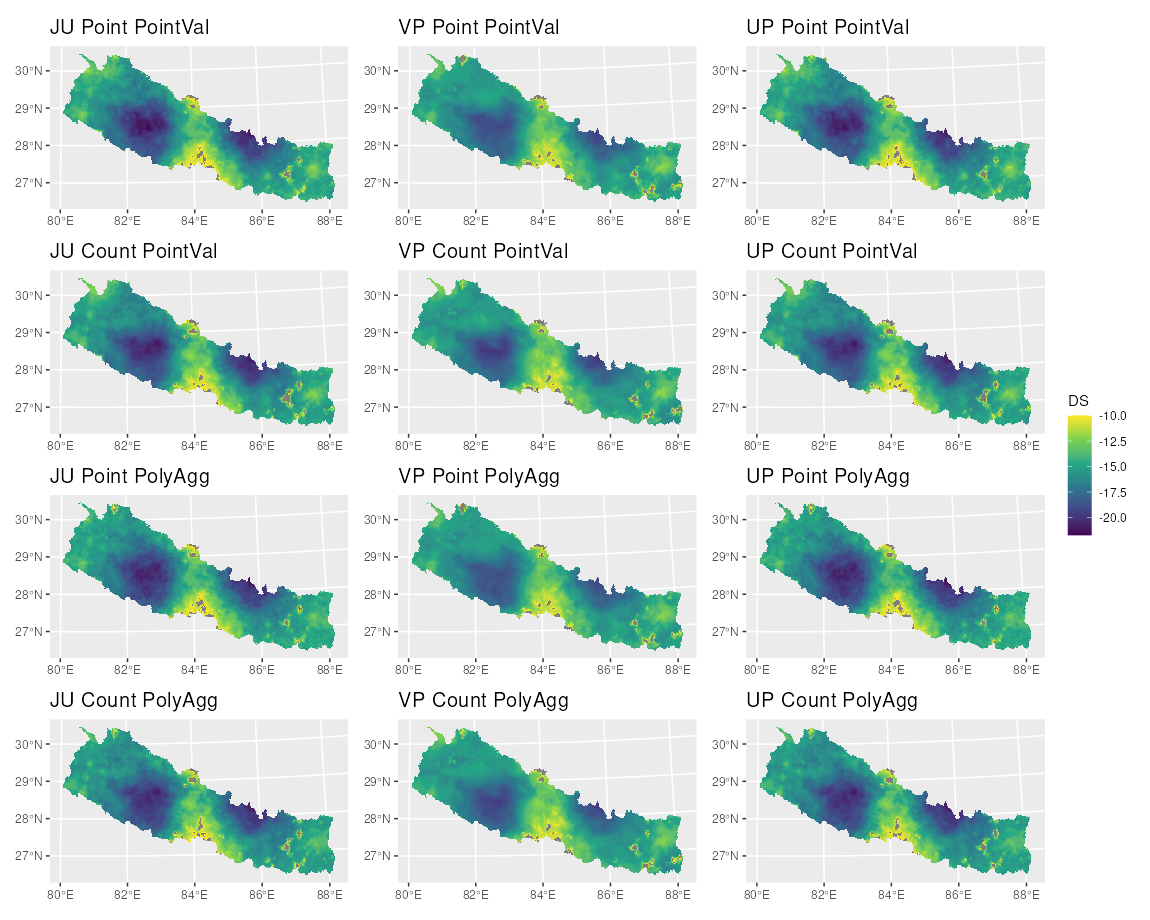}
    \caption{Dawid-Sebastiani (DS) score for the intensity field $\lambda$ across the models under incomplete covariate field scenario (truncated at $-10$). JU: Joint Uncertainty, VP: Value Plugin, UP: Uncertainty Plugin, Point: Point Pattern, Count: Aggregated Count, PolyAgg: Polygon Aggregation, PointVal: Point Values.}
    \label{fig:ds_ic}
\end{figure}

The performance of PolyAgg models surpasses than that of the corresponding PointVal models due to the additional noises in the PointVal covariate observations, as in the covariate field estimation in the previous subsection. Across all PolyAgg and PointVal categories, the JU models achieve the highest scores or come very close. Despite the under-performance in estimating the predicted covariate field in both PolyAgg and PointVal, the UP models perform reasonably well. In Figure \ref{fig:se_ic}, the areas with large SE (where high SE are truncated in grey for visualisation) for the VP models are larger compared to the JU and UP models in the regions with dense observations. The DS plots in Figure \ref{fig:ds_ic} show similar effects and the JU and UP models perform better in regions with fewer observations. Hence, this suggests the scores are consistent across Nepal. 

Overall, the PolyAgg and PointVal models with the JU and UP methods show significant improvements, compared to  the piecewise linear approach in the PolyAgg model in the previous subsection. The hotspots observed in DS plots for predicting the covariate field in the previous subsection are no longer observed in the DS plots for predicting the intensity field. This is likely because the Mat\'ern field compensates for the under-performance at those locations.

\subsection{Intensity Estimation under Nonlinear (NL) Misspecification }\label{sec:int_nl}

\begin{figure}[!hb]
    \centering
    \includegraphics[draft=FALSE,width=\textwidth, trim={0 0 0 0},clip]{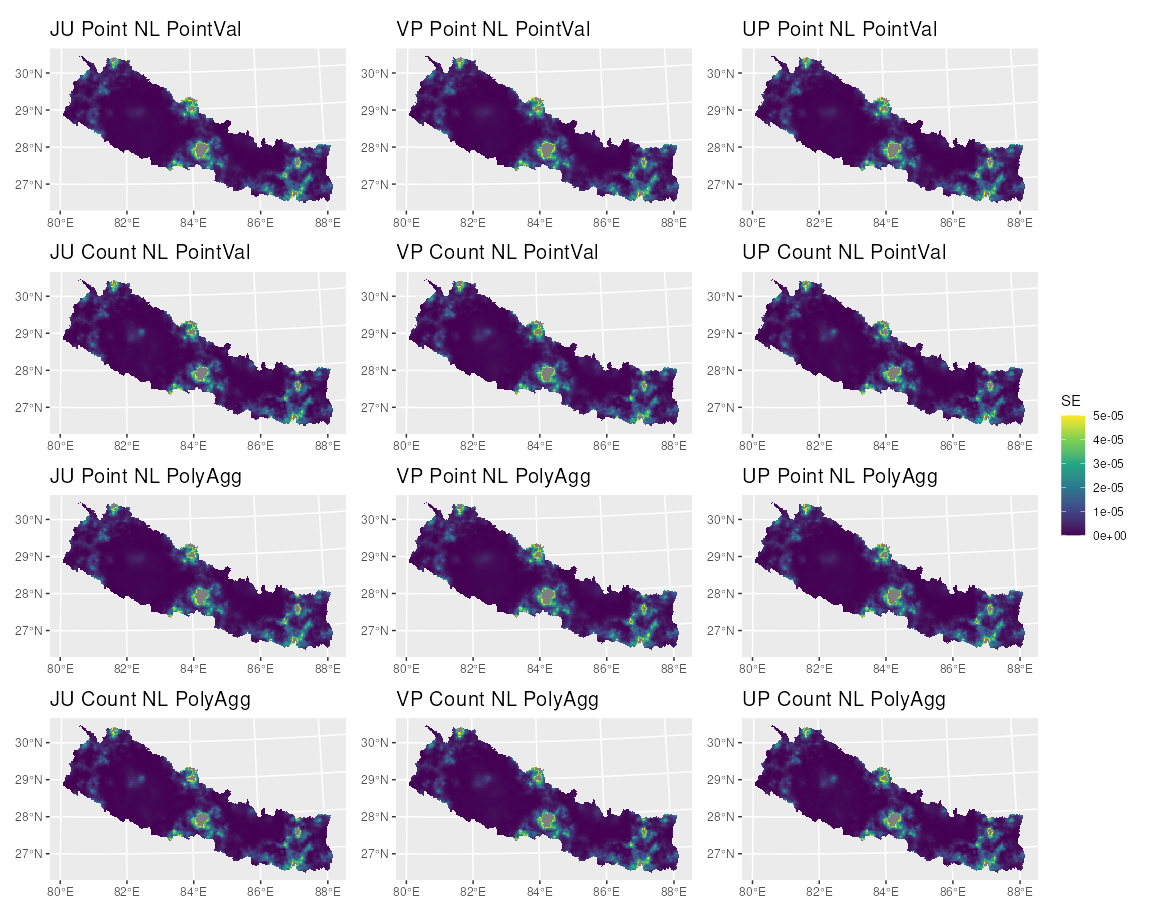}
    \caption{Squared Error (SE) for the intensity field $\lambda$ across the misspecified models under incomplete covariate field scenario (truncated at 5$\times 10^{-5}$). JU: Joint Uncertainty, VP: Value Plugin, UP: Uncertainty Plugin, Point: Point Pattern, Count: Aggregated Count, NL: Nonlinear misspecification, PolyAgg: Polygon Aggregation, PointVal: Point Values. $\Bar{\check{\lambda}} = 2.456 \times10^{-3}$.}
    \label{fig:se_ic_nl}
\end{figure}

\begin{figure}[!hb]
    \centering
    \includegraphics[draft=FALSE,width=\textwidth, trim={0 0 0 0},clip]{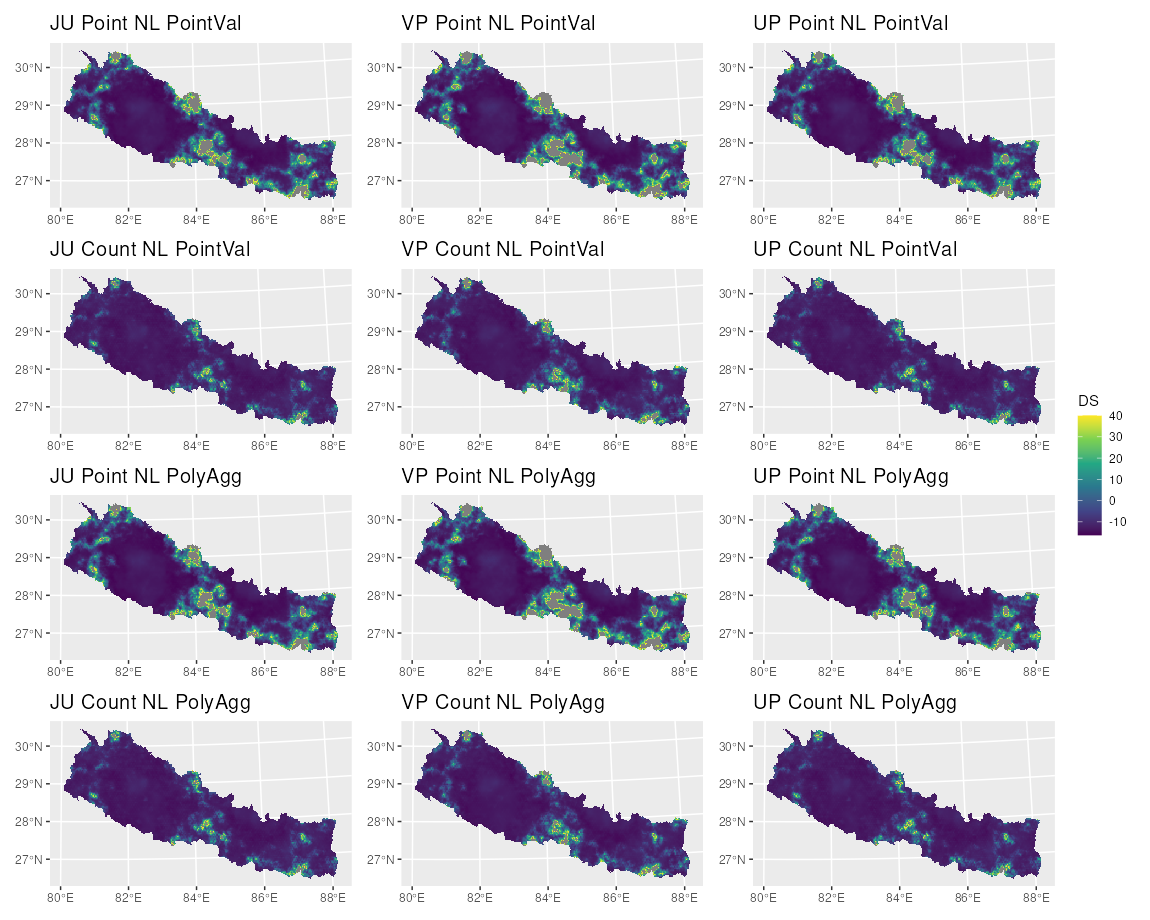}
    \caption{Dawid-Sebastiani (DS) score for the intensity field $\lambda$ across the misspecified models under incomplete covariate field scenario (truncated at 40). JU: Joint Uncertainty, VP: Value Plugin, UP: Uncertainty Plugin, Point: Point Pattern, Count: Aggregated Count, NL: Nonlinear misspecification, PolyAgg: Polygon Aggregation, PointVal: Point Values.}
    \label{fig:ds_ic_nl}
\end{figure}

We compare the performances of the PolyAgg and PointVal models with JU, VP and UP methods for point pattern and aggregated count observations under NL here. In Table \ref{tab:score_nl}, the JU and UP models clearly outperform the VP models due to the uncertainty propagation. The scores also are the measures of the robustness of these models under the NL misspecification towards model misspecification. The linearisation and Mat\'ern noise term $\epsilon_y$ are not flexible enough for the VP method to account for the model misspecification. Indeed, both JU and UP methods are better than the VP one in terms of MSE and, particularly, MDS scores. This is consistent with both the SE and DS plots in Figures \ref{fig:se_ic_nl} and \ref{fig:ds_ic_nl}.  This implies that the uncertainty propagation is better explained in terms of variance thanks to the extra noise term in the UP methods. 

Indeed, the VP method under NL misspecification is not as bad as expected in MSE scores because the Mat\'ern field term $u(\s)$ compensates for the model misspecification. Empirically, the JU and UP methods are more robust than the VP method. Again, we observe the PolyAgg models are better than the corresponding PointVal ones due to the additional noises in the observed covariate field. Surprisingly, the aggregated count models are better than point pattern ones in both scores. In Figure \ref{fig:ds_ic_nl}, the DS plots show the spread of the truncated area more clearly. This finding leads us to consider the projection of point observations into weights on mesh nodes to better explain the variance (see Section \ref{sec:ver} for discussion). 

\section{Conclusion and Discussion}\label{sec:ver}
% more clear conclusion 10 Jul 2025, restate the abstract
Aggregating data to align common geometric objects is a standard approach to addressing spatial misalignment, but it introduces systematic biases that weaken statistical inference. We developed a disaggregation framework with uncertainty quantification that enables continuous-domain modelling of both point pattern and aggregated count data while mitigating these biases. We applied this method to a landslide-motivated case study using synthetic data, demonstrating that it can recover the true underlying field under complex situations. These studies demonstrated that our approach achieved lower Mean Squared Error (MSE) and Dawid–Sebastiani (MDS) scores, with models based on point patterns and high-resolution raster covariates performing best, and traditional aggregated-count models performing worst. These results underscore the importance of preserving point-level information and propagating uncertainty. Under incomplete covariate fields and nonlinear model misspecification (i.e.\ modelling a nonlinear field as linear), Joint Uncertainty (JU) and Uncertainty Plugin (UP) models consistently outperformed the Value Plugin (VP) model, with UP models offering computational efficiency and improved control over covariate predictions. As the correct linear predictor is rarely known a priori, our framework provides a flexible and robust basis for inference under both uncertainty propagation and misspecification.

\subsection{Limitations and Future Research Directions}\label{sec:lim}
To streamline the paper’s focus, we did not explore alternative precision matrices for uncertainty propagation (see Section \ref{sec:2stage} and Appendix \ref{appendix:pmatrix}). The number of iterations needed for convergence is unpredictable and depends on factors such as tolerance levels, line search direction, mode assessment in \inla, the data, and the model specification (see convergence diagnostics in Appendix \ref{appendix:conv}). Further research is needed to determine the optimal settings. Implementing automatic differentiation can help accelerate computation \citep{carpenter2015stan}.

Observations, including those related to landslides and other fields, are manually registered, introducing inherent uncertainties regarding their exact spatial and temporal accuracy. To address this, we can split the observed point information and project it onto nearby mesh nodes with partial weights so that we introduce uncertainty in a manner akin to aggregated count models but with more controlled adjustments. This approach helps mitigate the covariate field uncertainties caused by model misspecification.  

The flexibility and \textit{degrees of freedom (DoF)} of the estimated intensity field is determined by the mesh resolution. For the integration schemes, we need to respect the function variability and degrees of freedom to all spatial random fields to ensure a bounded likelihood approximation. In particular, this applies to Mat\'ern effects defined on triangulation meshes and uncertainty propagation fields for high resolution covariates. If too few integration points were to be used, the nonlinear integral approximations would not sufficiently bound the likelihood. We address this by ensuring the integration scheme aligns with both mesh and covariate resolutions.
%, as demonstrated effectively in the simulation study.
Apart from the linearisation, the Mat\'ern random field in our models indeed accounts for the data misalignment, which can be controlled under stricter PC priors. 

The linearisation becomes more stable with an increase in integration points. A finer mesh by subdividing the elements would stabilise the linearisation further (see Appendix \ref{appendix:int}). There is a fundamental difference between increasing the integration points within each element and subdivision of elements. The former adds more equally weighted points, while the latter creates more piecewise linear mesh elements to account for the nonlinearity. In the future, we will explore alternative integration schemes, such as Simpson's rule or Gaussian quadrature. 

% In the future, we may also consider variational inference for linearisation approximation to improve the speed and accuracy of the linearisation approximation , where we look for a surrogate distribution $\mathcal{Q}(\bm{u})$ that approximates the true posterior distribution $\mathcal{P}(\bm{u}|\bm{y}, \bm{\theta})$. We can find a surrogate distribution that represents the approximate KL divergences in equations \eqref{eq:KL1} and \eqref{eq:KL2}. 
%This is done by maximising the evidence lower bound. 

\subsection{Connection to \inla \ Extension Packages}
% \subsubsection{Observed Point Uncertainty Propagation}
% One should describe it rather than simulate the scenario??? 20231201 

In various applications, including landslides, there is often uncertainty regarding the exact locations, as these are determined manually. In some instances, we cannot revert to the original trigger location over time. Hence, contour map can be useful in assessing the joint uncertainty excursion sets via \texttt{excursions} package \citep{bolin2017quantifying}. These maps add an extra dimension for evaluating contour credible risk regions and help propagate uncertainty arising from the aggregation.
% An alternative or combined approach is to split the point observations into vertices contribution. To this end, we split the uncertainty across the triangular mesh. Hence, the family \texttt{xpoisson} in \inla \ comes into play since this would allow non-integer count. This is to incorporate the uncertainty due to manual landslides mapping. To experiment, we can simulate by jittering our observation locations $s_i$.
% \subsubsection{Nonstationary and anisotropic model class in \texttt{aniso}}
Assuming stationarity can be quite restrictive. Extending the models to nonstationary and anisotropic scenarios would be complex, as it requires understanding how meshing and linearisation adapt to underlying unknown nonstationary and anisotropic features \citep{llamazares2024parameterization}. One potential workaround is to ensure that the mesh resolution is sufficiently fine to accommodate the shortest range of the anisotropic field.
% \subsubsection{Nonseparable model class in \texttt{spacetime}}
Another extension is to consider the nonseparable class of models using the \texttt{INLAspacetime} package \citep{lindgren2020diffusion}. The linearisation would work the same for the diffusion process. In principal, a drift term can be incorporated into the Mat\'ern field to account for the nonseparability. However, a caveat is that more data would be needed to accurately estimate these models.
% \subsubsection{Barrier model class}
% There is a natural extension with barrier models once we have an adaptive mesh setting. The barrier can have in a lower resolution while having higher resolution along the boundary of the barrier. In a worst case scenario where we have a model mismatch the model structure and the data do not account for the barrier. Barrier can interplay with network which the barrier can be interpreted as road or river network.
% \subsubsection{Network graph with \texttt{MetricGraph}}

We have extended the framework to incorporate river channel incision networks for landslide modelling \citep{suen2025influence}, an important development given that landslides frequently cluster along drainage pathways. Including channel data improves the representation of spatial dependencies and strengthens susceptibility predictions. Another promising direction is to link our aggregation framework with metric graph models implemented in the \texttt{MetricGraph} package \citep{bolin_gaussian_2024}. Integrating graph-based representations would enable modelling landslide risk along linear infrastructures such as road networks, supporting applications in road design, maintenance, and earthquake-driven evacuation planning.

\bibliography{landslide}

\newpage

\subsection*{Tables}\label{sec:tab}
% \subsubsection*{Abbreviations}
\textbf{Abbreviations: }Cov.: Covariate Observation, Obs.: Observation, $\Bar{\lambda}(\cdot)$: Mean of Intensity Field,  $\Bar{\check{\lambda}}$: Mean of Intensity Field under Nonlinear misspecification (NL), RastFull: Raster at Full Resolution, RastAgg: Aggregated Raster, PolyAgg: Polygon Aggregations, PointVal: Point Values, OP: Observation Plugin, JU: Joint Uncertainty, VP: Value Plugin, UP: Uncertainty Plugin, $\mathcal{Y}$: Point Pattern, $\mathcal{N}$: Aggregated Count, $\check{\mathcal{Y}}$: Point Pattern under Nonlinear misspecification (NL), $\check{\mathcal{N}}$: Aggregated Count under Nonlinear misspecification (NL).

\begin{table}[H]
\centering
  \begin{tabular}{|c|c|c|c|c|c|}
\hline
     & RastFull & RastAgg & Predicted raster & Mesh (i) & Mesh (ii) \\ \hline
    Area ($\text{km}^2$) & 0.737 & 73.70 & 2.61 & 6.785 & 26.1\\
    Edge length ($\text{km}$) & 0.858 & 8.58 & 1.62 & 3.96 & 7.76\\
    \hline
  \end{tabular}
  \caption{Comparison of the areas and edge lengths of raster cells and interior equilateral mesh triangles. See Section \ref{sec:data} for details.}
  \label{tab:cellsize}
\end{table}

 % Cov.: Covariate Observation, Obs.: Observation, RastFull: Raster at Full Resolution, RastAgg: Aggregated Raster, PolyAgg: Polygon Aggregation, PointVal: Point Values, $\mathcal{Y}$: Point Pattern, $\mathcal{N}$: Aggregated Count, OP: Observation Plugin, JU: Joint Uncertainty, VP: Value Plugin, UP: Uncertainty Plugin.
 
\begin{table}[H]
\centering
\renewcommand{\arraystretch}{1.1}
\begin{tabular}{|c|c|c|*{2}c|c|}
\hline
Cov. & Obs. & Method & MSE & MDS & $\Bar{\lambda}$\\ \hline
RastFull & \multirow{2}{*}{$\mathcal{Y}$} & \multirow{4}{*}{OP} 
       & \textbf{4.226}$\bm{\times 10^{-7}}$ & \textbf{-1.615}$\bm{\times 10^{1}}$ & \multirow{18}{*}{8.915$\times 10^{-4}$} \\
RastAgg & & & 4.559$\times 10^{-7}$ & -1.610$\times 10^{1}$ & \\ 
\cline{1-2} \cline{4-5}
RastFull & \multirow{2}{*}{$\mathcal{N}$} &  
       & \textbf{4.339}$\bm{\times 10^{-7}}$ & \textbf{-1.593}$\bm{\times 10^{1}}$ & \\ 
RastAgg & & & 4.706$\times 10^{-7}$ & -1.591$\times 10^{1}$ & \\
\cline{1-5}
\multirow{8}{*}{PolyAgg} 
  & \multirow{4}{*}{$\mathcal{Y}$} 
  & OP & 5.523$\times 10^{-7}$ & \textbf{-1.586}$\bm{\times 10^{1}}$ & \\
& & JU & \textbf{4.090}$\bm{\times 10^{-7}}$ & -1.573$\times 10^{1}$ & \\  
& & VP & 5.946$\times 10^{-7}$ & -1.521$\times 10^{1}$ & \\ 
& & UP & 4.092$\times 10^{-7}$ & -1.574$\times 10^{1}$ & \\  
\cline{2-5}
  & \multirow{4}{*}{$\mathcal{N}$}
  & OP & 5.995$\times 10^{-7}$ & \textbf{-1.579}$\bm{\times 10^{1}}$  & \\
& & JU & 4.541$\times 10^{-7}$ & -1.578$\times 10^{1}$ & \\  
& & VP & 7.228$\times 10^{-7}$ & -1.517$\times 10^{1}$ & \\ 
& & UP & \textbf{4.540}$\bm{\times 10^{-7}}$ & -1.578$\times 10^{1}$   & \\  
\cline{2-5}
\cline{1-5}
\multirow{6}{*}{PointVal} & \multirow{3}{*}{$\mathcal{Y}$} 
  & JU & \textbf{4.400}$\bm{\times 10^{-7}}$ & \textbf{-1.568}$\bm{\times 10^{1}}$ & \\  
& & VP & 6.019$\times 10^{-7}$ & -1.522$\times 10^{1}$ & \\ 
& & UP & 4.933$\times 10^{-7}$ & -1.560$\times 10^{1}$  & \\  
\cline{2-5}
& \multirow{3}{*}{$\mathcal{N}$} 
  & JU & \textbf{4.690}$\bm{\times 10^{-7}}$ & \textbf{-1.568}$\bm{\times 10^{1}}$ & \\  
& & VP & 7.308$\times 10^{-7}$ & -1.517$\times 10^{1}$ & \\ 
& & UP & 5.354$\times 10^{-7}$ & -1.561$\times 10^{1}$ & \\  
\cline{2-5}
\hline

\end{tabular}
    \caption{Table of the Mean Squared Error (MSE) and Mean Dawid-Sebastiani (MDS) scores for the intensity field $\lambda$ across the models under the aggregation and incomplete covariate field scenarios.}
    \label{tab:score}
\end{table} 

% \begin{table}[H]
% \centering
% % \footnotesize
% \renewcommand{\arraystretch}{1.1}
% \begin{tabular}{|c|c|c|*{2}c|c|}
% \hline
% Cov. & Obs. & Method & MSE & MDS & $\Bar{X}(\cdot)$\\ \hline
% \multirow{5}{*}{PolyAgg} & $\mathcal{Y}$, $\check{\mathcal{Y}}$, $\mathcal{N}$, $\check{\mathcal{N}}$ & VP and UP & 7.915$\times 10^{-6}$ & -1.115$\times 10^{1}$ & \multirow{10}{*}{0.18262}\\ 
% \cline{2-5} 
% & $\mathcal{Y}$ & \multirow{4}{*}{JU} & 7.571$\times 10^{-6}$ & \textbf{-1.116}$\bm{\times 10^{1}}$ & \\  
% & $\mathcal{N}$ & & 7.575$\times 10^{-6}$ & \textbf{-1.116}$\bm{\times 10^{1}}$ & \\ 
% & $\check{\mathcal{Y}}$ &  & 7.561$\times 10^{-6}$ & \textbf{-1.116}$\bm{\times 10^{1}}$ & \\  
% & $\check{\mathcal{N}}$ & & \textbf{7.513}$\bm{\times 10^{-6}}$ & \textbf{-1.116}$\bm{\times 10^{1}}$ & \\ 
% \cline{1-5}
% 
% \multirow{5}{*}{PointVal} & $\mathcal{Y}$, $\check{\mathcal{Y}}$, $\mathcal{N}$, $\check{\mathcal{N}}$ & VP and UP
%    & 1.309$\times 10^{-3}$ & -5.842 & \\ \cline{2-5} 
% &  $\mathcal{Y}$
%    & \multirow{4}{*}{JU} & \textbf{1.115}$\bm{\times 10^{-3}}$ & \textbf{-5.909}  & \\  
% & $\mathcal{N}$ & & 1.163$\times 10^{-3}$ & -5.892 & \\ 
% & $\check{\mathcal{Y}}$ & & 1.247$\times 10^{-3}$ & -5.861 & \\  
% & $\check{\mathcal{N}}$ & & 1.196$\times 10^{-3}$ & -5.882& \\ 
% \hline
% \end{tabular}
% 
% \caption{Table of the Mean Squared Error (MSE) and Mean Dawid-Sebastiani (MDS) scores for the covariate field $x$ across the models under incomplete covariate field scenario.}
% \label{tab:score_cov}
% \end{table} 

\begin{table}[H]
\centering
 \renewcommand{\arraystretch}{1.3}
\begin{tabular}{|c|c|c|*{2}c|c|}
\hline
Cov. & Obs. & Method & MSE & MDS & $\Bar{\check{\lambda}}$ \\ \hline
\multirow{6}{*}{PolyAgg} & \multirow{3}{*}{$\check{\mathcal{Y}}$} 
  & JU & 5.952$\times 10^{-6}$ & -2.690 & \multirow{12}{*}{2.456$\times10^{-3}$}\\  
& & VP & 6.150$\times 10^{-6}$ & 1.354  & \\ 
& & UP & \textbf{5.948}$\bm{\times 10^{-6}}$ & \textbf{-2.901} & \\  
\cline{2-5}
& \multirow{3}{*}{$\check{\mathcal{N}}$} 
  & JU & 5.401$\times 10^{-6}$ & -9.667 & \\  
& & VP & 5.682$\times 10^{-6}$ & -8.125 & \\ 
& & UP & \textbf{5.379}$\bm{\times 10^{-6}}$ & \textbf{-9.797} & \\  
\cline{1-5}

\multirow{6}{*}{PointVal} & \multirow{3}{*}{$\check{\mathcal{Y}}$} 
  & JU & \textbf{5.962}$\bm{\times 10^{-6}}$ & -2.124 & \\  
& & VP & 6.152$\times 10^{-6}$ & 1.504 & \\ 
& & UP & 5.964$\times 10^{-6}$ & \textbf{-2.170} & \\  
\cline{2-5}
& \multirow{3}{*}{$\check{\mathcal{N}}$} 
  & JU & 5.398$\times 10^{-6}$ & -9.723 & \\  
& & VP & 5.702$\times 10^{-6}$ & -7.940 & \\ 
& & UP & \textbf{5.379}$\bm{\times 10^{-6}}$ & \textbf{-9.797}  & \\  
\cline{2-5}
\hline

\end{tabular}
    \caption{Table of the Mean Squared Error (MSE) and Mean Dawid-Sebastiani (MDS) scores for the intensity field $\lambda$ across the models under the Nonlinear misspecification (NL). }
    \label{tab:score_nl}
\end{table}

% Cov.: Covariate Observation, Obs.: Observation, PolyAgg: Polygon Aggregations, PointVal: Point Values, $\mathcal{Y}$: Point Pattern, $\mathcal{N}$: Aggregated Count, $\check{\mathcal{Y}}$: Point Pattern under Nonlinear misspecification, $\check{\mathcal{N}}$: Aggregated Count under Nonlinear misspecification, JU: Joint Uncertainty, VP: Value Plugin, UP: Uncertainty Plugin.

\newpage
\appendix

\counterwithin*{equation}{section}
\renewcommand\theequation{\thesection\arabic{equation}}
\section*{Supplementary Information}
\begin{appendices}

\section{Derivations related to Poisson Point and Count Processes}
Here, we provide the derivations and implementations mentioned in Section \ref{sec:def}. 

\subsection{Derivation of the likelihood function}\label{appendix:intensity}
Here, we provide the derivation of the likelihood function for the Poisson point process specified in Definition \ref{def:loglik}.
Let $\mathcal{Y} = \{y_1,\dots,y_{N(\Omega)}\}$ be some observed point pattern, where $N(\cdot)$ is the random count measure on a bounded domain $\Omega \subset \R^d, d \in \mathbb{N}$, given some intensity function $\lambda$ that is defined on every Borel measurable set $A \subseteq \Omega$. This point process is locally finite such that $N(A) < \infty$ for all $A$;\ this implies the intensity functions are locally integrable (see \citet{baddeley2007spatial} for details). The likelihood function of the observed point pattern can be specified given the intensity function $\lambda$ and $N(\Omega)=n$,
% \begin{align*}
%     p ( \mathcal{Y} | \lambda) &= \mathbb{P} (N(A)=n, \{y_1,\dots,y_n\} | \lambda) \\
%     &= \mathbb{P} \left(N(A)=n | \lambda) \cdot \mathbb{P} (\{y_1,\dots,y_n\}|\lambda, N(A)=n \right)  \\
%     % \; (\text{law of total probability for conditional probability})\\
%     &= \mathbb{P} (N(A)=n | \lambda) \prod_{i=1}^n \mathbb{P} (y_i|\lambda) \\
%     &= e^{-\Lambda(A)} \frac{[\Lambda(A)]^n}{n!}\prod_{i=1}^n \frac{\lambda(y_i)}{\Lambda(A)} \\
%     &= \frac{ e^{-\Lambda(A)}}{n!}\prod_{i=1}^n \lambda(y_i). 
% \end{align*}
% \begin{align*}
%     p_{\mathcal{Y}} ( \mathcal{Y} | \lambda) &= \mathbb{P} (N(A)=n, \{y_1,\dots,y_n\} | \lambda)
% \end{align*}
% By law of total probability for conditional probability given the intensity function $\lambda$, 
\begin{align*}
    p_{Y} ( \mathcal{Y}, N(\Omega)=n  | \lambda) &:= \mathbb{P} (N(\Omega)=n | \lambda) \cdot p (\{y_1,\dots,y_n\}|\lambda, N(\Omega)=n), 
\end{align*}
where  $p (\{y_1,\dots,y_n\}|\lambda, N(\Omega)=n)$ is the joint density function of the observations. Since event locations are conditional independent,
\begin{align*}
    p_{Y} ( \mathcal{Y}, N(\Omega)=n | \lambda) &= \mathbb{P} (N(\Omega)=n | \lambda)\prod_{i=1}^n  p (y_i|\lambda, N(\Omega)=n) \\
    &= e^{-\Lambda(\Omega)} \frac{[\Lambda(\Omega)]^n}{n!} \prod_{i=1}^n \frac{\lambda(y_i)}{\Lambda(\Omega)} \\
    &=\frac{e^{-\Lambda(\Omega)}}{n!}\prod_{i=1}^n \lambda(y_i). 
\end{align*}
For Poisson count process, we have the set of counts $\{N_p := N(A_p)\}_{p=1}^{n_A}$, the likelihood function is 
\begin{align*}
        p_{Y} (\{N_p\}  | \lambda) &:= \mathbb{P} (\{N_p\} | \lambda) \\
        &= e^{-\sum_p \Lambda(A_p)} \prod_p\frac{ \Lambda(A_p)^{N_p}}{N_p!}
\end{align*}
The log-likelihood w.r.t a unit-rate count process is
\begin{equation*}
      \ell(\lambda; \{N_p\}) = |\Omega| -\Lambda(\Omega) + \underbrace{\sum_p N_p \log(\Lambda(A_p)) - \sum_p N_p \log(|A_p|)}_{:= \sum_p \log \lambda(A_p)}.
\end{equation*}

% or following \citep{baddeley2007spatial}, we introduce the space of $k$-tuples
% $$ S^{!k} = \{ (y_1, \dots, y_k): y_i \in S, y_i \neq y_j \text{ for all } i \neq j \}, $$ and define a mapping $I_k: S^{!k} \to \mathsf{N}_k$ by 
% $$I_k(y_1, \dots, y_k) = \delta_{y_1} + \dots + \delta_{y_k}.$$ 
% 
% Then the likelihood function is
% 
% \begin{align*}
% p_{\mathcal{Y}} ( \mathcal{Y} | \lambda) &= 
% \sum^n_{n=0} \mathbb{P} (N(A)=n) \mathbb{P} (I_n(\{y_1,\dots,y_n\}) | \lambda) \\ 
% &=  \sum^n_{n=0} \frac{e^{-\Lambda(A)}}{n!} 
% \int_{A} \dots \int_{A} \mathbf{1}\{ I_n(\{y_1,\dots,y_n\})\} \lambda (d y_1) \dots \lambda (d y_n) \\
% &= \frac{ e^{-\Lambda(A)}}{n!}\prod_{i=1}^n \lambda(y_i). 
% \end{align*}

\section{Derivations related to Computation}
Here, we provide the derivations mentioned in Section \ref{sec:computation}.

\subsection{Derivations of first-order Taylor approximation }\label{appendix:mj}
Derivations of $\ol{M}_j(\bm{u})$ and $\ol{m}_j(\bm{u})$ in equations \eqref{eq:M_j} and \eqref{eq:m_j} from Section \ref{sec:taylorint} are shown here. 
Assuming $\lambda$ is log-linear, 
\begin{align}
    -\int_{\Omega} \lambda (\s) \,d\s &= - \sum_{j=1}^J \int_{\Omega_j} \lambda (\s) \,d\s \nonumber \\
    &= - \sum_{j=1}^J \int_{\Omega_j} \exp[ \underbrace{\log \lambda (\s)}_{\approx \eta (\s)} ] \, d\s \label{eq:explog} \\ &\neq - \sum_{j=1}^J \exp\left(\int_{\Omega_j}  \eta (\s) \, d\s\right) \label{eq:expint}
\end{align}
A common problem shown in equation  \eqref{eq:expint} is that the observed point locations are lost when being aggregated with respect to the subset $\Omega_j$ to overcome the data misalignment. Mathematically speaking, the exponential function and the integral function are not commutative. This can lead to wrong estimation of intensity function thus the effects of the covariates. 

Following equation \eqref{eq:explog}, 
\begin{align}
    &\quad - \sum_{j=1}^J \int_{\Omega_j} \exp[ \eta (\s) ] \,d\s  \nonumber \\
    &= - \sum_{j=1}^J \exp  \underbrace{\left[\log \left( \int_{\Omega_j} \exp[ \eta (\s) ] \,d\s \right) \right] }_{=: M_j(\bm{u}) \approx m_j(\bm{u}_*) + \bm{J}^{(j)}(\bm{u}_*)^\intercal (\bm{u}-\bm{u}_*) =: \ol{M}_j(\bm{u})}  \nonumber \\
    &\approx - \sum_{j=1}^J \sum_{k=1}^{n_j} w_{jk} \exp( \eta (\s_{jk})) \nonumber \\
    &= - \sum_{j=1}^J \exp \left[ \underbrace{\log \left(\sum_{k=1}^{n_j} w_{jk} \exp[ \eta (\s_{jk})] \right)}_{\approx m_j(\bm{u}) \approx m_j(\bm{u}_*) + \bm{J}^{(j)}(\bm{u}_*)^\intercal (\bm{u}-\bm{u}_*) =: \ol{m}_j(\bm{u})} \right] \nonumber
\end{align}

It is worth noting that $w_{jk}$ do not have unique solutions. Thus $\sum_{k=1}^{n_j} w_{jk}$ do not have to sum up to 1 because these account for the contributions of the integration points towards an integral via a stable integration scheme (see details in Appendix \ref{appendix:int}).

When $J \to \infty$, then the function is closer to a piecewise linear function across $\Omega_j$ as $\Omega_j \to 0$, thus converges into the true field.

\subsection{Continuous Linearisation Case}\label{appendix:taylor_c}
Here, we provide the proof for the continuous linearisation case mentioned in Theorem \ref{thm:taylor}.
We assume the following:
\begin{enumerate}
 \item The function $M_j(\bm{u})$ is twice differentiable at some points $\bm{u}_* \in \R$; and
 \item The linear predictor $\eta(\cdot) = b + \bm{Au}$ is an affine function of $\bm{u}$.
\end{enumerate}

Given the subset $\Omega_j$ does not depend on $\bm{u}$ and by Leibniz's integral rule, the terms related to $\partial \Omega_j$ disappear. we spell out the derivation for the first and second-order partial derivatives of the $j$-th term with respect to the $l$ and $l'$, for $l,\, l' = 1, \dots, L$,
\begin{align}
    \frac{\partial M_j}{\partial u_l} &=  \frac{\int_{\Omega_j} \lambda (s) \nabla_u \eta (s) \; ds}{\int_{\Omega_j} \lambda (s) \; ds}  < \infty, \label{eq:1M}\\     
    \frac{\partial^2 M_j}{\partial u_l \partial u_{l'}} &=  \frac{\partial }{\partial u_{l'}} \frac{\int_{\Omega_j} \lambda (s) \nabla_u \eta (s) \; ds}{\int_{\Omega_j} \lambda (s) \; ds}  \nonumber \\
     &= \frac{\int_{\Omega_j} \lambda (s) \nabla_u \eta (s)\nabla_u^\intercal \eta(s) \; ds}{\int_{\Omega_j} \lambda(s) \; ds} - \frac{\int_{\Omega_j} \lambda (s) \nabla_u\eta(s) \; ds}{\int_{\Omega_j} \lambda(s) \; ds} \cdot \frac{ \int_{\Omega_j} \lambda (s) \nabla_u^\intercal \eta(s) \; ds}{\int_{\Omega_j} \lambda(s) \; ds} < \infty. \label{eq:2M}
\end{align}
Hence, equations \eqref{eq:1M} and \eqref{eq:2M} are bounded because $\int_{\Omega_j} \lambda(s) \; ds$ is bounded by definition and $\nabla_u \eta (\cdot) = \bm{A}$ is bounded. We interpret equation  \eqref{eq:2M} as the difference between the product of gradient average and a product of average gradient. Recall that the predictor $\eta(\cdot)$ comprises of covariate contribution here; without such contribution, there might be identifiability issue. 

The subsets can be specified as \texttt{block} argument with \texttt{ibm\_eval} function in the \inlabru \ package.

\begin{Schunk}
\begin{Sinput}
R> ibm_eval(..., input = list(block = .block, weights = weight))
\end{Sinput}
\end{Schunk}

\subsection{Discretised Linearisation Case}\label{appendix:taylor_d}
Here, we provide the proof for the discretised linearisation case mentioned in Theorem \ref{thm:taylor}.
We assume the following:
\begin{enumerate}
 \item The function $m_j(\bm{u})$ are twice differentiable at some points $\bm{u}_* \in \R$;
 \item The linear predictor $\eta(\cdot) = b + \bm{Au}$ is an affine function of $\bm{u}$.
\end{enumerate}

Given the subset $\Omega_j$ does not depend on $\bm{u}$ and by Leibniz's integral rule, the terms related to $\partial \Omega_j$ disappear, we spell out the derivation for the first and second-order partial derivatives of the $j$-th term with respect to the $l$ and $l'$ for $l,\, l' = 1, \dots, L$. 

We denote the following vectors and matrices, for $j = 1, \dots, J$, and $k = 1, \dots, n_j$, 
\begin{align*}
    \s_j &:= \begin{bmatrix} 
                    \vdots \\
                    \s_{jk} \\
                    \vdots 
                    \end{bmatrix},
                    \\
    \eta_{jk} &= \eta (\s_j ; \bm{u}) := \begin{bmatrix} 
                        \vdots \\
                        \eta(\s_{jk}) \\
                        \vdots 
                        \end{bmatrix} = b^{(j)} + \bm{A}^{(j)} \bm{u}, \\
        \bm{J}^{(j)} &= \begin{bmatrix} 
                    \frac{\partial  m_j}{\partial u_1} & \dots &  \frac{\partial  m_j}{\partial  u_L }
                    \end{bmatrix}, \\ 
        \bm{H}^{(j)} &= \begin{bmatrix} 
                    \frac{\partial^2 m_j}{\partial u_1^2} & \dots &  \frac{\partial^2 m_j}{\partial u_1 \partial u_{L}}\\
                    \vdots & \ddots & \vdots \\
                     \frac{\partial^2 m_j}{\partial u_L \partial u_1} & \dots &  \frac{\partial^2 m_j}{\partial u_L^2}
                    \end{bmatrix}.
\end{align*}

Then for the $j$-th component of the first-order derivative, we denote $q_{jk} =w_{jk} \exp (\eta_{jk} )$ and $ \bm{A}^{(j)}_{kl} = \frac{\partial \eta_{jk}}{\partial u_l}$ and we have 
\begin{align*}
    \frac{\partial m_j}{\partial u_l} &= \frac{\sum_k w_{jk} \exp (\eta_{jk} ) \frac{\partial \eta_{jk}}{\partial u_l}}{\sum_k w_{jk} \exp (\eta_{jk} )} \\ 
    &= \frac{\sum_k q_{jk} \frac{\partial \eta_{jk}}{\partial u_l}}{\sum_k q_{jk}} \\
    &= \frac{\sum_k q_{jk} \bm{A}^{(j)}_{kl}}{\sum_k q_{jk}} < \infty\\
    &= \frac{\bm{A}^{(j)^\intercal} \bm{q}^{(j)}}{\bm{1}^\intercal \bm{q}^{(j)}}.
    \end{align*}
Since the term $\bm{q}^{(j)}$ is bounded by definition given a stable integration scheme, thus the Jacobian term is bounded. For the second-order derivative, by the quotient rule $\frac{d}{dx} \frac{f}{g} = \frac{f' g - f g'}{g^2}$, and for some $l, l' = 1, \dots, L$, we have
\begin{align*}
    \frac{\partial^2 m_j}{\partial u_l \partial u_{l'}} &= \frac{(\sum_k q_{jk}) (\sum_k q_{jk} \bm{A}^{(j)}_{kl} \bm{A}^{(j)}_{kl'}) - (\sum_k q_{jk} \bm{A}^{(j)}_{kl} ) (\sum_k q_{jk} \bm{A}^{(j)}_{kl'})}{(\sum_k q_{jk})^2} < \infty.
\end{align*}
Hence, the matrix $\bm{J}^{(j)}$ is a ratio with nominator adjusted by the gradient of the linear predictor function at $\bm{u}$. $\bm{H}^{(j)}$ is the difference between the ratios. For $l,\, l' = 1, \dots, L$, $\frac{\partial m_j}{\partial u_l} < \infty$ and $\frac{\partial^2 m_j}{\partial u_l \partial u_{l'}} < \infty $ are bounded. 
To further simplify the notation, we denote 
\begin{align*}
  \bm{q}^{(j)} &= \bm{q}_{j\cdot}, \\
  \wh{\bm{q}}^{(j)} &:= \frac{\bm{q}^{(j)}}{\sum_k \bm{q}_{jk}} , \\
  \bm{d}^{(j)} &= \text{diag} (\bm{q}^{(j)}), \\
  \wh{\bm{d}}^{(j)} &= \text{diag}(\frac{\bm{q}^{(j)}}{\sum_k \bm{q}_{jk}}).
\end{align*}
Then we can simplify the expressions into,

\begin{align*}
    \bm{J}^{(j)} &= \bm{A}^{(j)^\intercal} \wh{\bm{q}}^{(j)} 
    \\
    \bm{H}^{(j)} &= \frac{(\bm{1}^{\intercal} \bm{q}^{(j)}) (\bm{A}^{(j)^\intercal} \bm{d}^{(j)} \bm{A}^{(j)}) - (\bm{A}^{(j)^\intercal} \bm{q}^{(j)})  (\bm{q}^{(j)^\intercal} \bm{A}^{(j)})}{(\bm{1}^\intercal \bm{q}^{(j)})^2} \\
    &= \bm{A}^{(j)^\intercal}  \wh{\bm{d}}^{(j)} \bm{A}^{(j)} - \bm{A}^{(j)^\intercal} \wh{\bm{q}}^{(j)} \wh{\bm{q}}^{(j)^\intercal} \bm{A}^{(j)}.
\end{align*}

\subsection{Aggregation Mapper}\label{appendix:agg}
Here, we provide the implementation of the logarithmic sum, i.e.\ $\ol{m}_j(\bm{\cdot})$ approximation mentioned in Section \ref{sec:taylorint}. 
% (double check) n\_block should be a vector of the same length as the state( sensibly speaking, or NULL, with NULL equivalent to all-1. If weights is NULL, it's interpreted as all-1.
The two mapper functions, \texttt{bm\_aggregate} and \\ \texttt{bm\_logsumexp}, in the \inlabru \ package, construct mapper objects to compute the blockwise sum and logarithmic sum of exponential terms in equation  \eqref{eq:mj}. Here the blocks (\texttt{sf\_obj} in the sample code below) are defined by $\Omega_j$. By default, \texttt{n\_block=NULL}, the maximum block index in the inputs is used, i.e.\ \texttt{nrow(sf\_obj)}. Otherwise,  \texttt{n\_block} can be defined as the predetermined number of output blocks by users.

A shortcut tool for aggregating these latent state vectors $\bm{u}$ to effect scalars $\ol{m}_j(\bm{u})$ for each \inlabru \ latent model component in equation  \eqref{eq:mj} are provided by the low level mapper \hfill \break \logsumexp. This mappers aggregates elements of \texttt{exp(state)}, where \texttt{state} are components, with optional non-negative weighting, and then takes the \texttt{log()}, so it can be used 
\begin{itemize}
    \item for \texttt{rescale=FALSE(default)}, $v_k=\log[\sum_{i\in I_k} w_k \exp(u_i)]$, and
    \item for \texttt{rescale=TRUE}, $v_k=\log[\sum_{i\in I_k} w_k \exp(u_i) / \sum_{i\in I_k} w_k]$.
\end{itemize}
The mapper \logsumexp \ relies on the input handling methods for \hfill \break \aggregate, but also allows the weights to be supplied on a logarithmic scale as \texttt{log\_weights}.

To avoid numerical overflow, it uses the common method of internally
shifting the state blockwise via $\Omega_j$ with 
\hfill \break  \texttt{(state-log\_weights)[block] - max((state-log\_weights)[block])}, \\ 
and shifts the result back afterwards.

% \section{Subset}
% \subsection{Perimeter-to-area ratio}
% \label{appendix:p2a}
% We wish to minimise the perimeter-to-area ratio so that the mesh is less exposed to discretisation effect. Equilateral triangle is shown here since they are most ideal triangular mesh, it is extendable to other types of triangle. Let the area be unit $1$ and denote $a$ and $r$ as the edge length of the polygon and radius of the circle respectively. 
% \begin{center}
%     \begin{tabular}{|c|c|c|c|c|}
%     \hline
%     Shape & Equilateral Triangle &  Square & Hexagon & Circle\\\hline
%     Area formula & $\frac{a^2}{2}$ & $a^2$ & $\frac{3\sqrt{3}a^2}{2}$ & $\pi r^2$ \\
%     Perimeter-to-area ratio  & $3\sqrt{2} \approx 4.24$ & $4$ &  $2\sqrt{2\sqrt{3}} \approx 3.72$ & $2\sqrt{\pi}\approx3.54$\\ \hline
%     \end{tabular}
% \end{center}

The reconstruction of the continuous covariate field described in Section \ref{sec:icap} involves evaluating the aggregated field using a piecewise linear mesh via the \texttt{bm\_aggregate(.., rescale = TRUE)} mapper. The option \texttt{rescale = TRUE} specifies that $z_p$ is computed as the mean of the covariate field within the $p$-th polygon. A sample code snippet is provided below.

\begin{Schunk}
\begin{Sinput}
R> # sf_obj: the aggregated polygon sf object
R> # matern: the model defined by inla.spde2.pcmatern here
R> cmp <- ~ x(main = geometry, model = matern)
R> agg_mapper <- bm_aggregate(
+    rescale = TRUE,
+    n_block = NULL
+  )
R> fml <- ~ ibm_eval(agg_mapper,
+    input = list(
+      block = .block,
+      weights = weight
+    ),
+    state = x
+  )
\end{Sinput}
\end{Schunk}

In \inlabru{} version 2.12.0.9021, the standard mapper class names were shortened from \texttt{bru\_mapper\_<type>} to \texttt{bm\_<type>} for readability, with old objects automatically converted and legacy constructors redirected to the new ones.

\subsection{Nonlinear Taylor Expansion}\label{appendix:nonlinear}
Here, we provide the derivation of a further Taylor expansion on a nonlinear predictor expression mentioned in Section \ref{sec:com}. This is an extension from the linearisation case in Appendices \ref{appendix:taylor_c} and \ref{appendix:taylor_d}.
Let $f$ be some known non-linear function and once differentiable, 
% not verified yet 
\begin{align*}
    \eta_{jk} &= b^{(j)} + \bm{A}^{(j)} f(\bm{u}) \text{, and }
    \frac{\partial \eta_{jk}}{\partial u_l} = \bm{A}^{(j)} \frac{\partial f(\bm{u})}{ \partial u_l} < \infty.
\end{align*}
With first-order Taylor expansion, we have
\begin{equation*}
    f(\bm{u}) = f(\bm{u}_*) + \bm{J}(\bm{u}_*)^\intercal (\bm{u} - \bm{u}_*) + \mathcal{O}(\| \bm{u} - \bm{u}_*\|^2).
\end{equation*}

\subsection{Linearisation Difference}
\label{appendix:lin_diff}
Here, we provide the derivation of the difference between the continuous and discretised linearisation cases mentioned in Theorem \ref{thm:exp_diff}.

By definition, $\exp(x) := \sum_{k=0}^\infty \frac{x^k}{k!} = 1+x+\frac{x^2}{2}+\dots$, we ignore the higher-order terms that are more than or equal to cubic. Then we have the difference of the domain contribution terms between the continuous and discretised linearisation cases is
\begin{align*}
\log \frac{ \wt{\mathbb{P}} ( \bm{y} | \bm{u}, \bm{\theta})}{\ol{\mathbb{P}} ( \bm{y} | \bm{u}, \bm{\theta}) } 
&= \sum_j^J e^{\ol{m}_j} - \sum_j^J e^{m_j} \\ 
&= \sum_j^J ( e^{\ol{m}_j} -  e^{m_j} ) \\
&= \sum_j^J e^{\ol{m}_j} (1 - e^{m_j-\ol{m}_j}) \\
&= \sum_j^J \left[e^{ \ol{m}_j(\bm{u}_*) + \bm{J}_j (\bm{u}_*)^\intercal (\bm{u} - \bm{u}_*)}\right] \\
&\phantom{= }\cdot \left[1-\exp{\left\{\frac{1}{2} (\bm{u} - \bm{u}_*)^\intercal \bm{H}^{(j)} (\bm{u} - \bm{u}_*)+\mathcal{O}(\| \bm{u} - \bm{u}_*\|^3)\right\}} \right]  \\
&= -\sum_j^J \left[e^{ \ol{m}_j(\bm{u}_*) + \bm{J}_j (\bm{u}_*)^\intercal (\bm{u} - \bm{u}_*)}\right] \\
&\phantom{= }\cdot \left[\frac{1}{2} (\bm{u} - \bm{u}_*)^\intercal \bm{H}^{(j)} (\bm{u} - \bm{u}_*)+\mathcal{O}(\| \bm{u} - \bm{u}_*\|^3)\right] \\ 
&= -\sum_j^J \left[ e^{\ol{m}_j(\bm{u}_*)} \left(1+\bm{J}_j (\bm{u}_*)^\intercal (\bm{u} - \bm{u}_*) + \mathcal{O}(\| \bm{u} - \bm{u}_*\|^2)\right) \right] \\ 
&\phantom{= } \cdot \left[\frac{1}{2} (\bm{u} - \bm{u}_*)^\intercal \bm{H}^{(j)} (\bm{u} - \bm{u}_*)+\mathcal{O}(\| \bm{u} - \bm{u}_*\|^3)\right] \\
&= - \sum_j^J e^{\ol{m}_j(\bm{u}_*)}\left[\frac{1}{2} (\bm{u} - \bm{u}_*)^\intercal \bm{H}^{(j)} (\bm{u} - \bm{u}_*) + \mathcal{O}(\| \bm{u} - \bm{u}_*\|^3) \right] 
\end{align*}

\subsection{Expectation Difference}
\label{appendix:exp_diff}
Here, we provide the derivation of the expectation difference between the continuous and discretised linearisation cases mentioned in Theorem \ref{thm:exp_diff}.
We denote $\bm{\mu}_{\bm{\theta}}=\E(\bm{u}|\bm{y}, \bm{\theta})$ and $\bm{Q_\theta}^{-1} = \textsf{Var} \left(\bm{u}|\bm{y}, \bm{\theta}\right) = \E\left[(\bm{u} - \bm{\mu})(\bm{u} - \bm{\mu})^\intercal | \bm{y} \right]$.

\begin{align*}
\E_{\bm{u} \sim \mathsf{N}(\bm{\mu}, \bm{Q}^{-1})}\left(\log \frac{ \wt{\mathbb{P}} ( \bm{y} | \bm{u}, \bm{\theta})}{ \ol{\mathbb{P}} ( \bm{y} | \bm{u}, \bm{\theta}) }\right) 
&= \E_{\bm{u} \sim \mathsf{N}(\bm{\mu}, \bm{Q}^{-1})}(\sum_{j=1}^{J} e^{\ol{m}_j} - \sum_{j=1}^{J} e^{m_j}) \\
&= \E\left(- \sum_j^J e^{\ol{m}_j(\bm{u}_*)}\left[\frac{1}{2} (\bm{u}-\bm{u}_*)^\intercal \bm{H}^{(j)} (\bm{u}-\bm{u}_*) + \mathcal{O}(\| \bm{u} - \bm{u}_*\|^3) \right] \right) \\
&= - \frac{1}{2}\sum_j^J e^{\ol{m}_j(\bm{u}_*)} \E \left( \tr \left[ (\bm{u}-\bm{u}_*)^\intercal \bm{H}^{(j)} (\bm{u}-\bm{u}_*) \right]\right)  \\
&\quad + \mathcal{O}\left(\E \left[\| \bm{u} - \bm{u}_*\|^3\middle|\bm{y}\right] \right) \\
&= - \frac{1}{2}\sum_j^J e^{\ol{m}_j(\bm{u}_*)}  \\
&\quad \cdot \E \left( \tr \left\{  \bm{H}^{(j)} \left[(\bm{u}-\bm{\mu}) + (\bm{\mu} -\bm{u}_*)\right] \left[(\bm{u}-\bm{\mu}) + (\bm{\mu} -\bm{u}_*)\right]^\intercal \right\}\right) \\
&\phantom{= }  + \mathcal{O}\left(\E \left[\| \bm{u} - \bm{u}_*\|^3\middle|\bm{y}\right] \right) \\
&= - \frac{1}{2}\sum_j e^{\ol{m}_j(\bm{u}_*)} \left[\mathsf{tr} (\bm{H}^{(j)}\bm{Q}^{-1}) + (\bm{\mu}-\bm{u}_*)^\intercal \bm{H}^{(j)} (\bm{\mu}-\bm{u}_*)\right] \nonumber \\ 
&\phantom{= } + \mathcal{O}\left(\E \left[\| \bm{u} - \bm{u}_*\|^3\middle|\bm{y}\right] \right)
\end{align*}

\subsection{Kullback-Leibler (KL) Divergence}
\label{appendix:KL}
Here, we provide the derivation of the KL divergence between the continuous and discretised linearisation cases mentioned in Theorem \ref{thm:KL}.
We follow the proof of Theorem 1 in \citet{lindgren2024inlabru} and denote 
\begin{equation}
\bm{G} = \sum_i \bm{J}^{(i)}|_{\bm{u}_*} \bm{H}^{(i)}|_{\bm{u}_*} = - \sum_i e^{\ol{m}_i} \bm{H}^{(i)}.
\end{equation}
We denote $\ol{\bm{m}}_{\bm{\theta}}=\pE_{\ol{p}}(\bm{u}|\bm{y},\bm{\theta})$ and
$\ol{\bm{Q}}^{-1}_{\bm{\theta}}=\pCov_{\ol{p}}(\bm{u},\bm{u}|\bm{y},\bm{\theta})$. Hence, the first term in the KL divergence is
\begin{align}
\E_{\ol{p}}\left[    \log \wt{p}(\bm{y}|\bm{u},\bm{\theta}) -
    \log \ol{p}(\bm{y}|\bm{u},\bm{\theta})\right]
    \nonumber
 &=
    \frac{1}{2}
    \tr(\bm{G}\ol{\bm{Q}}_{\bm{\theta}}^{-1}) + \frac{1}{2} (\ol{\bm{m}}_{\bm{\theta}}-\bm{u}_*)^\top\bm{G}(\ol{\bm{m}}_{\bm{\theta}}-\bm{u}_*) \\
&\phantom{= } + \mathcal{O}\left(\pE_{\ol{p}}\left[\|\bm{u}-\bm{u}_*\|^3\middle|\bm{y},\bm{\theta}\right]\right)
.
\end{align}

\begin{proof}
The KL divergence for the first and second Taylor expansion can be expressed as
\begin{equation}
\mathsf{KL}\left(\wt{\mathbb{P}}(\boldsymbol{u}|\bm{y},\boldsymbol{\theta}) \,\middle\|\,\ol{\mathbb{P}}(\boldsymbol{u}|\bm{y},\boldsymbol{\theta})\right) = E_{\wt{\mathbb{P}}}\left[
\log\frac{\wt{\mathbb{P}}(\bm{y}|\boldsymbol{u},\boldsymbol{\theta})}{\ol{\mathbb{P}}(\bm{y}|\boldsymbol{u},\boldsymbol{\theta})}
\right]
-
\log\frac{\ol{\mathbb{P}}(\bm{y}|\boldsymbol{\theta})}{\wt{\mathbb{P}}(\bm{y}|\boldsymbol{\theta})}
\end{equation}
The details of the first term and corresponding derivation see \citet{lindgren2024inlabru}. 

\begin{align*}
\pE_{\ol{p}}\left[ \log \wt{\mathbb{P}}(\bm{y}|\bm{u},\bm{\theta}) -
    \log \ol{\mathbb{P}}(\bm{y}|\bm{u},\bm{\theta})\right]
    \nonumber
 &=
 - \frac{1}{2}
    \tr(\bm{G}\ol{\bm{Q_\theta}}^{-1}) - \frac{1}{2} (\ol{\bm{\mu}}_{\bm{\theta}}-\bm{u}_*)^\top\bm{G}(\ol{\bm{\mu}}_{\bm{\theta}}-\bm{u}_*) \\
&\phantom{= } + \mathcal{O}\left(\pE_{\ol{p}}\left[\|\bm{u}-\bm{u}_*\|^3\middle|\bm{y},\bm{\theta}\right]\right)
\end{align*}
The second term above can be expressed as
\begin{equation}
    - \mathsf{tr} \left[(\sum_j e^{\ol{m}_j(\bm{u}_*)}\bm{H}^{(j)})(\bm{Q_\theta} + \sum_j e^{\ol{m}_j(\bm{u}_*)}\bm{H}^{(j)})^{-1} \right],
    % TODO add the E former part (ln det) from the paper G matrix = \left[(\sum_j e^{\ol{m}_j(\bm{u}_*)}\bm{H}^{(j)}), (Q bar - G)=\bm{Q_\theta}_{prior} + \sum_j e^{\ol{m}_j(\bm{u}_*)}\bm{H}^{(j)})^{-1}
\end{equation}
where $\bm{Q_\theta}$ is the Gaussian prior on the latent variable $\bm{u}$.
\end{proof}
\subsection{Stable Integration Scheme}\label{appendix:int}
Here, we provide the technical details of a stable integration scheme mentioned in Sections \ref{sec:taylorint} and \ref{sec:data}. We present the stable integration scheme for $\R^2$ implemented in \inla \ and \inlabru \ packages. There are alternative stable integration schemes to be implemented in the pipeline. 

We refer to a stable integration scheme as a bounded estimate of the likelihood which is the combination of the domain and the observed contribution, as shown in equation \eqref{eq:loglik}, is bounded, i.e.\ $\ell(\lambda|\mathcal{Y}) < \infty$. The calculation of observed contribution is evaluated at the observed point which further details available in \citet{simpson_going_2016}. 

We denote a function on a triangulation of space $f(\cdot)$ with $t = 1,\dots,n_t$ and $v=1,2,3$, as the indices of the triangular mesh and the vertex respectively and $\mathcal{T}_t$ referring to the $t$-th element. We further define $f_{(t,v)}:= f(s_{(t,v)})$ and $\sum_{t}^{n_t}\sum_{v \in \mathcal{T}_t } w_{(t,v)} = 1$, for some weights $\{w_{(t,v)}\}$, and some vertices $\{s_{(t,v)}\} \subset \Omega$ inside the sample space. The domain contribution can be approximated via integration points as,
\begin{equation}
    \int_{\Omega} f(s) \,ds
    \approx \sum_{t}^{n_t} \sum_{v \in \mathcal{T}_t } w_{(t,v)}f_{(t,v)}; 
    \label{eq:intsch}
\end{equation}
A 3-column integer matrix with 1-based vertex indices for each triangle $(t,v)$ (with reference to \texttt{mesh\$loc}) is stored under \texttt{mesh\$graph\$tv} in the mesh object under the \texttt{fmesher}\ package.  

\begin{figure}[H]
    \centering
    \begin{tikzpicture}
  \coordinate (s0) at (3,4/3);
  \coordinate (s1) at (0,0);
  \coordinate (s2) at (2,4);
  \coordinate (s3) at (7,0);

  \node [above right] at (s0) {$s_0$};
  \node [below left] at (s1) {$s_1$};
  \node [above]  at (s2) {$s_3$};
  \node [below right] at (s3) {$s_2$};

    \draw (3.9,1.6) coordinate[label=$T_1$] ;
    \draw (1.8,1.5) coordinate[label=$T_2$] ;
    \draw (3.1,0.4) coordinate[label=$T_3$] ;
    
    \draw[-] (s3)--(s0)--(s1)--(s3)--(s2)--(s0)--(s1)--(s2);
    
    \end{tikzpicture}
    \caption{Barycentric coordinates of a triangle. The volume of $T_v$ for $v=1,2,3$ is $|T_{(t,v)}|$.}
    \label{fig:bary}
\end{figure}
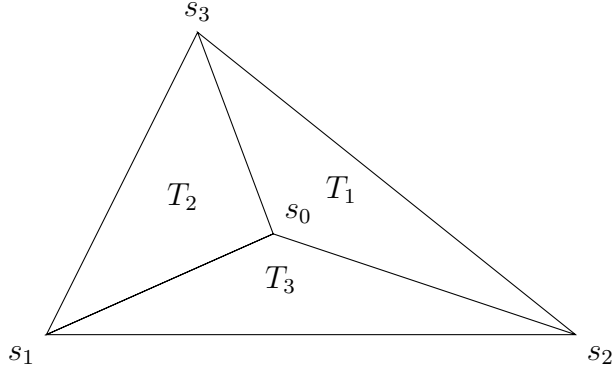

We illustrate with a single triangle, i.e.\ $t=1$, as in Figure \ref{fig:bary}. We denote the volume of the triangle $|T_{(t,\cdot)}| = \sum_{v=1}^{3} |T_{(t,v)}|$ where $|T_{(t,v)}|$ are the volume of the sub-triangles through barycentric subdivision. We assume a piecewise linear function within the triangle $\Delta s_1 s_2 s_3$. The barycentre can be computed 
\begin{equation}
    s_{t_0} = \sum_{v=1}^{3} w_{(t,v)} s_{(t,v)}; \quad w_{(t,v)} = \frac{|T_{(t,v)}|}{|T_{(t,\cdot)}|},    
\end{equation}
where $w_{t,v}$ is known as the barycentric coordinates with respect to $s_{t_0}$. Since the triangle is piecewise linear, $ w_{t,v} = \psi_{k(t,v)}(s_{t_0})$ specified in equation \eqref{eq:psi}. 
 The weighted sum of a function evaluated at these points is the integral of this function approximated by linear basis functions $\psi_{k(t,v)}$ as in equation \eqref{eq:psi}. Hence, the function evaluated at the barycentre
\begin{equation}
        f(s_{t_0}) = \sum_{v=1}^{3} w_{(t,v)} f_{(t,v)}.
\end{equation}

\begin{figure}[H]
    \centering
    \begin{tikzpicture}
\coordinate (s1o) at (3/2,4/3);
 \coordinate (s1) at (0,0);
 \coordinate (s1') at (9,4);
 \coordinate (s13) at (1,2);
 \coordinate (s2) at (2,4);
 \coordinate (s23) at (9/2,2);
 \coordinate (s3) at (7,0);
 \coordinate (s12) at (7/2,0);
 \coordinate (s1'2) at (11/2,4);
 \coordinate (s1'3) at (8,2);

 \node [below left] at (s1) {$s_1$};
 \node [below left] at (2.8,0) {$s_{12_c}$};
 \node [above left] at (1.7,3.2) {$s_{13_c}$};
 \node [left] at (s13) {};
 \node [above]  at (s2) {$s_3$};
 \node [above right]  at (s23) {};
 \node [below right] at (s3) {$s_2$};
 \node [above]  at (s1') {$s_{1'}$};
 \node [below ] at (s12) {};
 \node [above]  at (s1'2) {};
 \node [right]  at (s1'3) {};

 \draw[-] (s1)-- node[sloped] {$|$}(s13) -- node[sloped] {$|$} (s2)--node[sloped] {$||$}(s23)--node[sloped] {$||$}(s3)--node[sloped] {$|||$}(s12)--node[sloped] {$|||$}(s1);
 \draw[-] (s12)--(s13)--(s23)--(s12);
 \draw[dashed] (s1')--(s2)--(s1')--(s3);
 \draw[dashed] (s1'3)--(s1'2)--(s23)--(s1'3)--(s23);
 \draw[dashdotted] (3/2,4)--(3,-1);

 \draw (3/2,2/3) coordinate[label=$s_{1_0}$] (s1o) circle (1.5pt);
 \draw[densely dashed] (15/2,10/3) coordinate[label=$s_{1'_0}$] (s1'o) circle (1.5pt);
 \draw (5,2/3) coordinate[label=$s_{3_0}$] (s3o) circle (1.5pt);
 \draw[densely dashed] (13/2,4/3) coordinate[label=$s_{2'_0}$] (s2'o) circle (1.5pt);
 \draw (5/2,8/3) coordinate[label=$s_{2_0}$] (s2o) circle (1.5pt);
 \draw[densely dashed] (4,10/3) coordinate[label=$s_{3'_0}$] (s3'o) circle (1.5pt);
 \draw (3,4/3) coordinate[label=$s_{0_0}$] (s0o) circle (1.5pt);
 \draw[densely dashed] (6,8/3) coordinate[label=$s_{0'_0}$] (s0'o) circle (1.5pt);

    \end{tikzpicture}
    \caption{Illustration of integration points with mirror triangle and mid-edge subdivision. Circle: integration points; Points above the diagonal ($s_2 s_3$) should be reflected into the lower triangle. Solid line: triangular mesh; Dashed line: mirror image; Dashdotted line: cut.}
    \label{fig:mir}
\end{figure}
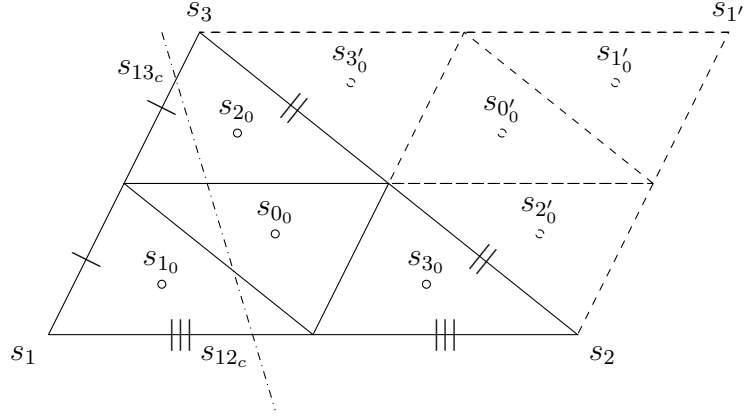

\begin{figure}[H]
% https://tex.stackexchange.com/questions/403112/isosceles-triangle-equal-side-symbol
    \centering
    \begin{tikzpicture}[arrow/.style = {draw=#1,-{Stealth[]}, 
                shorten >=1mm, shorten <=1mm}, % styles of arrows
arrow/.default = black,]
    \coordinate (s0) at (3,4/3);
    \coordinate (s1o) at (3/2,4/3);
    \coordinate (s1) at (0,0);
    \coordinate (s1') at (9,4);
    \coordinate (s12) at (1,2);
    \coordinate (s2) at (2,4);
    \coordinate (s23) at (9/2,2);
    \coordinate (s3) at (7,0);
    \coordinate (s0') at (6,8/3);

    \coordinate (s121) at ($(s1)!1/3!(s2)$);
    \coordinate (s122) at ($(s1)!2/3!(s2)$);
    \coordinate (s131) at ($(s1)!1/3!(s3)$);
    \coordinate (s132) at ($(s1)!2/3!(s3)$);
    \coordinate (s231) at ($(s2)!2/3!(s3)$);
    \coordinate (s232) at ($(s2)!1/3!(s3)$);
    \coordinate (s1'31) at ($(s1')!2/3!(s3)$);
    \coordinate (s1'32) at ($(s1')!1/3!(s3)$);
    \coordinate (s1'21) at ($(s1')!2/3!(s2)$);
    \coordinate (s1'22) at ($(s1')!1/3!(s2)$);

    \node [below left] at (s1) {$s_1$};
    \node [above] at (s2) {$s_3$};
    \node [below right] at (s3) {$s_2$};
    \node [above]  at (s1') {$s_{1'}$};

    \draw[-] (s1)--(s2)--(s3)--(s1);
    
    \draw[-] (s232)--(s131)--(s121)--(s231);
    \draw[-] (s231)--(s132)--(s122)--(s232);
    
    \draw[dashed] (s1'22)--(s1'32)--(s232);
    \draw[dashed] (s1'22)--(s231);
    \draw[dashed] (s1'21)--(s232);
    \draw[dashed] (s1'21)--(s1'31)--(s231);
    \draw[dashed] (s1')--(s2)--(s1')--(s3);
    \draw[dashed] (s1')--(s2)--(s1')--(s3);
    \node[circle,draw=black, fill=white, inner sep=0pt,minimum size=3pt] at (barycentric cs:s1=1,s131=1,s121=1) {};
    \node[circle,draw=black, fill=white, inner sep=0pt,minimum size=3pt] at (barycentric cs:s1=1,s131=1,s121=1) {};
    \node[circle,draw=black, fill=white, inner sep=0pt,minimum size=3pt] at (barycentric cs:s121=1,s0=1,s122=1) {};
    \node[circle,draw=black, fill=white, inner sep=0pt,minimum size=3pt] at (barycentric cs:s131=1,s0=1,s132=1) {};
    \node[circle,draw=black, fill=white, inner sep=0pt,minimum size=3pt] at (barycentric cs:s231=1,s0=1,s232=1) {};
    \node[circle,draw=black, fill=white, inner sep=0pt,minimum size=3pt] at (barycentric cs:s122=1,s232=1,s2=1) {};
    \node[circle,draw=black, fill=white, inner sep=0pt,minimum size=3pt] at (barycentric cs:s231=1,s3=1,s132=1) () {};
    
    \node[circle,draw=black, fill=white, inner sep=0pt,minimum size=3pt] at (barycentric cs:s122=1,s0=1,s232=1) (m1) {};
    \node[circle,draw=black, fill=white, inner sep=0pt,minimum size=3pt] at (barycentric cs:s132=1,s0=1,s231=1) (m2) {};
    \node[circle,draw=black, fill=white, inner sep=0pt,minimum size=3pt] at (barycentric cs:s131=1,s0=1,s121=1) (m3) {};
    
    \node[circle,densely dashed,draw=black, fill=white, inner sep=0pt,minimum size=3pt] at (barycentric cs:s1'21=1,s0'=1,s232=1) (m1') {};
    \node[circle,densely dashed,draw=black, fill=white, inner sep=0pt,minimum size=3pt] at (barycentric cs:s231=1,s0'=1,s1'31=1) (m2') {};
    \node[circle,densely dashed,draw=black, fill=white, inner sep=0pt,minimum size=3pt] at (barycentric cs:s1'32=1,s0'=1,s1'22=1) (m3') {};

    \draw[arrow, black] (m1'.north) to [out=180, in=45] (m1);
    \draw[arrow, black] (m2'.north) to [out=180, in=45] (m2);
    \draw[arrow, black] (m3'.north) to [out=180, in=45] (m3);
    \end{tikzpicture}
    \caption{Illustration of integration points with mirror triangle and mid-edge subdivision. Circle: integration points; Solid line: triangular mesh; Dashed line: mirror image}
    \label{fig:dual}
\end{figure}
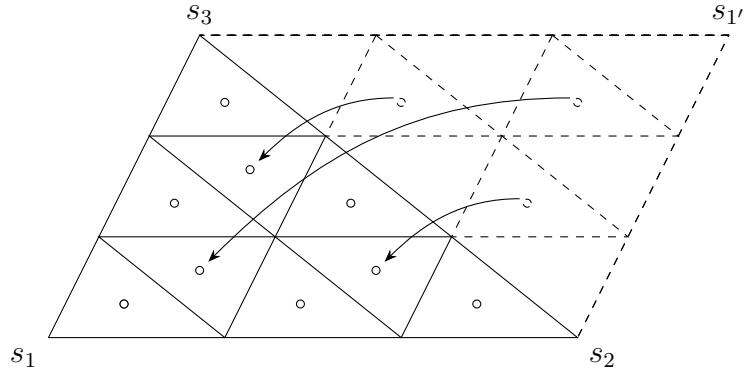

We mirror the triangular element to create a parallelogram and distribute integration points $s_{i_0}$ for $i=1,\dots,n$ over the parallelogram in Figure \ref{fig:mir} and halve the integration in the end. Generally speaking, the more integration points are, the more accurate the approximation is. It is an implementation trick to avoid considering whether the integration points are inside or outside the triangle.  The integration points are the barycentres of the sub-triangles generated via mid-edge subdivision. Let $w_{(t_i,v)}$ be the barycentric weight with respect to the $v$-th vertex of the $i$-th sub-triangle in the $t$-th triangle and $s_{t_{i_0}}$ be the barycentre of the sub-triangle. The integral of the function for the triangle can be approximated as

 \begin{align}
    \int_{|T_{(t,\cdot)}|} f(s_{t_0}) \, ds &\approx \frac{|T_{(t,\cdot)}|}{n}  \sum_{i=1}^n f(s_{t_{i_0}}) \nonumber \\
    &= \frac{|T_{(t,\cdot)}|}{n} \sum_{i=1}^n \sum_{v=1}^3 f_{(t_i,v)} w_{(t_i,v)}  \nonumber \\
    &= \sum_{v=1}^3 f_{(t,v)} \underbrace{\frac{|T_{(t,\cdot)}|}{n}\sum_{i=1}^n w_{(t_i,v)}}_{:=w_{(t,v)}} \label{eq:fw}
\end{align}
In the scenario of removing a part of the integration via triangle dissection, say due to out of the boundary $\partial \Omega$, we simply set the weights to zero. For instance, the line segment $s_{13_c}s_{12_c}$ cuts through the mesh, thus $\Delta s_{1}s_{13_c}s_{12_c}$ is removed and the weight of $s_{1_0}$ set to zero in Figure \ref{fig:mir}. Refer to \citet{hjelle_triangulations_2006} for further details of the algorithm.

\begin{figure}[htp]
    \centering
    \includegraphics[width=\textwidth,
    trim={ 1cm 0cm 1cm 0cm},clip]{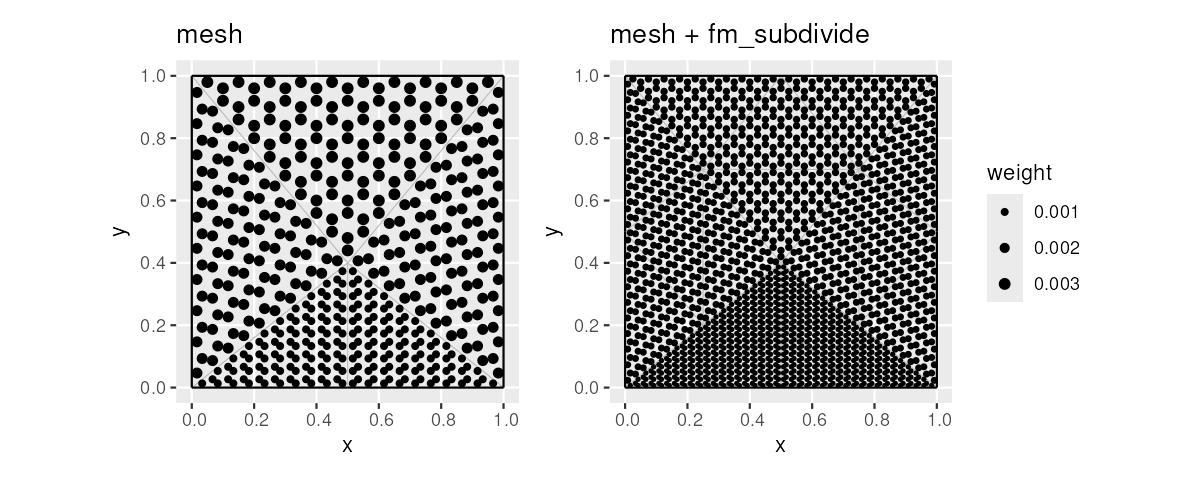}
    \caption{An example of integration points weights given a triangulation over a unit square}
    \label{fig:intpts}
\end{figure}

If we consider the function $f(\cdot)$ to be nonlinear, we can improve the approximation by subdividing each mesh through the mid-edge subdivision thus with more integration points, each with less weight (via \texttt{fmesher::fm\_subdivide} function, see Figure \ref{fig:intpts}). When the sub-triangles are sufficiently fine in scale, we can assume piecewise linear sub-triangles and approximate the integral. Gaussian quadrature rule is considered as well but reweighing integration points on the corners complicates the rule when there is a cut. 

A sample code of the \logsumexp \ mapper is shown below. 

\begin{Schunk}
\begin{Sinput}
R> logsumexp_mapper <- bm_logsumexp(
+    rescale = FALSE,
+    n_block = nrow(sf_object)
+  )
R> formula <- ~ ibm_eval(logsumexp_mapper,
+    input = list(block = .block, weights = weight),
+    state = state
+  )
\end{Sinput}
\end{Schunk}

% <<int_pts_print, include=FALSE, fig=TRUE, echo=FALSE>>=
%# print(p1 + p2 + patchwork::plot_layout(ncol = 2, guides = "collect") & scale_size_area(max_size = 2, limits = c(range[1],range[2])))
% @

\subsection{Degree of Freedom (DoF)}\label{appendix:dof}
Here, we provide the technical details of the degree of freedom (Dof) mentioned in Section \ref{sec:discret}. We use the notations from Section \ref{appendix:int}.
The DoF is a measure of the flexibility of a function space. We simplify and rewrite equation \eqref{eq:psi} as
 \begin{equation}
         \log \lambda(\s) = \eta(\s) := u(\s) = \sum^{K}_{k=1} \psi_k(\s) u_k, \label{eq:varpsi} 
 \end{equation}
for some weight $u_k$. 
Given the volume of the bounded domain $|\Omega|$ is fixed, we only consider the latter two terms in equation \eqref{eq:loglik}, and combining with equation \eqref{eq:intsch},  
\begin{equation}
  -\int_{\Omega} \lambda (\s) \,d\s + \sum_{i=1}^n \log \lambda (y_i) \approx 
  - \sum_{t}\sum_{v \in \mathcal{T}_t } w_{(t,v)} \lambda_{(t,v)} + \sum_{i=1}^n \sum_{k=1}^K u_k \psi_k (y_i),
\end{equation} 
where $\{u_1, \dots, u_k\}$ is defined in equation \eqref{eq:psi}. We can see observations are pulling the likelihood upwards, hence the locations of the integration points are crucial to bound the estimation of the likelihood.

By placing the integration points at proper locations in the domain space, we can bound the likelihood specified in equations \eqref{eq:loglik} and \eqref{eq:loglik_count} . We now illustrate how the construction in equation \eqref{eq:fw} is crucial to obtain stable estimation for the numerical integration of the likelihood function. 

\begin{figure}[H]
    \centering
    \includegraphics[width=\textwidth]{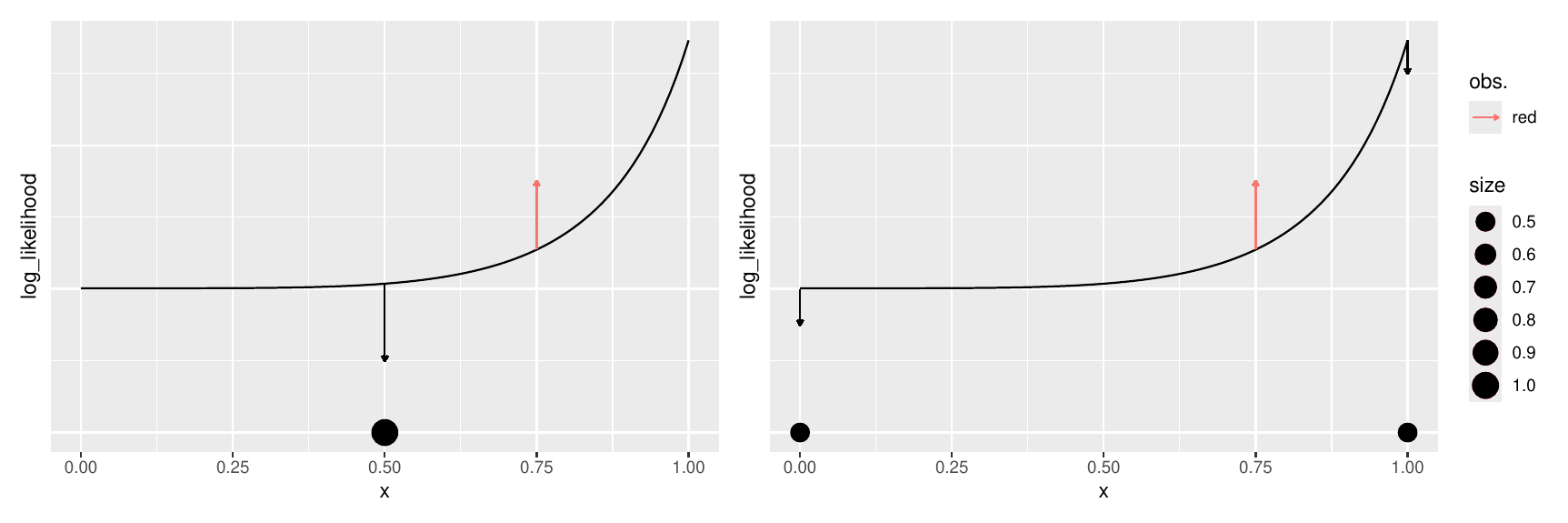}
    \caption{Unstable integration point (left) and stable integration point (right) with respect to log-likelihood for a single observation point (red). Arrow: observed contribution (red), domain contribution (black).}
    \label{fig:int}
\end{figure}

We illustrate unstable and stable integration schemes using a one-dimensional mesh on the range $[0,1]$ (see Figure \ref{fig:int}). For the unstable scheme, a single integration point is placed at $x=0.5$, whereas the stable scheme uses two points at $x=0$ and $x=1$. In both cases, there is one observation point (indicated by the red arrow in the figure, which pulls the log-likelihood upwards). As shown in Figure \ref{fig:int}(left), the single-point integration scheme results in unbounded integration. Since we put the integration point at the end points, we can stabilise the integration. This one dimensional mesh example translates to the integration points in triangular mesh formulation. The message that we want to convey here is with \texttt{fmesher::fm\_subdivide} which subdivides an existing mesh, we can improve the approximation of likelihood by placing the more integration points evaluated at more locations.

\section{Simulation}
Here, we provide the implementation details and plots of the simulation study mentioned in Sections \ref{sec:computation}, \ref{sec:mod} and \ref{sec:sim}. 
\subsection{Implementation Details of \texttt{fmesher::fm\_extensions}}\label{appendix:fmext}
Here, we provide the implementation details of the function \texttt{fmesher::fm\_extensions} mentioned in Section \ref{sec:data}.
The function \texttt{fmesher::fm\_extensions} constructs inner and outer boundaries of a mesh. internally calls \texttt{fmesher::fm\_nonconvex\_hull} to create nonconvex hull over the domain. 

\begin{Schunk}
\begin{Sinput}
R> # d1, d2: some distances from the boundary
R> mesh <- fm_mesh_2d_inla(...,
+    boundary = fm_extensions(
+      bnd,
+      c(d1, d2)
+    )
+  )
\end{Sinput}
\end{Schunk}

For this reason, the covariate field is simulated over the mesh domain for the purpose of easy handling. In real application, we might not have data at the buffered region that \hfill \break \texttt{inlabru::bru\_fill\_missing} can help fill missing data with nearest neighbour values.

\subsection{Sampling from an Inhomogeneous Poisson Process} \label{appendix:sample.lgcp}
Here, we provide the implementation details of the function \texttt{inlabru::sample.lgcp} mentioned in Section \ref{sec:data}. 
The function \texttt{inlabru::sample.lgcp} is a function for sampling from an inhomogeneous Poisson process. The log intensity has to be provided via its values at the vertices of the mesh elements. In between mesh nodes the log intensity is assumed to be linear. The function is called as follows:

\begin{Schunk}
\begin{Sinput}
R> # mesh: An INLA::inla.mesh object
R> # loglambda: A vector of log intensities at the mesh vertices
R> # samplers: A sf or inla.mesh object where simulated points
R> #           that fall outside these polygons are discarded.
R> sample.lgcp(mesh, loglambda, samplers)
\end{Sinput}
\end{Schunk}

The algorithm is based on rejection sampling. The maximum of the \texttt{loglambda}, denoted by \texttt{wmax}, is input to simulate a Poisson random variable based on the product of the areas of the mesh elements and the maximal for each mesh element. If this random variable is larger than zero, another uniform random variable is simulated based on this Poisson random variable. The point(s) is only accepted with a probability that the uniform random variable is smaller than $\exp(\texttt{loglambda} - \texttt{wmax})$. This is repeated for all mesh elements within the samplers. 

\subsection{Simulated Point Pattern}\label{appendix:ptsobs}
Here, we provide the plots of the simulated point patterns mentioned in Section \ref{sec:data}. The point pattern observations over Nepal are simulated via \texttt{sample.lgcp} (see Appendix \ref{appendix:sample.lgcp}) according to equation \eqref{eq:point} with the linear predictor $\eta(\s) = \beta_0 + \beta_x x(\s) + u(\s),$
where $x(\s)$ is a continuous covariate field formulated as $({s_1}^2 - {s_2}^2)\exp\left[-\frac{1}{2}({s_1}^2 +{s_2}^2)\right]; \; s_1 \in [-4,4],\, s_2 \in [-2,2]$ projected to Nepal, and $u(\s)$ is defined in equation \eqref{eq:psi}. We set $\beta_0 = -7$, $\beta_x = -6$ and $u(\s)$ with $\rho = 50$ and $\sigma = 0.5$. The nonlinear transformation of the covariate field is via a nonlinear function $f(\cdot)$, i.e.\ $\log \check{\lambda}(\cdot) := \check{\eta}(\cdot) = \beta_0 + \beta_x  f\left[x(\cdot)\right] + u(\cdot)$. We construct $f(\cdot)$ as a nonlinear function, $f\left[x(\cdot)\right] = b^{-1} \exp{\left[a x(\cdot)\right]}-c,$  where we set $a = 3$, $b = 9$ and $c = 3$.

\begin{figure}[H]
    \centering
        \includegraphics[width=\textwidth, trim={ 0 0cm 0 0cm},clip]{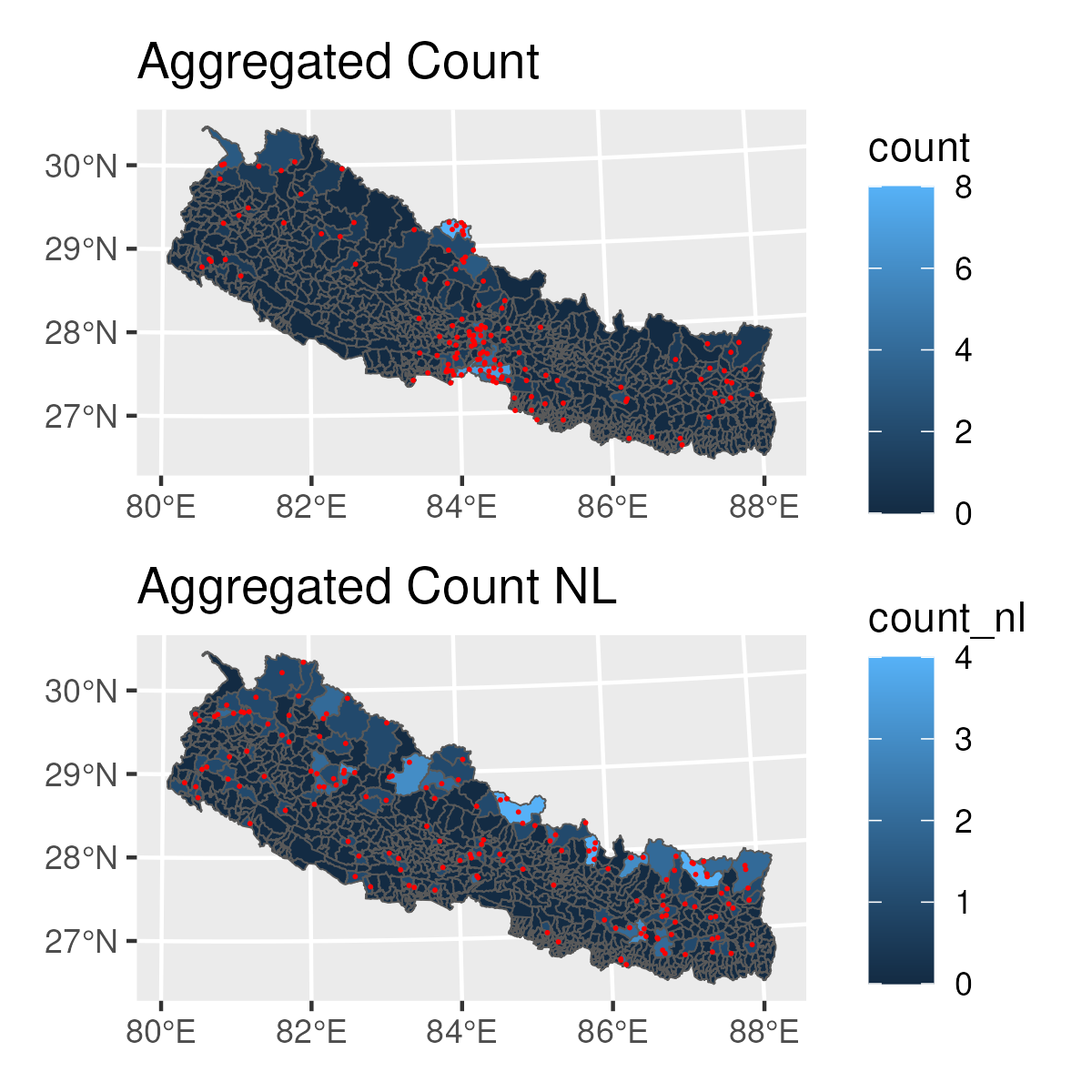}
    \caption{Simulated point pattern (red dots) and aggregated count observations from covariate field (top) and covariate field after nonlinear transformation (bottom). The underlying intensity field of the top panel follows equation \eqref{eq:eta_linear} in the main text while the bottom follows equation \label{eq:mm} in the main text.}
    \label{fig:ptsobs}
\end{figure}

% \subsection{lgcp.sample algorithm}
% The \texttt{lgcp.sample}

\subsection{Sampled Covariate Points}\label{appendix:z_locs}
Here, we illustrate the plot of Incomplete covariate field for Point Values (PointVal) mentioned in Section \ref{sec:ic}. We sample one point from each administrative district and evaluate the covariate field at that point.

\begin{figure}[H]
    \centering
        \includegraphics[draft=FALSE, width=\textwidth, trim={ 0 0cm 0 0cm},clip]{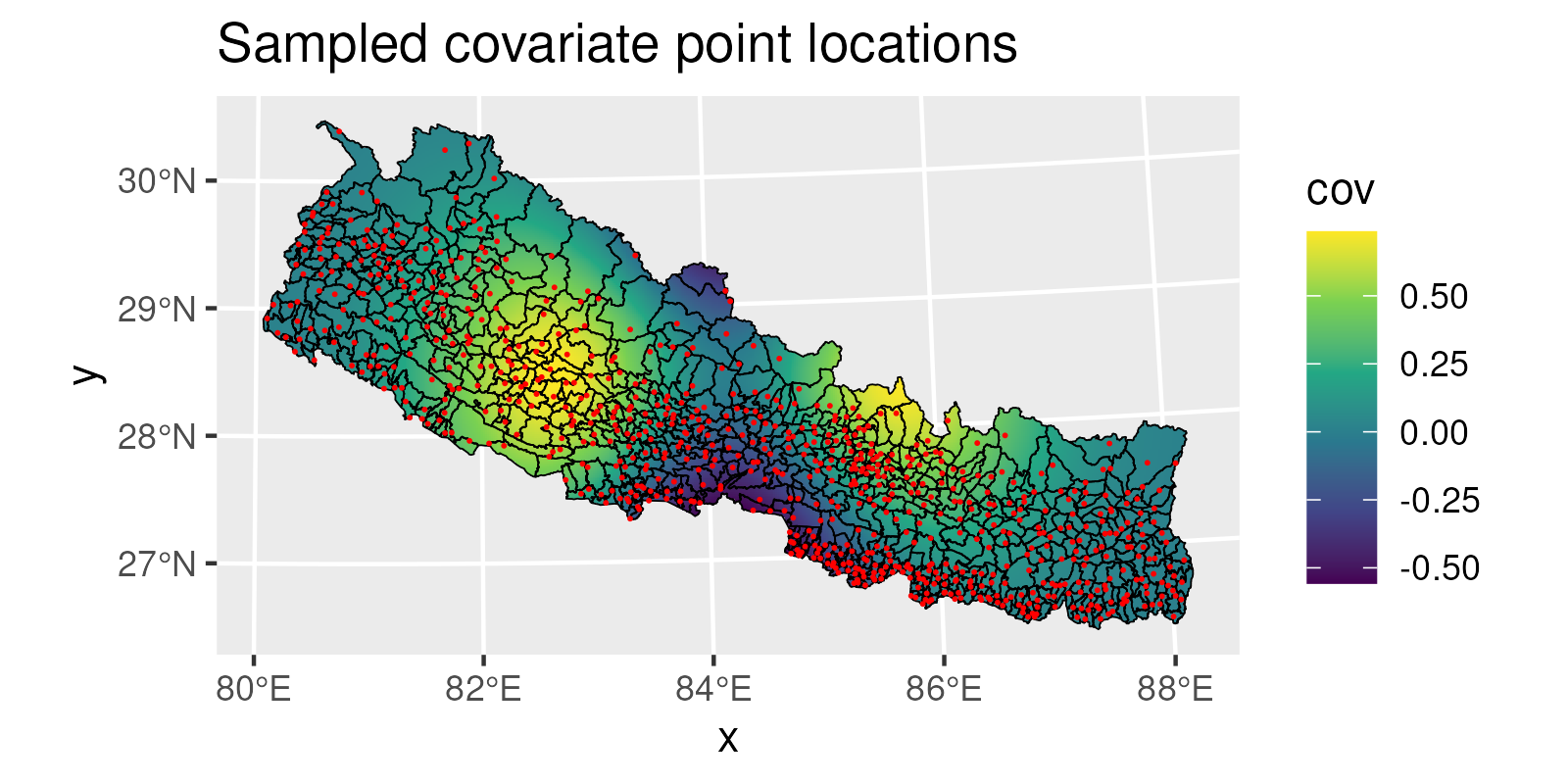}
    \caption{Sampled covariate points (red dots) with the covariate field.}
    \label{fig:z_locs}
\end{figure}

\subsection{PC Priors}\label{appendix:pcprior}
Here, we provide the implementation code for the Penalised Complexity (PC) priors mentioned in Sections \ref{sec:mod}.
The PC priors are implemented in the \\ \texttt{INLA::inla.spde2.pcmatern} function.
The PC priors are used to set the range and standard deviation of the Matérn covariance function, following equation \eqref{eq:matern2}. 

In Sections \ref{sec:agg} and \ref{sec:icap}, the PC priors for \textit{mesh (i)} are set to be $\mathbb{P}(\rho < 4) = 0.1$ and $\mathbb{P}(\sigma > 2) = 0.1$. 

In Section \ref{sec:2stage}, the PC priors for \textit{mesh (ii)} are set to be $\mathbb{P}(\rho < 8) = 0.1$ and $\mathbb{P}(\sigma > 2) = 0.1$.

The PC priors are implemented for \textit{mesh (i)} as follows:

\begin{Schunk}
\begin{Sinput}
R> # mesh_fm: fm_mesh_2d object; mesh (i) in Sec 5
R> INLA::inla.spde2.pcmatern(mesh_fm,
+    prior.range = c(4, 0.1),
+    prior.sigma = c(2, 0.1)
+  )
\end{Sinput}
\end{Schunk}

%% FL: Text added to append.Rnw in July, but wasn't in appendice.Rnw
% We estimate both the covariate field $\hat{x}(\cdot)$ with \textit{mesh (i)}
% and Mat\'ern random field $u(\cdot)$ with \textit{mesh (ii)} from
% Section~\ref{sec:data} for all the models under the incomplete covariate field
% scenarios respectively (See Section \ref{sec:data} for the mesh setup).
% This significantly reduces the computation time by having a coarser
% \textit{mesh (ii)} and improves the model accuracy by reducing the
% identifiability issue. The finer mesh resolution for the covariate
% field reduces the uncertainty in the second stage. We can impose
% stronger PC priors by setting different sensible values for the
% prior parameters of both fields. Our Bayesian framework grants the
% uncertainty quantification for the estimated covariate field. However,
% it takes longer to fit a one-stage model than a two-stage model
% (see Appendix \ref{appendix:comptime}) due to these confounding spatial terms.

\subsection{Rasterisation and Meshing}\label{appendix:rast}
Here, we provide the implementation details of the rasterisation of covariate field and meshing, mentioned in Sections \ref{sec:datares} and \ref{sec:data}. The R scripts of full implementation are mesh.R and covariate\_rast.R on \url{https://github.com/enoch26/misalignedata}. We illustrate the outline of the code below. 

\begin{Schunk}
\begin{Sinput}
R> # INPUT
R> # df: data.frame of covariate field with x, y, and
R> #     covariate values columns
R> # bnd: sf object of boundary
R> # OUTPUT
R> # nepal_rast: SpatRaster object of covariate field;
R> #             RastFull in Sec 5
R> # nepal_rast_agg: SpatRaster object of aggregated (10x)
R> #                 covariate field; RastAgg in Sec 5
R> 
R> nepal_rast <- terra::rast(df,
+    type = "xyz",
+    crs = fm_crs(bnd)
+  )
R> 
R> nepal_rast_agg <- terra::aggregate(nepal_rast,
+    fact = 10, fun = mean,
+    na.rm = TRUE
+  )
R> 
R> # INPUT
R> # lattice: sf object of lattice points for mesh (i);
R> #           the resolution of the lattice points has taken
R> #           the resolution of nepal_rast into consideration;
R> #           see comparison of mesh and raster in Table 1
R> # lattice2: sf object of lattice points for mesh (ii);
R> #           1.961 times apart from lattice in terms of
R> #           edge_len
R> # edge_len(2): numeric; edge length of the mesh (i)/(ii) element
R> # bnd_out: sf object of outer boundary extension
R> # OUTPUT
R> # mesh_fm: fm_mesh_2d object; mesh (i) in Sec 5
R> # mesh_fmc: fm_mesh_2d object; mesh (ii) in Sec 5
R> 
R> mesh_fm <- fm_mesh_2d_inla(
+    loc = lattice,
+    max.edge = c(
+      1.1 * edge_len, # 1.1* to avoid interfering with the lattice
+      5 * edge_len
+    ),
+    boundary = fm_extensions(bnd, bnd_out),
+    crs = fm_crs(bnd)
+  )
R> 
R> mesh_fmc <- fm_mesh_2d_inla(
+    loc = lattice2,
+    max.edge = c(
+      1.1 * edge_len2, # 1.1* to avoid interfering with the lattice
+      5 * edge_len2
+    ),
+    boundary = fm_extensions(bnd, bnd_out),
+    crs = fm_crs(bnd)
+  )
\end{Sinput}
\end{Schunk}

% A rule of thumb is that the mesh resolution should not be $10$ times larger than the data resolution in length, i.e.\ \texttt{max.edge} $< \frac{1}{10} \cdot$ correlation range, in the \texttt{fmesher::fm\_mesh\_2d()} argument. % \citep{Hakkon rule of thumb paper somewhere}.

\subsection{Precision Matrix (\texttt{CMatrix})}\label{appendix:pmatrix}
Here we illustrate how to convert upper triangular storage precision matrix $\bm{U}_{\epsilon_x}$ generated from \inla \ into the full symmetric matrix $\hat{\bm{Q}}_{\epsilon_x}$ mentioned in Sections \ref{sec:2stage} and \ref{sec:lim}. 
\begin{align*}
    \adiag (\hat{\bm{Q}}_{\epsilon_x}) &= 
\adiag \left(\bm{U}_{\epsilon_x} + \bm{U}_{\epsilon_x}^\intercal \right) \\ 
\diag(\hat{\bm{Q}}_{\epsilon_x}) &= \diag(\bm{U}_{\epsilon_x}),
\end{align*}
where $\adiag$ denotes the anti-diagonal matrix. The implementation code for fitting Type 0 generic model and making the full symmetric precision matrix is shown below.

\begin{Schunk}
\begin{Sinput}
R> # fit: a INLA/inlabru::bru object
R> # cmp: inlabru model component object
R> # make_sym function to make the precision matrix symmetric
R> make_sym <- function(Q) {
+    d <- diag(Q)
+    Q <- (Q + t(Q))
+    diag(Q) <- d
+    return(Q)
+  }
R> Q_hat <- make_sym(fit$misc$configs$config[[1]]$Q)
R> cmp <- ~ covariate(geometry,
+    mapper = bru_get_mapper(matern),
+    model = "generic0", Cmatrix = Q_hat
+  )
\end{Sinput}
\end{Schunk}

Type 0 generic model implements a precision matrix $\bm{Q} = \tau \hat{\bm{Q}}_{\epsilon_x}$ as in equation \ref{eq:xhat}, where $\tau$ is a precision parameter.

We did not explore different priors for the precision matrix in this paper. However, different initial values can be set to explore the effect of the priors on the estimation.

\subsubsection{Creating a Mesh for Prediction \texttt{fm\_pixels}}\label{appendix:fm_pixels}
In Section \ref{sec:sim_scenarios}, we precompute the point locations for rasterised mesh elements at the outset. Details are provided in the GitHub repository (see Section \ref{appendix:reproducibility}).
Here, we present an alternative method to create a rasterised mesh for prediction mentioned in Section \ref{sec:sim_scenarios}. The function \texttt{fmesher::fm\_pixels} can generate lattice locations from a rasterised mesh for prediction. A sample code is shown below. 

\begin{Schunk}
\begin{Sinput}
R> # mesh: the mesh object
R> # dims: the dimensions of the raster
R> # mask: the boundary sf object
R> fm_pixels(mesh,
+    dims = rep(250, 2),
+    mask = bnd_buff, format = "sf"
+  )
\end{Sinput}
\end{Schunk}

\subsection{One- and Two-stage Uncertainty Quantification}\label{appendix:uncertainty}

Here, we illustrate the analytical and computational approaches for accessing the stability of the model mentioned in Section \ref{sec:plikup}.

\subsubsection{Analytical Approach for Uncertainty Quantification}\label{sec:analytical}
Here, we try to find a closed form solution for the derivatives of the density with respect to the coefficient and the uncertainty term. 
\begin{align*}
    \frac{\partial \log p(\theta|x)} {\partial \epsilon_{x_j}} &= \sum_j (-e^{\eta_j} +y_j) (\beta_x) - \frac{\epsilon_{x_j}}{ \sigma^2_{x_j|z_p}} = 0  \\
    \frac{\partial^2 \log p(\theta|x)} {\partial \epsilon_{x_j}^2} &= \sum_j -e^{\eta_j} \beta_x^2 - \frac{1}{ \sigma^2_{x_j|z_p}} <0   \\
    \frac{\partial \log p(\theta|x)}{\partial \beta_x} &= \sum_i (-e^{\eta_i} +y_i) (\hat{x}_i + \epsilon_{x_i}) = 0  \\
    \frac{\partial^2 \log p(\theta|x)}{\partial \beta_x^2} &= \sum_i (-e^{\eta_i} ) (\hat{x}_i + \epsilon_{x_i})^2 \leq 0  \\
    &\Rightarrow \sum_i \frac{\epsilon_{x_i}(\hat{x}_i + \epsilon_{x_i})}{\beta_x \sigma^2_{x_i|z_i}}  = 0 \\
&\Rightarrow \sum_i \frac{(\hat{x}_i \epsilon_{x_i} + \epsilon^2_{x_i})}{\sigma^2_{x_i|z_i}}  = 0 \\
&\Rightarrow \sum_i \left(\frac{\hat{x}_i + 2\epsilon_{x_i}}{2\sigma_{x_i|z_i}}\right)^2  = \sum_i \left(\frac{\hat{x}_i}{2\sigma_{x_i|z_i}}\right)^2   
\end{align*}

Hence, we cannot derive a closed form solution for the derivatives.

\subsubsection{Profile Likelihood for Uncertainty Quantification}\label{sec:plik}
\begin{figure}[H]
    \centering
    \includegraphics[width=\textwidth]{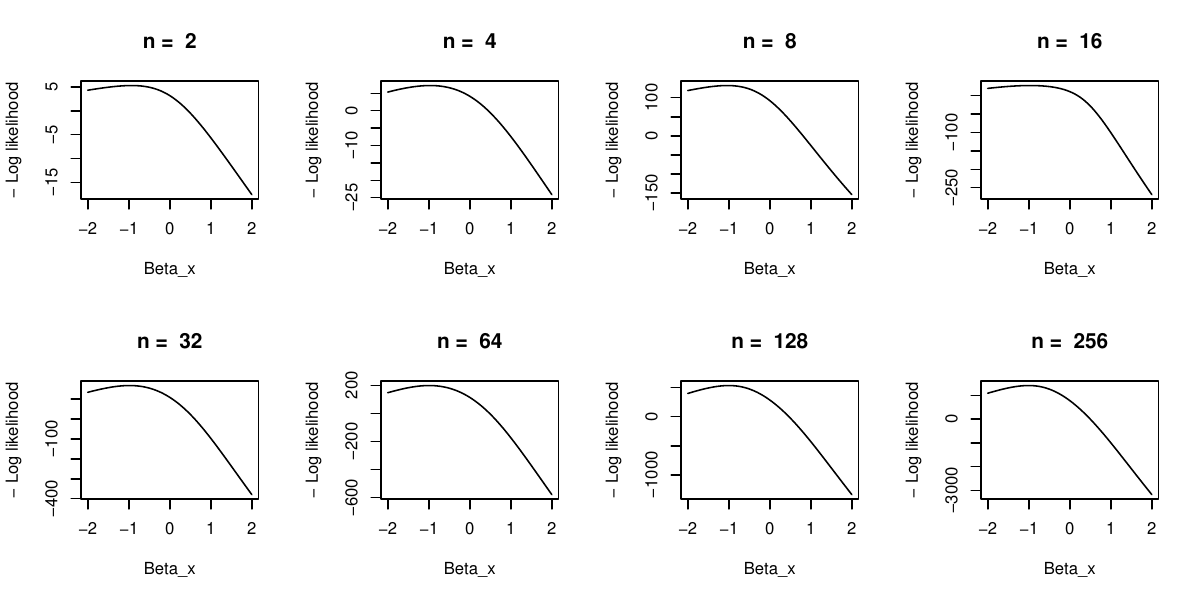}
    \caption{Profile likelihood optimisation of $\beta_x$ for single-stage Joint Uncertainty (JU) approach.}
    \label{fig:plik_single}
\end{figure}

\begin{figure}[H]
    \centering
    \includegraphics[width=\textwidth]{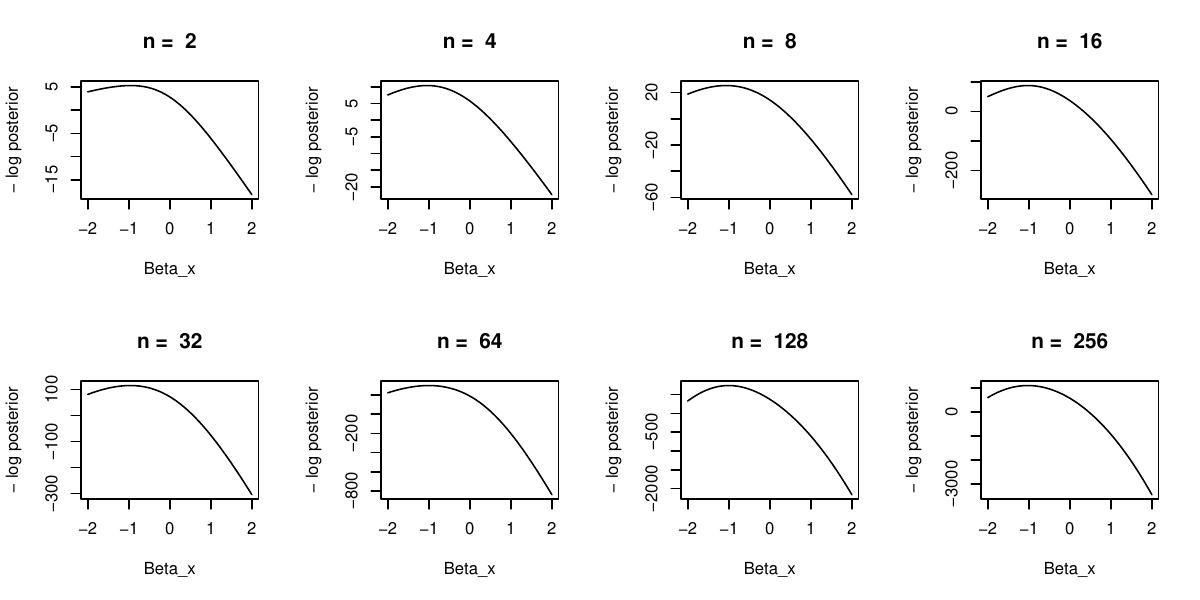}
    \caption{Profile likelihood optimisation of $\beta_x$ for two-stage Value Plugin (VP) approach.}
    \label{fig:plik_plugin}
\end{figure}

\begin{figure}[H]
    \centering
    \includegraphics[width=\textwidth]{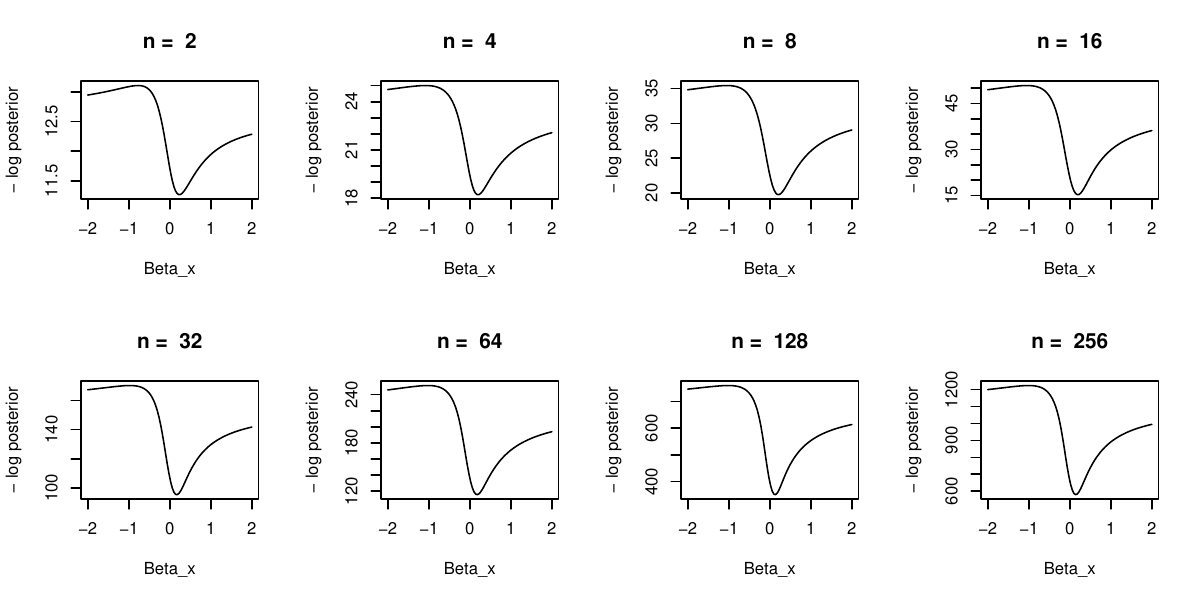}
    \caption{Profile likelihood optimisation of $\beta_x$ for two-stage Uncertainty Plugin (UP) approach.}
    \label{fig:plik_uncertainty}
\end{figure}

Since we cannot derive a closed form solution, we seek to optimise the profile likelihood for the coefficient. Here, we illustrate one dimensional toy example for JU, VP and UP methods. We use the profile likelihood to optimise $\beta_x$ with $n_z \in \{2, 4, 8, 16, 32, 64, 128, 256\}$, $ \argmax_{\beta_x} \left(\argmax_{(\beta_0, \epsilon_x)}(\mathfrak{p})\right)$ as in equation \eqref{eq:logpost} in Section \ref{sec:plikup}. In Figures \ref{fig:plik_single} and \ref{fig:plik_plugin}, there are only one global maximum for JU and VP methods. However, in Figure \ref{fig:plik_uncertainty}, there is another one local maximum for $\beta_x$ approaching infinity for UP method. This shows that one should be careful about the initial starting points when using the UP method. Nonetheless, we did not encounter optimisation issues in the simulation.

\subsection{Covariate Estimation under Incomplete Covariate Field Scenario}\label{appendix:score_cov}
The table and plots refer to Section \ref{sec:cov_ic}. 

\begin{table}[H]
\centering
% \footnotesize
\renewcommand{\arraystretch}{1.2}
\begin{tabular}{|c|c|c|*{2}c|c|}
\hline
Cov. & Obs. & Method & MSE & MDS & $\Bar{X}(\cdot)$\\ \hline
\multirow{5}{*}{PolyAgg} & $\mathcal{Y}$, $\check{\mathcal{Y}}$, $\mathcal{N}$, $\check{\mathcal{N}}$ & VP and UP & 
7.915$\times 10^{-6}$ & -1.115$\times 10^{1}$ & \multirow{10}{*}{0.18262}\\ 
\cline{2-5} 
& $\mathcal{Y}$ & \multirow{4}{*}{JU} & 7.571$\times 10^{-6}$ & \textbf{-1.116}$\bm{\times 10^{1}}$ & \\  
& $\mathcal{N}$ & & 7.575$\times 10^{-6}$ & \textbf{-1.116}$\bm{\times 10^{1}}$ & \\ 
& $\check{\mathcal{Y}}$ &  & 7.562$\times 10^{-6}$ & \textbf{-1.116}$\bm{\times 10^{1}}$ & \\  
& $\check{\mathcal{N}}$ & & \textbf{7.513}$\bm{\times 10^{-6}}$ & \textbf{-1.116}$\bm{\times 10^{1}}$ & \\ 
\cline{1-5}

\multirow{5}{*}{PointVal} & $\mathcal{Y}$, $\check{\mathcal{Y}}$, $\mathcal{N}$, $\check{\mathcal{N}}$ & VP and UP
   & 1.309$\times 10^{-3}$ & -5.842 & \\ \cline{2-5} 
&  $\mathcal{Y}$
   & \multirow{4}{*}{JU} & \textbf{1.116}$\bm{\times 10^{-3}}$ & \textbf{-5.908}  & \\  
& $\mathcal{N}$ & & 1.163$\times 10^{-3}$ & -5.891 & \\ 
& $\check{\mathcal{Y}}$ & & 1.247$\times 10^{-3}$ & -5.861 & \\  
& $\check{\mathcal{N}}$ & & 1.196$\times 10^{-3}$ & -5.883& \\ 
\hline
\end{tabular}

\caption{Table of the Mean Squared Error (MSE) and Mean Dawid-Sebastiani (MDS) scores for the covariate field $x$ across the models under incomplete covariate field scenario. PointVal: Point Values, PolyAgg: Polygon Aggregation, $\Bar{X}(\cdot)$: Mean of Covariate field.}
\label{tab:score_cov}
\end{table} 

\begin{figure}[H]
    \centering
    \includegraphics[draft=FALSE,width=\textwidth]{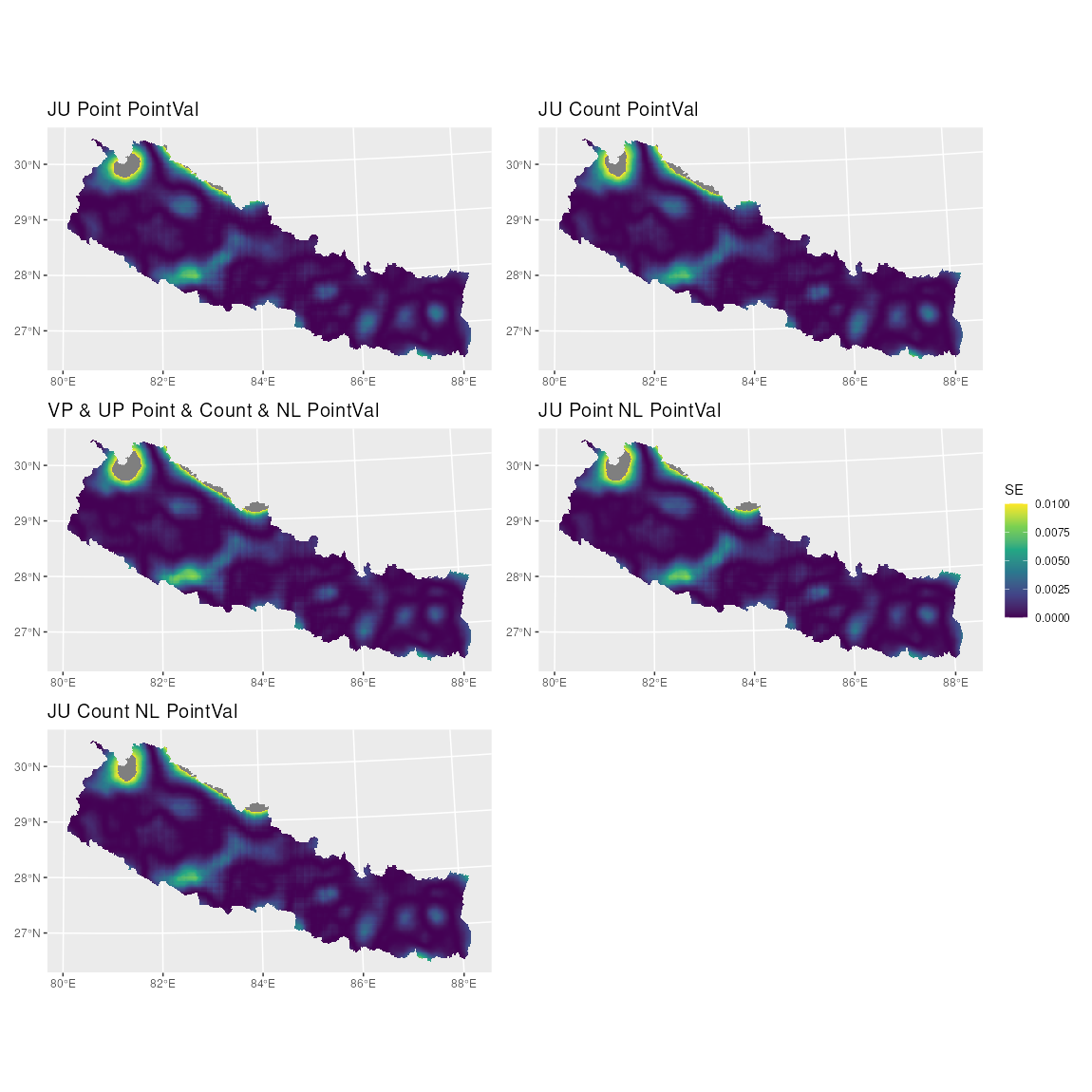}
    \caption{Squared Error (SE) for the covariate field $X$ across the models under incomplete covariate field scenario with observed noises $\sigma_{\epsilon_x}$ (truncated at 0.01). JU: Joint Uncertainty, VP: Value Plugin, UP: Uncertainty Plugin, Point: Observed Point, Count: Aggregated Count, PointVal: Point Values. $\Bar{X}(\cdot) = 0.18262$}
    \label{fig:se_cov}
\end{figure}

\begin{figure}[H]
    \centering
    \includegraphics[draft=FALSE,width=\textwidth]{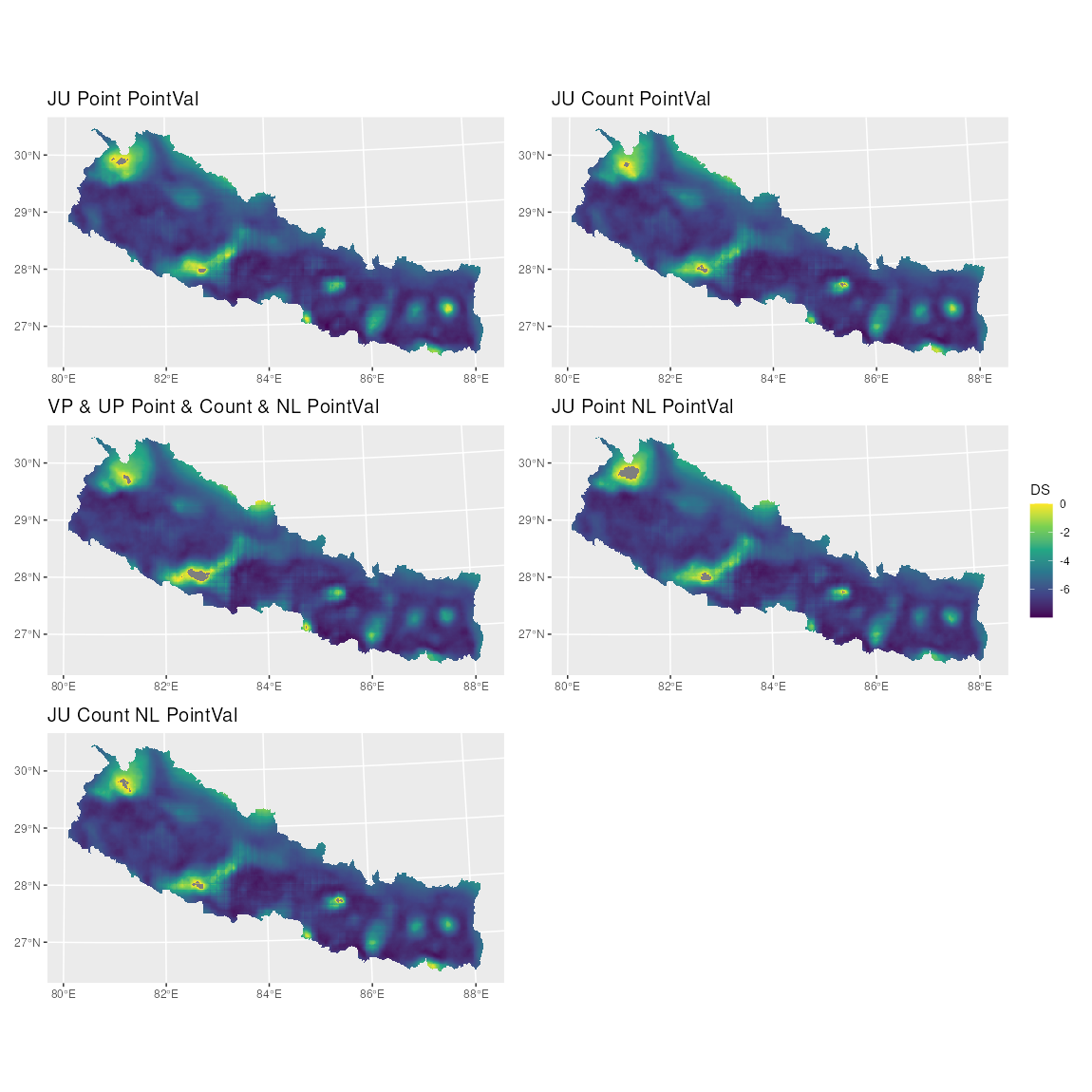}
    \caption{Dawid-Sebastiani (DS) score for the covariate field $X$ across the models under incomplete covariate field scenario with observed noises $\sigma_{\epsilon_x}$ (truncated at 0). JU: Joint Uncertainty, VP: Value Plugin, UP: Uncertainty Plugin, Point: Observed Point, Count: Aggregated Count, PointVal: Point Values. } 
    \label{fig:ds_cov}
\end{figure}

\begin{figure}[H]
    \centering
    \includegraphics[draft=FALSE,width=\textwidth]{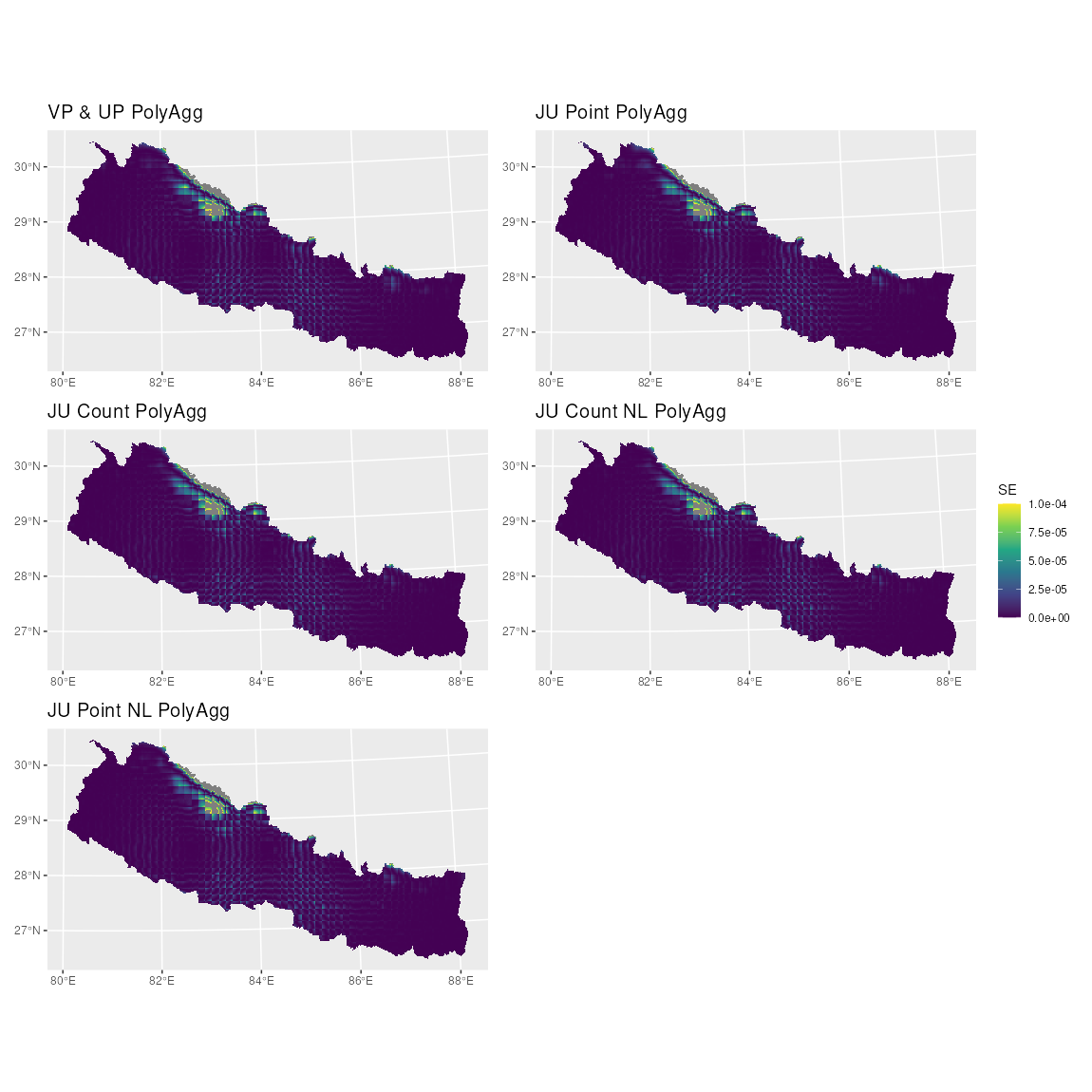}
    \caption{Squared Error (SE) for the covariate field $X$ across the models under incomplete covariate field scenario with observed noises $\sigma_{\epsilon_x}$ (truncated at 1$\times 10^{-4}$). JU: Joint Uncertainty, VP: Value Plugin, UP: Uncertainty Plugin, Point: Observed Point, Count: Aggregated Count, PolyAgg: Polygon Aggregation. $\Bar{X}(\cdot) = 0.18262$}
    \label{fig:se_cov_ap}
\end{figure}

\begin{figure}[H]
    \centering
    \includegraphics[draft=FALSE,width=\textwidth]{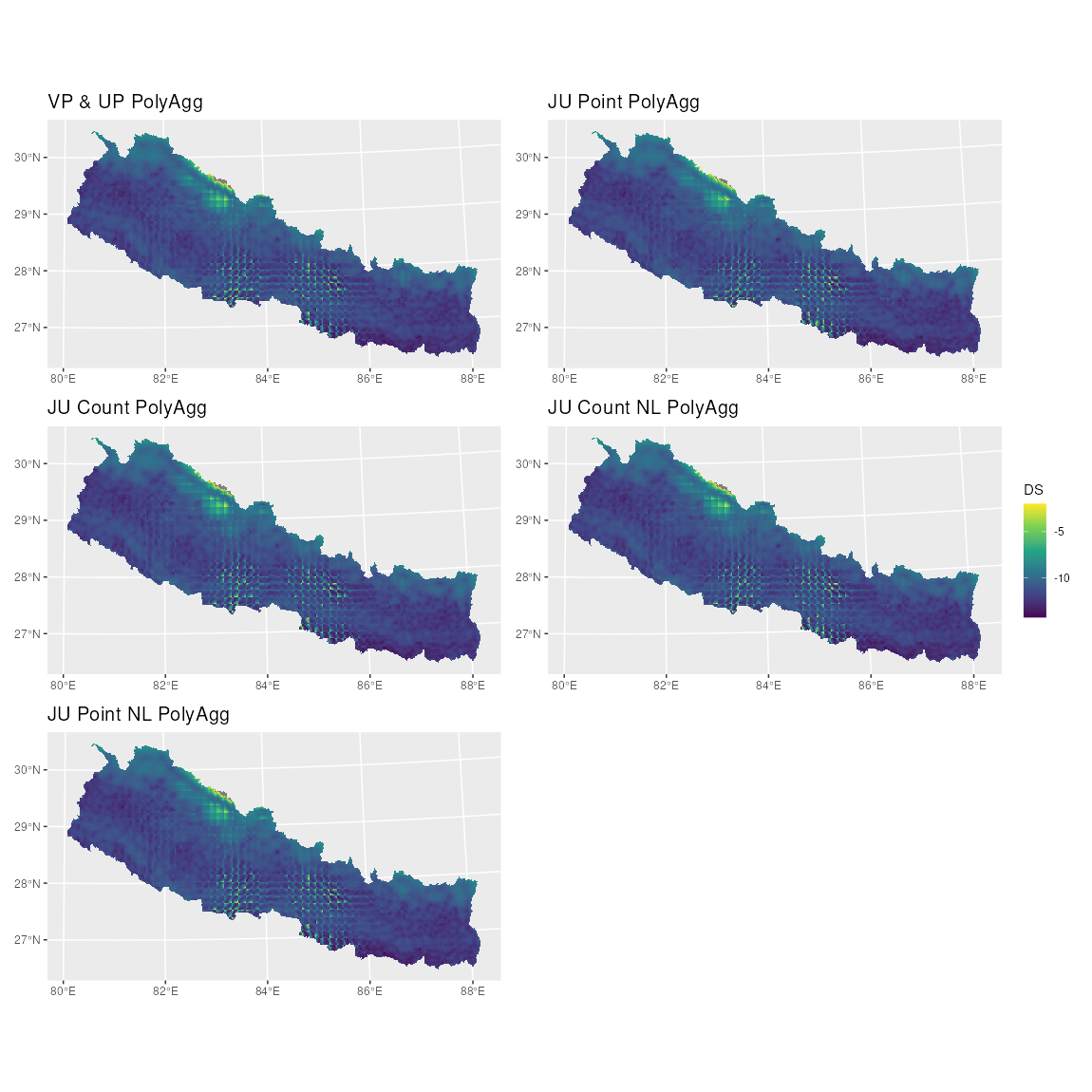}
    \caption{Dawid-Sebastiani (DS) score for the covariate field $X$ across the models under incomplete covariate field scenario with observed noises $\sigma_{\epsilon_x}$ (truncated at 0). JU: Joint Uncertainty, VP: Value Plugin, UP: Uncertainty Plugin, Point: Observed Point, Count: Aggregated Count, PolyAgg: Polygon Aggregation.} 
    \label{fig:ds_cov_ap}
\end{figure}

\subsection{Computation Time}\label{appendix:comptime}
Here refers to Sections \ref{sec:1stage} and \ref{sec:2stage} to compare the computation time across different models.

All numerical experiments for the case study/simulation were performed limited to 10 threads on a Four Intel Xeon E5-2680 v3 2.5GHz, 30M Cache, 9.6 GT/s QPI 192 GB RAM machine (i.e.\ 48 cores in total and 2 threads each core). 
% \textcolor{red}{TODO: collected bar chart to show the computation time in details wit prep process time breakup with different colours in each bar and possibly iterations.}

% latex table generated in R 4.4.1 by xtable 1.8-4 package
% Wed Jul  3 16:13:28 2024
\begin{table}[H]
\centering
\begin{tabular}{rlr}
  \hline
  Model name & Description & Time (sec) \\
  \hline
fit\_pts & Point RastFull & 413.67 \\ 
  fit\_pts\_agg & Point RastAgg & 387.74 \\ 
  fit\_pts\_poly & Point PolyAgg & 412.79 \\ 
  fit\_count & Count RastFull & 1223.40 \\ 
  fit\_agg & Count RastAgg & 1327.17 \\ 
  fitly & Count PolyAgg & 1383.79 \\ 
  fit\_icov\_sd01 & VP and UP point+count NL PointVal & 474.51 \\ 
  fit\_ap & VP and UP PolyAgg & 973.82 \\ 
  fit\_ic\_pts\_sd01 & JU Point PointVal & 5373.32 \\ 
  fit\_ic\_sd01 & JU Count PointVal & 6549.09 \\ 
  fit\_ic\_ap\_pts & JU Point PolyAgg & 10479.75 \\ 
  fit\_ic\_ap & JU Count PolyAgg & 13643.07 \\ 
  fit\_ic\_pts\_sd01\_nl & JU Point NL PointVal & 4822.12 \\ 
  fit\_ic\_sd01\_nl & JU Count NL PointVal & 4950.74 \\ 
  fit\_ic\_ap\_pts\_nl & JU Point NL PolyAgg & 5374.53 \\ 
  fit\_ic\_ap\_nl & JU Count NL PolyAgg & 8100.82 \\ 
  fit\_icov2\_pts\_sd01 & VP Point PointVal & 335.58 \\ 
  fit\_icov2\_sd01 & VP Count PointVal & 475.57 \\ 
  fit\_icov2\_ap\_pts & VP Point PolyAgg & 421.68 \\ 
  fit\_icov2\_ap & VP Count PolyAgg & 487.53 \\ 
  fit\_icov2\_pts\_nl\_sd01 & VP Point NL PointVal & 308.22 \\ 
  fit\_icov2\_nl\_sd01 & VP Count NL PointVal & 609.28 \\ 
  fit\_icov2\_ap\_pts\_nl & VP Point NL PolyAgg & 286.63 \\ 
  fit\_icov2\_nl\_ap & VP Count NL PolyAgg & 613.25 \\ 
  fit\_icov3\_pts\_sd01 & UP Point PointVal & 1540.68 \\ 
  fit\_icov3\_sd01 & UP Count PointVal & 2899.59 \\ 
  fit\_icovap\_pts & UP Point PolyAgg & 2255.58 \\ 
  fit\_icovap & UP Count PolyAgg & 3189.34 \\ 
  fit\_icov3\_pts\_nl\_sd01 & UP Point NL PointVal & 2605.85 \\ 
  fit\_icov3\_nl\_sd01 & UP Count NL PointVal & 2840.81 \\ 
  fit\_icovap\_nl & UP Point NL PolyAgg & 3485.97 \\ 
  fit\_icovap\_pts\_nl & UP Count NL PolyAgg & 3670.17 \\ 
   \hline
\end{tabular}

\caption{Table of computation times for different models. JU: Joint Uncertainty, VP: Value Plugin, UP: Uncertainty Plugin, Point: Observed Point, Count: Aggregated Count, PointVal: Point Values, PolyAgg: Polygon Aggregation.}\label{tab:mod}
\end{table}

\newpage
% \afterpage{\clearpage}

% \begin{sidewaysfigure}[H]
    \begin{figure}[H]
    \centering
    \includegraphics[width=\textwidth]{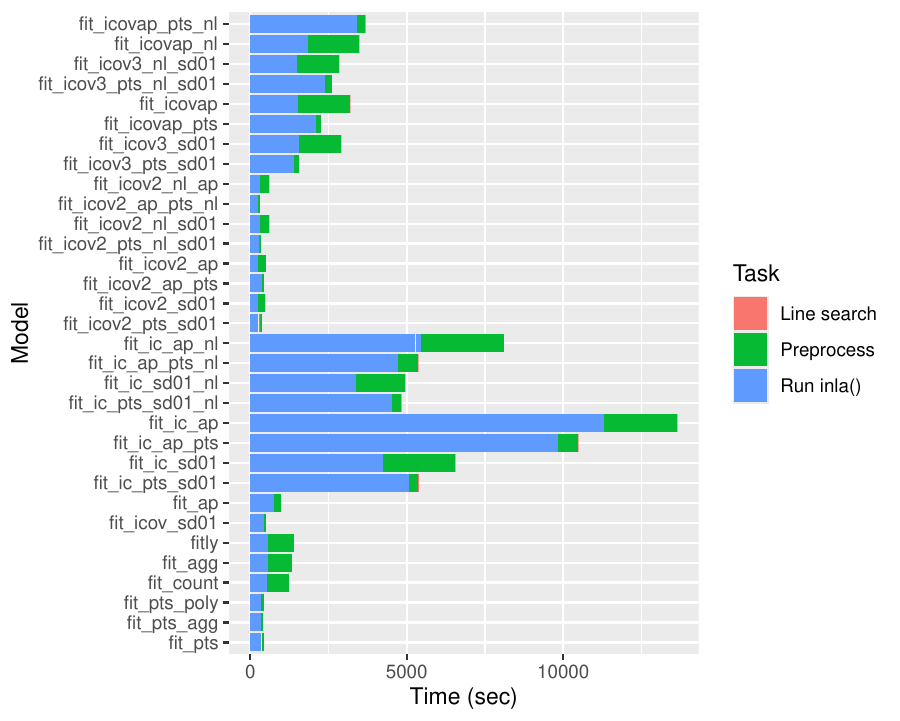}
    \caption{Computation times for different models.}
    \label{fig:ctime}
    \end{figure}
% \end{sidewaysfigure}

\section{Convergence Diagnostic for Linearisation}\label{appendix:conv}
  % "fit_agg"               ,"fit_ap"                ,"fit_count"            ,
  % "fit_ic_ap"             ,"fit_ic_ap_nl"          ,"fit_ic_ap_pts"        ,
  % "fit_ic_ap_pts_nl"      ,"fit_ic_pts_sd01"       ,"fit_ic_pts_sd01_nl"   ,
  % "fit_ic_sd01"           ,"fit_ic_sd01_nl"        ,"fit_icov2_ap"         ,
  % "fit_icov2_nl_ap"       ,"fit_icov2_nl_sd01"     ,"fit_icov2_ap_pts"     ,
  % "fit_icov2_ap_pts_nl"   ,"fit_icov2_pts_nl_sd01" ,"fit_icov2_pts_sd01"   ,
  % "fit_icov2_sd01"         ,"fit_icov3_nl_sd01"     ,"fit_icov3_pts_nl_sd01",
  % "fit_icov3_pts_sd01"     ,"fit_icov3_sd01"         ,"fit_icovap"           ,
  % "fit_icovap_nl"         ,"fit_icovap_pts"        ,"fit_icovap_pts_nl"    ,
  % "fit_icov_sd01"          ,"fit_pts"               ,"fit_pts_agg"          ,
  % "fit_pts_poly"          ,"fitly"  
In Section \ref{sec:lim}, we discussed the limitations on convergence iterations. The convergence plots of the models across iterations are shown below. The four panels of convergence diagnostics:
\begin{itemize}
  \item Tracks: Mode and linearisation values for each effect;
  \item Mode - Lin: Difference between mode and linearisation values for each effect;
  \item $|$Change$|$/sd: Absolute change in mode and linearisation values divided by the standard deviation for each effect;
  \item Change \& sd: Absolute change in mode and linearisation values and standard deviation for each effect.
\end{itemize}
Refer to Table \ref{tab:mod} in Appendix \ref{appendix:comptime} for the model names and descriptions.
\newpage
% \begin{figure}[H]
\foreach \x in {fit_agg,fit_ap,fit_count,fit_ic_ap,fit_ic_ap_nl,fit_ic_ap_pts,fit_ic_ap_pts_nl,fit_ic_pts_sd01,fit_ic_pts_sd01_nl,fit_ic_sd01,fit_ic_sd01_nl,fit_icov2_ap,fit_icov2_nl_ap,fit_icov2_nl_sd01,fit_icov2_ap_pts,fit_icov2_ap_pts_nl,fit_icov2_pts_nl_sd01,fit_icov2_pts_sd01,fit_icov2_sd01,fit_icov3_nl_sd01,fit_icov3_pts_nl_sd01,fit_icov3_pts_sd01,fit_icov3_sd01,fit_icovap,fit_icovap_nl,fit_icovap_pts,fit_icovap_pts_nl,fit_icov_sd01,fit_pts,fit_pts_agg,fit_pts_poly,fitly} { 
    % \centering
    
    \includegraphics[width=1.25\textwidth,angle=90,origin=c]{append_figure/bru_plot_\x.eps}
    % \caption{Convergence plots for model \x.}
    % \label{fig:\x}
    % \clearpage
    
}
% \end{figure}
 % \afterpage{\input{ctable}}
 
\section{Reproducibility}\label{appendix:reproducibility}
The results were generated with \texttt{INLA} 25.01.23, \inlabru \ 2.12.0.9005 and \texttt{fmesher} 0.2.0. The shape file of Nepal's municipalities can be downloaded: \url{https://opendatanepal.com/dataset/nepal-municipalities-wise-geographic-data-shp-geojson-topojson-kml/resource/06b90abc-1380-46ed-b529-e455de6d794d}. The reproducible code for the simulation study can be found in the github repository: \url{https://github.com/enoch26/misalignedata}.

\end{appendices}

\end{document}